\documentclass[11pt,a4paper]{article}
\pdfoutput=1
\usepackage{jheppub}

\usepackage{epsfig}
\usepackage{amsmath,amsfonts,mathtools}
\usepackage{amssymb}
\usepackage{verbatim}
\usepackage{bm}
\usepackage{dsfont}
\usepackage[footnotesize]{caption}
\usepackage{subfigure}
\usepackage{textcomp}
\usepackage{calc}
\usepackage{bbm}
\usepackage{xcolor}
\usepackage[utf8]{inputenc}
\usepackage{slashed}
\usepackage{upgreek}
\usepackage{lipsum}
\usepackage{enumerate}
\usepackage[autostyle, english = american]{csquotes}
\usepackage{tikz}
\usepackage{bbold}
\usepackage{placeins}
\MakeOuterQuote{"}
\newcommand{\non}[0]{\nonumber \\}
\newcommand{\bee}[0]{\begin{eqnarray}}
\newcommand{\eee}[0]{\end{eqnarray}}
\newcommand{\comm}[2]{\left[ #1 ,#2 \right]}
\newcommand{\acomm}[2]{\left\{ #1 ,#2 \right\}}

\newcommand{\be}[0]{\begin{equation}}
\newcommand{\ee}[0]{\end{equation}}

\renewcommand{\Re}{\mathfrak{Re}}

\renewcommand{\tilde}{\widetilde}

\newcommand{\M}{\bar{M}}
\newcommand{\ii}{{\rm i}}
\newcommand{\Mu}{\upmu}
\newcommand{\flav}{\mathfrak{f}}
\newcommand{\PP}{p}
\renewcommand{\Im}{{\rm Im}}
\usepackage{hyperref}
\hypersetup{colorlinks=true,urlcolor=blue,linkcolor=red,citecolor=green!60!black}

\title{Low-scale leptogenesis with three heavy neutrinos}

\author[a]{Asmaa Abada,}
\author[b]{Giorgio Arcadi,}
\author[c]{Valerie Domcke,}
\author[d]{Marco Drewes,}
\author[e,f]{Juraj Klaric,}
\author[d]{and Michele Lucente}

\affiliation[a]{Laboratoire de Physique Th\'eorique (UMR8627), CNRS\\ 
Univ. Paris-Sud, Universit\'e Paris-Saclay, 91405 Orsay, France}
\affiliation[b]{Max-Planck-Institut f\"ur Kernphysik (MPIK), 69117 Heidelberg, Germany}
\affiliation[c]{Deutsches Elektronen-Synchrotron (DESY), 22607 Hamburg, Germany}
\affiliation[d]{Centre for Cosmology, Particle Physics and Phenomenology,
Universit\'{e} catholique de Louvain\\
Chemin du Cyclotron 2,
B-1348 Louvain-la-Neuve,
Belgium}
\affiliation[e]{Technische Universit\"at M\"unchen (TUM), Garching, Germany}
\affiliation[f]{Institute of Physics, Laboratory for Particle Physics and Cosmology (LPPC),\\
Ecole Polytechnique F\'{e}d\'{e}rale de Lausanne (EPFL), CH-1015 Lausanne, Switzerland}

\emailAdd{asmaa.abada@th.u-psud.fr}
\emailAdd{arcadi@mpi-hd.mpg.de}
\emailAdd{valerie.domcke@desy.de}
\emailAdd{marco.drewes@uclouvain.be}
\emailAdd{juraj.klaric@epfl.ch}
\emailAdd{michele.lucente@uclouvain.be}

\abstract{
Leptogenesis induced by the oscillations of GeV-scale neutrinos provides a minimal and testable explanation of the baryon asymmetry of the Universe. 
 In this work we extend previous studies invoking only two heavy neutrinos to the case of three heavy neutrinos. We find qualitatively new behaviour as a result of lepton number violating oscillations and decays, strong flavour effects in the washout and a resonant enhancement due to matter effects.
 An approximate global $B - \bar L$ symmetry 
  (representing the difference of baryon and a generalised lepton number) 
 can protect the light neutrino masses from large radiative corrections, while simultaneously providing the ingredients for the resonant enhancement of the lepton asymmetry due to thermal contributions to the heavy neutrino dispersion relations.
 This mechanism is particularly efficient for large heavy neutrino mixing angles near the current experimental limits, a regime in which leptogenesis is not feasible in the minimal scenario with two heavy neutrinos. In this new parameter regime, low-scale leptogenesis is testable by the LHC and other existing experiments.
 }


\arxivnumber{1810.12463}
                                           
\begin{document}
\maketitle



\renewcommand*{\thefootnote}{\arabic{footnote}}
\setcounter{footnote}{0}

\section{Introduction}\label{sec:intro}


All elementary fermions  with the exception of neutrinos are known to exist with both chiralities, left-handed and right-handed, in the Standard Model (SM) of particle physics.
Right-handed neutrinos could, if they exist, explain a number of open puzzles in particle physics as well as in cosmology, cf.~e.g.~\cite{Drewes:2013gca} for an overview.
Most importantly, they can generate non-zero neutrino masses $m_i$ that explain the light neutrino flavour oscillations via the \emph{type-I seesaw mechanism}~\cite{Minkowski:1977sc,GellMann:1980vs,Mohapatra:1979ia,Yanagida:1980xy,Schechter:1980gr,Schechter:1981cv}. 
A key prediction of the seesaw mechanism is the existence of heavy neutrino mass states $N_i$ with masses $M_i\gg m_i$ and weak interactions with the SM leptons $\ell_a$ (with  $a=e,\mu,\tau$) which are suppressed by small mixing angles $\theta_{a i}$.
For $M_i$ below the TeV scale, the $N_i$ can be searched for experimentally. 
The experiments ATLAS \cite{Aad:2011vj,Aad:2015xaa,Aaboud:2018jbr}, CMS \cite{Khachatryan:2015gha,Khachatryan:2016olu,Sirunyan:2018mtv} and LHCb \cite{Aaij:2014aba,Ossowska:2018ybk} at the LHC currently perform such searches in the mass range $M_i>5$ GeV.  
For the $M_i$ below the $W$ gauge  boson mass considered in this work, the sensitivity is expected to improve significantly by using a wider range of signatures and improved triggers \cite{Helo:2013esa,Izaguirre:2015pga,Gago:2015vma,Dib:2015oka,Dib:2016wge,Cottin:2018nms,Abada:2018sfh,Drewes:2018xma}.
Further improvement could be achieved with additional detectors \cite{Kling:2018wct,Helo:2018qej,Curtin:2018mvb}.
In the future, a lepton collider could offer an ideal tool to search for heavy neutrinos with masses below the W mass \cite{Blondel:2014bra,Antusch:2015mia,Abada:2014cca,Asaka:2015oia,Graverini:2015dka,Abada:2015zea,Antusch:2016vyf,Antusch:2016ejd,Caputo:2016ojx,Antusch:2017pkq}.
Searches at smaller masses $M_i < 5$ GeV are preformed at the NA62 experiment \cite{Abada:2016plb,CortinaGil:2017mqf,Drewes:2018gkc} as well as at T2K \cite{lamoureux_mathieu_2018_1300449}, and in the future at SHiP \cite{Alekhin:2015byh,Graverini:2015dka}.

Further motivation for the existence of heavy neutrinos comes from cosmology.
Their Yukawa interactions $F_{a i}$ with the SM flavours $a=e,\mu,\tau$ generally violate $CP$ and can potentially generate a matter-antimatter asymmetry in the primordial plasma that filled the early Universe, which can be converted into a baryon asymmetry by weak sphaleron processes~\cite{Kuzmin:1985mm}.
This mechanism is known as \emph{leptogenesis}~\cite{Fukugita:1986hr} and provides an attractive explanation for the \emph{baryon asymmetry of the Universe} (BAU), which is believed to be the origin of baryonic matter in the Universe at present time (cf.~e.g.~\cite{Canetti:2012zc} for a discussion).
Leptogenesis can either be realised during the freeze-out and decay of the heavy neutrinos~\cite{Fukugita:1986hr} ("freeze-out scenario") or during their production~\cite{Akhmedov:1998qx,Asaka:2005pn} ("freeze-in scenario").
The freeze-in scenario is particularly interesting from a phenomenological viewpoint because it is feasible for masses $M_i$ below the electroweak scale~\cite{Canetti:2012kh}, which are within reach of experiments~\cite{Chun:2017spz}.

The number $n$ of right-handed neutrinos is not constrained by theoretical arguments within the SM.
However, in the context of many gauge extensions of the SM it should equal the number of SM generations ($n=3$) to ensure the anomaly freedom of the theory.
From an experimental viewpoint $n\geq 2$ is needed to explain the two observed light neutrino mass splittings if the type-I seesaw is the sole origin of the light neutrino masses.

Most phenomenological studies of low-scale leptogenesis in the past have focused on the minimal model with $n=2$.
This effectively also describes neutrino mass generation and leptogenesis in the \emph{Neutrino Minimal Standard Model} ($\nu$MSM)~\cite{Asaka:2005pn,Asaka:2005an}, where the third right-handed neutrino is a Dark Matter candidate, and the observational constraints on its properties \cite{Adhikari:2016bei,Boyarsky:2018tvu} imply that it practically decouples.
First estimates of the $N_i$ properties in the $\nu$MSM were made \cite{Shaposhnikov:2008pf} shortly after the viability of the freeze-in mechanism in the minimal setup had been shown \cite{Asaka:2005pn}.
Following a number of conceptual treatments \cite{Anisimov:2010aq,Gagnon:2010kt,Anisimov:2010dk,Garny:2011hg,Garbrecht:2011aw}, the parameter space was first systematically studied in Refs.~\cite{Canetti:2010aw,Canetti:2012vf,Canetti:2012kh}.
Following this, several authors have investigated details of the problem, such as 
the momentum averaging in the kinetic equations \cite{Asaka:2011wq,Ghiglieri:2017gjz,Ghiglieri:2017csp},
the thermal production rates \cite{Anisimov:2010gy,Besak:2012qm,Garbrecht:2013gd,Garbrecht:2013urw,Ghisoiu:2014ena,Ghiglieri:2016xye,Ghiglieri:2017gjz,Ghiglieri:2017csp,Ghiglieri:2018wbs}, 
the gradual sphaleron freeze-out \cite{Eijima:2017cxr},
lepton number violating (LNV) effects in the decay and scattering rates \cite{Hambye:2016sby,Ghiglieri:2017gjz,Ghiglieri:2017csp,Antusch:2017pkq} and from mixing \cite{Eijima:2017anv},
the dependence on the initial conditions \cite{Asaka:2017rdj}
and the connection to neutrinoless double $\beta$ decay experiments \cite{Hernandez:2016kel,Drewes:2016gmt,Asaka:2016zib}. Recent parameter scans of the minimal $n=2$ model that have implemented some of this progress can be found in Refs.~\cite{Hernandez:2015wna,Drewes:2016lqo,Drewes:2016jae,Hernandez:2016kel,Antusch:2017pkq,Eijima:2018qke} for the minimal seesaw and
 for its embeddings in inverse and linear seesaw models \cite{Abada:2015rta,Abada:2017ieq}. 
While this minimal model is extremely predictable and in principle fully testable \cite{Hernandez:2016kel,Drewes:2016jae}, a key disadvantage is that the requirement to protect the generated asymmetries in the early Universe from washout limits the feasibility of the freeze-in leptogenesis mechanism to values of the mixing angles $\theta_{a i}$ that are so small that it will be very challenging to produce the particles in sizeable numbers at the LHC.

The scenario with $n=3$ has a much larger parameter space, which makes a phenomenological exploration more difficult. In Ref.~\cite{Canetti:2014dka} it has been pointed out that this additional freedom can make leptogenesis with much larger mixing angles possible because it allows to make strong hierarchies 
$|F_{a i}|\ll |F_{b i}|$ 
amongst the Yukawa couplings 
consistent with light neutrino oscillation data \cite{Drewes:2015iva}.
This allows to protect the asymmetry in the flavour $a$ from washout while the coupling $|F_{b i}|$ can be large enough to yield observable event rates at the LHC. The numerical analysis in Ref.~\cite{Canetti:2014dka} is by now known to be incorrect because it neglected the early equilibration of one of the interaction eigenstates, see e.g.~\cite{Drewes:2016gmt}. However, the physical argument can still be expected to be true.
Further studies of the model with $n=3$ \cite{Drewes:2012ma,Khoze:2013oga,Canetti:2014dka,Shuve:2014zua,Hernandez:2015wna,Drewes:2016lqo} have not systematically explored the parameter space, so that the range of heavy neutrino couplings that can be made consistent with leptogenesis and with light neutrino oscillation data in this scenario is not yet known.
With the present work we want to address this issue and systematically scan the parameter space of the low-scale seesaw model with three right-handed neutrinos.
While we perform an agnostic scan of the entire parameter space, we pay special attention on the region where the seesaw model approximately respects a generalised $B-L$ symmetry \cite{Shaposhnikov:2006nn}.
In this parameter region the symmetry protects the light neutrino masses in a way that observable $N_i$ production rates at colliders can be made consistent with the observed neutrino oscillation data in a technically natural way \cite{Kersten:2007vk}.

Our systematic analysis demonstrates significant quantitative and qualitative differences with respect to the $n = 2$ scenario. On the one hand, we confirm that the parameter space which simultaneously accounts for the neutrino oscillation data and the observed baryon asymmetry of the Universe, projected onto experimentally accessible quantities such as the active-sterile mixings and neutrinoless double beta decay effective mass, is significantly enlarged, implying significant discovery space for experiments such as NA62, T2K, Belle II and the LHC. On the other hand, we find qualitatively new dynamical processes in the kinetic equations describing leptogenesis, such as a dynamically generated resonant enhancement, providing new channels to generate the baryon asymmetry of the Universe.

The remainder of the paper is organised as follows. In Section~\ref{sec:seesaw_model} we introduce our setup, with a particular focus on the role of (approximate) global symmetries. The kinetic equations governing leptogenesis are introduced in Section~\ref{sec:KineticEquations}, emphasising the subtleties associated with a temperature dependent mass eigenbasis. We come back to this point in Section~\ref{sec:mechanisms}, where we discuss the different physical processes involved in the generation of the lepton asymmetry both for $n = 2$ and $n = 3$. The details of our parameter scan are given in Section~\ref{sec:scanstrategy}, with the resulting experimental prospects discussed in Section~\ref{sec:Results}. We illustrate the different dynamical processes contributing to leptogenesis by means of some representative benchmark points in Section~\ref{sec:benchmarks} before concluding in Section~\ref{sec:conclusions}. Further technical details can be found in the three appendices.

\section{The seesaw model \label{sec:seesaw_model}}

\subsection{Review of the model and notation.}

The most general renormalizable Lagrangian that contains only SM fields and $n$ flavours of right-handed neutrinos $\nu_{R i}$ reads 
\begin{equation}
    \mathcal{L}
  = \mathcal{L}_\text{SM} + i \, \overline{\nu_{R i}}\slashed\partial\nu_{R i}
  - \frac{1}{2} \left( \overline{\nu_{R i}^c}(M_M)_{ij}\nu_{R j} + \overline{\nu_{R i}}(M_M^\dagger)_{ij}\nu_{R j}^c \right)
  - F_{a i}\overline{\ell_{L a}}\varepsilon\phi^* \nu_{R i}
  - F_{a i}^*\overline{\nu_{R i}}\phi^T \varepsilon^\dagger \ell_{L a}
\ . \label{eq:Lagrangian}
\end{equation}
Here we have suppressed SU(2) indices; $\varepsilon$ is the totally antisymmetric SU(2) tensor.
The $F_{a i}$ are Yukawa couplings between the $\nu_{R i}$ and the SM leptons $\ell_a$, $M_M$ is a Majorana mass matrix for the singlet fields $\nu_{R i}$.\footnote{Here we use four-component spinor notation. Since spinors $\nu_R$ and $\ell_L$ are chiral, i.e., have only two non-zero components ($P_R\nu_R=\nu_R$ and $P_L\ell_L=\ell_L$), no explicit chiral projectors are required in the weak interaction term \eqref{WeakWW}.}
In the following we work in the flavour basis where $M_M$ is diagonal unless a different basis is explicitly specified.
The breaking of electroweak symmetry by the Higgs expectation value $\phi=(0, v)^T$ (with $v=174$ GeV at $T=0$) generates a Dirac mass term $\overline{\nu_{L}} m_D\nu_{R}$ with $m_D=vF$ from the Yukawa interaction term {$F\overline{\ell_L}\varepsilon\phi^*\nu_{ R }$}.

After electroweak symmetry breaking (EWSB), the complete neutrino mass term reads
\begin{equation}
\frac{1}{2}(\bar{\nu_L} \ \bar{\nu_R^c})
\underbrace{\begin{pmatrix}{\delta}m_{\nu}^{1loop} & m_D \\ m_D^T & M_M \end{pmatrix}}_{ \equiv \, {\mathcal M}}
\begin{pmatrix}
\nu_L^c
\\
\nu_R
\end{pmatrix}\ .
\label{eq:auxlabel}
\end{equation}
Here we have added the 1-loop correction
$\delta m_{\nu}^{1loop}$ \cite{Pilaftsis:1991ug} since we aim to perform an analysis that is consistent at second order in the Yukawa couplings $F$.
The mass matrix (\ref{eq:auxlabel}) can be diagonalised as
\begin{equation}
\mathcal{U}^{\dagger}\mathcal{M}\mathcal{U}^{\ast}=\begin{pmatrix}m_{\nu}^{\rm diag} & \\ & M_N^{\rm diag} \end{pmatrix}\ ,
\label{eq:Mdiagonalization}
\end{equation}
where $m_{\nu}^{\rm diag}$ and $M_N^{\rm diag}$ are diagonal $3\times3$ matrices.
It is convenient to parametrise $\mathcal{U}$ as \cite{Fernandez-Martinez:2015hxa}
\begin{equation}
\mathcal{U}= \begin{pmatrix} \cos(\theta) & \sin(\theta) \\ -\sin(\theta^\dagger) & \cos(\theta^\dagger)  \end{pmatrix}
\begin{pmatrix} U_{\nu} & \\ & U_N^{\ast} \end{pmatrix}\ ,
\end{equation}
with 
\begin{align}
\cos(\theta)=\sum_{n=0}^\infty \frac{(-\theta\theta^\dagger)^n}{(2n)!} \quad , \quad
 \sin(\theta)=\sum_{n=0}^\infty \frac{(-\theta\theta^\dagger)^n\theta}{(2n+1)!}.
 \end{align} 
 In the parameterisation (\ref{eq:Mdiagonalization}) $\mathcal{M}$ is first block-diagonalised by a complex $3 \times n$ matrix $\theta$ that mediates the mixing between the active neutrinos $\nu_L$ and the sterile neutrinos $\nu_R$. 
The unitary matrices $U_\nu$ and $U_N^*$ then diagonalise the $3\times3$ and $n\times n$ blocks $m_\nu$ and $M_N$ in the upper left and lower right corners, respectively, as $U_\nu^\dagger m_\nu U_\nu^* = {\rm diag}(m_1,m_2,m_3)\equiv m_\nu^{\rm diag}$ and $U_N^T M_N U_N = {\rm diag}(M_1,M_2,\ldots,M_n)\equiv M_N^{\rm diag}$.

In the \emph{seesaw limit} $|\theta_{a i}|\ll 1$, one can approximate
\begin{equation}\label{eq:Uexpand}
\theta \simeq m_D M_M^{-1} = v F M_M^{-1}\,,\quad  \quad \cos(\theta) = 1-\frac{1}{2}\theta\theta^{\dagger} + {\cal O}(\theta^4) \,, \quad  \quad   \sin(\theta) = \theta  + {\cal O}(\theta^3) \,,
\end{equation}
and
\begin{eqnarray}\label{eq:blocks_mass_matrix}
\delta m_\nu^{\rm 1loop} &=&  F M_N^{\rm diag} l(M_N^{\rm diag})F^T
= \frac{1}{v^2} \theta M_M M_N^{\rm diag} l(M_N^{\rm diag})M_M\theta^T\ ,\label{1loopquantities}
\\
    m_\nu
 & =& - \theta \tilde{M} \theta^T ,\label{eq:mnu_full_low}\\
M_N &=& M_M + \frac{1}{2} (\theta^\dagger \theta M_M + M_M^T \theta^T \theta^{*})\label{MN_Def} \,,
\end{eqnarray}
with  \cite{Pilaftsis:1991ug,Grimus:2002nk,AristizabalSierra:2011mn,Dev:2012sg,LopezPavon:2012zg,Lopez-Pavon:2015cga}
\begin{eqnarray}
\tilde{M} &=& \big[ 1 - \frac{1}{v^2} M_M M_N^{\rm diag} l(M_N^{\rm diag})
\big] M_M \,, \label{eq:tildeM}\\   
l(M_i) &=& \frac{1}{(4\pi)^2}\left[\frac{3\text{ln}[(M_i/m_Z)^2]}{(M_i/m_Z)^2 -
  1} + \frac{\text{ln}[(M_i/m_H)^2]}{(M_i/m_H)^2 - 1}\right]\,\label{loopfunction}.
\end{eqnarray}
By splitting
\begin{equation}
    m_\nu
  = m_\nu^{\rm tree} + \delta m_\nu^{\rm 1loop}\ ,
\label{seesaw1loop}
\end{equation}
we can recover the well-known tree-level result
\begin{equation}
    m_\nu^{\rm tree}
  = - m_D M_M^{-1} m_D^T
  = - \theta M_M \theta^T. 
\label{seesaw}
\end{equation}
The spectrum of neutrino mass states is clearly separated into three light and $n$ heavy mass eigenstates which can be expressed in terms of the Majorana spinors
\begin{align}
    \upnu_i &
  = \left[
  V_{\nu}^{\dagger}\nu_L-U_{\nu}^{\dagger}\theta \nu_R^c+V_{\nu}^T\nu_L^c-U_{\nu}^T\theta^{\ast} \nu_R
  \right]_i
\ ,
  & N_i &
  = \left[
  V_N^\dagger\nu_R+\Theta^T \nu_L^c + V_N^T\nu_R^c+\Theta^{\dagger}\nu_L
  \right]_i
\ , \label{HeavyMassEigenstates}
\end{align}
respectively, with $V_\nu = (1 - \frac{1}{2}\theta\theta^\dagger ) U_\nu$,  $V_N = (1 - \frac{1}{2} \theta^T \theta^*) U_N$ and $\Theta = U_N^*\theta$.

The mixing matrix $\Theta$ quantifies the misalignment of the mass eigenstates $\upnu_i$ and $N_i$ with the original "active" and "sterile" neutrinos $\nu_L$ and $\nu_R$.
It leads to a $\theta$-suppressed weak interaction of the heavy mass eigenstates $N_i$,
\begin{align}
    \mathcal L
  \supset&
  - \frac{g}{\sqrt{2}}\overline{N}_i \Theta^\dagger_{i a}\gamma^\mu e_{L a} W^+_\mu
 - \frac{g}{2\cos\theta_W}\overline{N_i} \Theta^\dagger_{i a}\gamma^\mu \nu_{L a} Z_\mu
 - \frac{g}{\sqrt{2}}\frac{M_i}{m_W}\Theta_{a i} h \overline{\nu_{L a}}N_i
+ \text{h.c.}
\ . \label{WeakWW}
\end{align}
The first two terms represent the $\theta$-suppressed interactions of the $N_i$ via the weak currents.  
Through these interactions the heavy neutrinos $N_i$ replace ordinary neutrinos in all processes if this is kinematically allowed, but with amplitudes suppressed by the angles~$\Theta_{a i}$.
The third term represents the Yukawa coupling to the physical Higgs field $h$ in the unitary gauge.
Here we have employed the relation $m_W = \frac{1}{2} g v$ involving the weak gauge coupling constant $g$.
It is convenient to introduce the quantities
\begin{align}
U_{a i}^2 &= |\Theta_{a i}|^2
\ ,
 &  U_a^2
 &= \sum_i U_{a i}^2
\ ,
 &  U_i^2
 &= \sum_a U_{a i}^2
\ ,
 &  U^2
 &= \sum_i U_i^2 \,,
\end{align}
which practically govern the event rates for processes involving the $N_i$.

\subsection{Approximate lepton number conservation.}\label{sec:B-L}

In absence of any special structure in the matrices $F$ and $M_M$, the seesaw relation~\eqref{seesaw} suggests that 
\begin{equation}\label{NaiveSeesaw}
U_i^2 \sim \sqrt{\Delta m_\text{atm}^2 + m_{\rm lightest}^2} / M_i < 10^{-10}~{\rm {GeV}}/M_i, 
\end{equation}
which would imply unobservably tiny branching ratios in collider experiments.
However, the seesaw relation is a matrix valued equation, and the light neutrino mass squares $m_i^2$ are the eigenvalues of the matrix $m_\nu^\dagger m_\nu$.
If there are cancellations in $m_\nu^\dagger m_\nu$, then small $m_i^2$ can be made consistent with arbitrarily large $U_{a i}^2$. 
Constraints from experiments other than neutrino oscillation ones are comparably weak in most of the mass range between the kaon and $W$ boson masses, cf.\ e.g.\ \cite{Drewes:2015iva,Abada:2017jjx} and Section~\ref{sec:ExpConstr}, so that $U_{a i}^2 \sim 10^{-5}$ are phenomenologically viable.
The cancellations could either occur accidentally (which would require a tuning of at least five orders of magnitude to achieve  $U_{a i}^2$ near the current LHC bounds \cite{Sirunyan:2018mtv}) or be owed to a symmetry. 
Indeed, if the Lagrangian~\eqref{eq:Lagrangian} approximately respects a generalised $B-\bar{L}$ symmetry~\cite{Shaposhnikov:2006nn,Kersten:2007vk,Moffat:2017feq}, 
then the cancellations in $m_\nu^\dagger m_\nu$ occur in a technically natural way because the light neutrino masses must be proportional to small parameters that quantify the amount of $B-\bar{L}$ violation.
Here $B$ denotes the usual SM baryon number and $\bar{L}$ is a generalised lepton number,
\begin{equation}
\bar{L}=L + L_{\nu_R} \,,
\end{equation}
that is composed of the SM lepton number $L$ and some charge associated with the $\nu_R$ (see below).
Specific models that realise an approximate $B-\bar{L}$ symmetry include 
models with $R$-parity violation~\cite{Ellis:1984gi,Ross:1984yg,Romao:1999up,Abada:1999ai,Hirsch:2000ef,Abada:2001zh}, 
"inverse seesaw" type scenarios~\cite{Wyler:1982dd,Mohapatra:1986aw,Mohapatra:1986bd,Bernabeu:1987gr,GonzalezGarcia:1988rw} (cf.\ also \cite{Deppisch:2004fa,Abada:2014vea}), the "linear seesaw"~\cite{Akhmedov:1995ip,Akhmedov:1995vm} (cf.\ also \cite{Barr:2003nn,Malinsky:2005bi,Gavela:2009cd}), scale invariant models~\cite{Khoze:2013oga}, some technicolor-inspired models~\cite{Appelquist:2002me,Appelquist:2003uu},  the $\nu$MSM~\cite{Shaposhnikov:2006nn} and other low-scale seesaw realisations~\cite{Ibarra:2010xw,Ibarra:2011xn,Dinh:2012bp}. 
Low-scale leptogenesis in connection to an approximate $B-\bar{L}$ symmetry has previously been studied in the framework of linear and inverse seesaw scenarios in Refs.~\cite{Abada:2015rta,Abada:2017ieq}, while the $\nu$MSM parameter space has been studied in Refs.~\cite{Shaposhnikov:2008pf,Canetti:2010aw,Canetti:2012vf,Canetti:2012kh} and numerous follow up publications (cf.\ references given in Section~\ref{sec:intro}).

The $B-\bar{L}$ symmetry enforces that the $\nu_{R i}$ must either $i)$ decouple, $ii)$ have vanishing Majorana masses or $iii)$ arrange themselves in pairs that form (pseudo-)Dirac spinors. 
For the $n=3$ case this can be made explicit with the parameterisation 
\begin{equation}
M_M=\M\begin{pmatrix} 1 - \Mu  &0  & 0\\ 0 & 1 + \Mu & 0 \\ 0 & 0 &\Mu'\end{pmatrix} \quad , \quad
F=\frac{1}{\sqrt{2}}\begin{pmatrix}  F_e(1 + \epsilon_e) & \ii F_e (1 - \epsilon_e) & F_e\epsilon'_e \\ F_\mu(1 + \epsilon_\mu) & \ii F_\mu(1 - \epsilon_\mu) & F_\mu \epsilon'_\mu \\ F_\tau(1 + \epsilon_\tau) & \ii F_\tau(1 - \epsilon_\tau) & F_\tau\epsilon'_\tau \end{pmatrix}\label{FullNeutrinoMass}.
\end{equation}
In the limit $\epsilon_a, \epsilon'_a, \Mu, \Mu' \to 0$ the quantity $B-\bar{L}$ is conserved. In terms of the original $\nu_{R i}$, one can make the following assignment of charges,
\begin{eqnarray}\label{ChargeAssignments}
\begin{tabular}{c c c }
${\rm spinor}$ & $\quad$ & $\bar{L}$-${\rm charge}$ \\
\hline
$\nu_{R {\rm s}}\equiv \frac{1}{\sqrt{2}}(\nu_{R 1} + i \nu_{R 2})$ & $\quad$ & $+1$ \\
$\nu_{R {\rm w}}\equiv \frac{1}{\sqrt{2}}(\nu_{R 1} - i \nu_{R 2})$ & $\quad$ & $-1$ \\
$\nu_{R 3}$ & $\quad$ & $0$ \\
\end{tabular} 
\end{eqnarray}
where the subscripts $s,w$ indicate that the corresponding states $\nu_{R {\rm s,w}}$ are strongly/weakly coupled to the SM states in the high temperature limit, $T \gg \bar M$.
We can now write the Lagrangian in the form 
\begin{eqnarray}\label{PseudoDiracL}
\mathcal{L}&=&\mathcal{L}_{\rm SM} + \overline{\psi_N}(\ii\slashed{\partial} - \M)\psi_N 
+\overline{\nu_{R 3}}\ii\slashed{\partial}\nu_{R 3}
- F_a^*\overline{\psi_N}\phi^T\varepsilon^\dagger \ell_{L a} - F_a \overline{\ell}_{L a} \varepsilon\phi^* \psi_N\nonumber\\
&&-\epsilon_a^* F_a^*\overline{\psi^c_N}\phi^T\varepsilon^\dagger \ell_{L a} 
- \epsilon_a F_a\overline{\ell}_{L a} \varepsilon\phi^* \psi_N^c
  - \epsilon'_{a}F_a\overline{\ell_{L a}}\varepsilon\phi^* \nu_{R 3}
  - \epsilon_a^{' *}F_a^*\overline{\nu_{R 3}}\phi^T \varepsilon^\dagger \ell_{L a}\nonumber\\
  &&-\Mu\M \frac{1}{2}\left(\overline{\psi_N^c}\psi_N + \overline{\psi_N}\psi_N^c\right) - \Mu' \M\overline{\nu_{R 3}^c}\nu_{R 3} \,,
\end{eqnarray}
where we have introduced the (pseudo-)Dirac spinor $\psi_N=\frac{1}{\sqrt{2}}(\nu_{R {\rm s}} + \nu_{R {\rm w}}^c)$ and we sum over multiple occurrences of the flavour index ``$a$''.

The generalised lepton number $\bar{L}$ is significantly violated by the oscillations amongst the heavy neutrinos even if $\epsilon_a, \epsilon'_a, \Mu, \Mu' \ll 1$. This means that it is in general not a suitable quantity to describe the time evolution in the early Universe, at least not if the rate of the oscillations is faster than the expansion of the Universe ($T< (M_0  |M_i^2 -M_j^2 |)^{1/3}$) and faster than the equilibration time scale of the heavy neutrinos (${\rm max}(|(F^\dagger F)_{ij}|)\gamma_{av} M_0^{2/3}/(M_i^2-M_j^2)^{2/3}  \ll 1$) \cite{Drewes:2016gmt}, where  $\gamma_{av}\sim 10^{-2}$ is a generic numerical coefficient appearing in the rate~(\ref{eq:gammas_av}) and
$M_0\equiv m_P(45/(4\pi^3g_*))^{1/2}=T^2/H$ 
can be interpreted as the comoving temperature in a radiation dominated Universe with Hubble parameter
$H$, $m_P = 1.22 \times10^{19}$~GeV and  $g_*$ counting the relativistic degrees of freedom in the thermal bath.
In the regime $T\gg M_i$ the helicity states of the heavy Majorana neutrinos $N_i$ practically act as "particle" and "antiparticle" states. 
One can use this fact to define another generalised lepton number 
\begin{equation}\label{HelicityChargeDef}
\tilde{L} = L + L_N\ ,
\end{equation}
that is approximately conserved even if the parameters $\epsilon_a, \epsilon'_a, \Mu, \Mu'$ are not small.
Here $L_N$ is a quantum number that can be assigned to the helicity states $P_h N_i$, where $P_h=1/2\times[1 + h \gamma^0\gamma^i\gamma^5 (\textbf{p}_i/|\textbf{p}|)]$ is the helicity projector with momentum $\textbf{p}$, as\footnote{In the ultra-relativistic limit, positive (negative) helicity corresponds to right-handed (left-handed) chirality.}
\begin{eqnarray}\label{HelicityChargeAssignments}
\begin{tabular}{c c c }
${\rm spinors}$ & $\quad$ & $\tilde{L}$-${\rm charge}$ \\
\hline
$P_+ N_i , \quad \bar{N_i}P_+$ & $\quad$ & $+1$ \\
$P_- N_i , \quad \bar{N_i}P_-$ & $\quad$ & $-1$ \\
\end{tabular} 
\end{eqnarray}
It is indeed $\tilde{L}$ (not $L$ or $\bar{L}$, both of which are violated) that is usually being referred to when distinguishing between "lepton number conserving" and "lepton number violating" terms in the low-scale leptogenesis literature.

\subsection{The LHC testable scenario. \label{subsec:LHCtestable}}

The $\nu$MSM realises the $B-\bar{L}$ conservation by choosing the seesaw scale $\Lambda$ 
below the electroweak scale and keeping all parameters $\epsilon_a, \epsilon'_a, \Mu, \Mu'$ tiny. 
In that model, $\epsilon_a, \Mu \ll 1$ is required to make sizeable $F_a$ consistent with light neutrino oscillation data and for successful leptogenesis (which requires $\Mu\ll1$ as only two heavy neutrinos participate in the process).
Regarding the third heavy neutrino, which serves as a Dark Matter candidate, $\epsilon'_a \ll 1$  is required to ensure its longevity for any mass allowed by structure formation considerations, while  $\Mu'\ll1$ is in addition needed for consistency with indirect Dark Matter searches if one assumes $\epsilon_a'$ to be sizeable enough that the Dark Matter is produced thermally via weak interactions, cf.\ e.g.\ \cite{Adhikari:2016bei,Boyarsky:2018tvu}.
The parameter space of the $\nu$MSM is the subject of many past and ongoing studies and will not be further addressed here because it is, from the viewpoint of neutrino mass generation and leptogenesis, practically a scenario with $n=2$ due to the extreme smallness of the $\epsilon'_a, \Mu'$ required for the stability of the Dark Matter candidate. 
Instead we focus on scenarios where all three heavy neutrinos have masses $M_i$ of roughly the same magnitude.
In this context it is worthwhile noting that it is sufficient for the $B-\bar{L}$ conservation that \emph{either} $\Mu'=0$ \emph{or} $\epsilon'_a=0$ {(as well as $\epsilon_a = 0 = \Mu$), since in this case the third right-handed neutrino either decouples or obtains only a Dirac mass term.} 
Hence, scenarios with moderately small $\epsilon'_a$  and $\Mu'$ of order unity are technically natural. It turns out that the choice $\epsilon_a, \epsilon'_a, \Mu$ $\ll 1$  with $\Mu'\gtrsim 1$ allows for successful leptogenesis with $|F_{a i}|$ larger than the electron Yukawa coupling.
This leads to mixings $U_{a 1}^2$ and $U_{a 2}^2$ that are well within reach of current experiments.


\section{Kinetic equations for leptogenesis \label{sec:KineticEquations}}

\subsection{Quantum kinetic equations.}
The quantum kinetic equations for freeze-in leptogenesis~\cite{Akhmedov:1998qx} in the density matrix formalism~\cite{Sigl:1992fn} read (see e.g.\ Refs.~\cite{Asaka:2005pn,Asaka:2011wq,Canetti:2012kh}):
\begin{align}
\frac{d R_N}{dt}= & -i \left[\langle H \rangle ,R_N\right]-\frac{1}{2}\langle \gamma^{(0)} \rangle \left \{ F^{\dagger}F, R_N-I\right \}- \frac{1}{2} \langle \gamma^{(1b)} \rangle \left \{ F^{\dagger} \mu F, R_N \right \}+\langle \gamma^{(1a)} \rangle F^{\dagger} \mu F + \nonumber \\
&
-\frac{1}{2}\langle \tilde{\gamma}^{(0)} \rangle \left \{ M_M F^{T}F^*M_M , R_N-I\right \} + \frac{1}{2} \langle \tilde{\gamma}^{(1b)} \rangle \left \{ M_M F^T \mu F^*M_M, R_N \right \} + \nonumber \\
& -
\langle \tilde{\gamma}^{(1a)} \rangle M_M F^T \mu F^*M_M
\,, \label{eq:Rnstart} \\
\frac{d {\mu_\Delta}_{a}}{dt} = & \,  - \frac{9 \,  \zeta (3)}{2 N_D\, \pi^2} \left \{
	\langle \gamma^{(0)} \rangle \left(F R_N F^{\dagger}-F^{*} R_{\bar N} F^T\right)
	-2  \langle \gamma^{(1a)}  \rangle \mu F F^{\dagger} + \right.  \nonumber \\   
 & +  \langle \gamma^{(1b)} \rangle \mu  \left(F R_N F^{\dagger}+F^{*} R_{\bar N}F^T\right)\nonumber \\
 &{
 + \langle \tilde{\gamma}^{(0)} \rangle \left(F^* M_M R_{\bar N} M_M F^T -F M_M R_N M_M F^\dagger\right)
 -2  \langle \tilde{\gamma}^{(1a)}  \rangle \mu F^* M_M^2 F^T} \nonumber \\
 &\left.
 {+ \langle \tilde{\gamma}^{(1b)} \rangle \mu  \left(F^* M_M R_{\bar N} M_M F^T+F M_M R_{N} M_M F^\dagger \right)}
 \right\}_{a a} \,, \label{eq:mustart}
\end{align}
where the $n \times n$ matrix $R_N$ encodes the density matrix of the three heavy neutrinos in kinetic equilibrium normalised to the entropy density, $(\rho_N(k,T))_{ij} = (R_N(T))_{ij} f_F(k/T)$ with $f_F$ denoting the Fermi-Dirac distribution with vanishing chemical potential, {$f_F(k/T) = \left[1 +  \exp \left( k/T \right) \right]^{-1}$}. 
The SM sector is taken to be in thermal equilibrium, and is thus fully characterised by the chemical potentials 
\begin{equation}
{\mu}_a = \mu_{L_a} + \mu_\phi,
\end{equation}
where $\mu_{L_a}$ are the flavoured left-chiral lepton chemical potentials and $\mu_\phi$ is the Higgs chemical potential, which appear in the corresponding distribution functions. 
These are connected to the chemical potential associated with $B-L$, ${\mu_\Delta}=\sum_a {\mu_{\Delta_a}}$ by the relation
\begin{equation}
\mu = \text{diag}({\mu}_a) \,, \quad  {\mu}_a = N_D \chi_{a b} {\mu_\Delta}_b \,, \quad \quad \chi = - \frac{1}{711} \begin{pmatrix}
257  & 20 &  20 \\
20 & 257 & 20 \\
20 & 20 & 257 
\end{pmatrix} \,. \label{eq:muDelta}
\end{equation}
 ${\mu_\Delta}$ is  invariant with respect to the SM $B+L$ violating processes.
$\langle H \rangle$ is the momentum averaged effective Hamiltonian for the heavy neutrinos (see Appendix~\ref{sec:equations}), 
\begin{equation}\label{eq:Heff}
\langle H \rangle =  {\langle H_0  + V_N \rangle = \frac{\pi^2}{36 \, \zeta(3)} \left( \frac{\text{diag}(0, M_2^2 - M_1^2, M_3^2 - M_1^2)}{T} + \frac{T}{8} F^\dagger F\right) \,,}
\end{equation}
while the terms involving coefficients $\langle \gamma^{(i)} \rangle$ and $\langle \tilde{\gamma}^{(i)} \rangle$ represent $\tilde{L}$-conserving and $\tilde{L}$-violating dissipation rates, respectively. 
Note that we neglect the $\tilde{L}$-violating part of $\langle H \rangle$.  $N_D = 2$ is an SU(2) factor. 
The thermally averaged rates
\begin{equation}
\langle \gamma (T) \rangle =\frac{\int d^3 p \, \gamma(p,T) f_F(p/T)}{\int d^3 p \,f_F(p/T)} \,, \label{eq:gammas_av}
\end{equation}
are given by \cite{Hernandez:2016kel} (cf.\ also \cite{Besak:2012qm,Anisimov:2010gy,Garbrecht:2013urw,Ghisoiu:2014ena,Ghiglieri:2017gjz}),
\begin{align}
&\langle \gamma^{(i)} \rangle = A_i \left[ c^{(i)}_\text{LPM} + y_t^2 c_Q^{(i)} + (3 g^2 + g'^2) \left( c_V^{(i)} - \ln(3 g^2 + g'^2) \right) \right]\ ,\\\notag
\end{align}
and \cite{Antusch:2017pkq}
\begin{align}\label{eq:gamma-def}
&\langle \tilde{\gamma}^{(i)} \rangle = A_i c^{(i)}_{1\rightarrow2}\ ,
\end{align}
where $g, g'$ denote the SM $SU(2)$ and $U(1)$ gauge couplings (which are temperature dependent due to their renormalisation group running), $y_t$ is the top Yukawa coupling, and
\begin{equation}
A_0 = 2 A_{1a} = -4 A_{1b} = \frac{\pi T}{2304 \, \zeta(3)} \,.
\end{equation}
The numerical values of $c^{(i)}_{LPM,Q,V}$ are reported in Table~\ref{tab:parameters} of Ref.~\cite{Hernandez:2016kel}, while those of $c^{(i)}_{1\rightarrow2}$ are determined following Ref.~\cite{Antusch:2017pkq}.\footnote{
What was computed in Ref.~\cite{Antusch:2017pkq} is in fact the quantity $\langle\tilde\gamma^{(1b)} \rangle  - \langle \tilde\gamma^{(1a)} \rangle$ that appears in front of the term $ \mu F F^{\dagger}$ in Eq.~(\ref{eq:mustart}) after rewriting
\begin{eqnarray}
-2  \langle \tilde\gamma^{(1a)}  \rangle \mu F F^{\dagger}      
  +  \langle \tilde\gamma^{(1b)} \rangle \mu  \left(F R_N F^{\dagger}+F^{*} R_{\bar N}F^T\right)
  &=&
 2\left(\langle \tilde\gamma^{(1b)} \rangle  - \langle \tilde\gamma^{(1a)}  \rangle \right) \mu F F^{\dagger}      \\
 && +  \langle \tilde\gamma^{(1b)} \rangle \mu  \left(F (R_N-1) F^{\dagger}+F^{*} (R_{\bar N} - 1)F^T\right).\nonumber
  \end{eqnarray}
  We extracted $\langle\tilde\gamma^{(1b)} \rangle$ and  $\langle \tilde\gamma^{(1a)} \rangle$ from this by assuming that these coefficients have equal values. 
  Practically this means that we guessed the coefficient in front of the term $\mu(F (R_N-1) F^{\dagger}+F^{*} (R_{\bar N} - 1)F^T)$. We do not expect this to have any phenomenological consequences because that term is small at all times: at early times $\mu$ is small, and at late times the heavy neutrinos are close to equilibrium.
}  
Both $c^{(i)}_Q$ and $c^{(i)}_V$ are found to be $T$-independent, 
the temperature dependence of $c^{(i)}_{LPM}$ is so mild that we will neglect it in the following. 
Using $c^{(i)}_{LPM}(T = 10^4~\text{GeV})$ as a reference value, this yields 
\begin{align}
	c^{(0)}_{LPM}  &= 4.22\,,  &c^{(0)}_{Q}  = 2.57\,, && c^{(0)}_{V}  &= 3.17\,, & c^{(0)}_{1\rightarrow2}  &=  0.86/T^2 \,,\nonumber \\
	c^{(1a)}_{LPM} &= 3.56\,,  &c^{(1a)}_{Q} = 3.10\,, && c^{(1a)}_{V} &= 3.83\,, & c^{(1a)}_{1\rightarrow2} &= 20.4/T^2 \,,\nonumber \\
	c^{(1b)}_{LPM} &= 4.77\,,  &c^{(1b)}_{Q} = 2.27\,, && c^{(1b)}_{V} &= 2.89\,, & c^{(1b)}_{1\rightarrow2} &= 20.4/T^2 \,.\label{tab:ci}
\end{align}
The running of the SM gauge couplings (included in the $\tilde L$-conserving rates) is given by
\begin{align}
 g(\Lambda) & = \left(\frac{1}{g_0^2}  + \frac{19}{48 \, \pi^2} \ln \frac{\Lambda}{m_Z}\right)^{-1/2}, \\
  g'(\Lambda) & = \left(\frac{1}{(g'_0)^2}  + \frac{41}{48 \, \pi^2} \ln \frac{\Lambda}{m_Z}\right)^{-1/2} \,,
\end{align}
where $g_0 = 0.652$ and $g'_0 = 0.357$ denote the values of the corresponding gauge couplings at $\Lambda = \pi T = m_Z$.  For the purpose of our numerical scan, we find it convenient to switch the time variable from cosmic time $t$ to $x=T_\text{EW}/T$, leading to the system of differential equations \eqref{eq:RNscan} - \eqref{eq:muscan}. See Ref.~\cite{Ghisoiu:2014ena} for further details.

The final $B$$-$$L$ asymmetry is obtained by evaluating the chemical potentials $\mu_\Delta$ at the scale of electroweak symmetry breaking,
\begin{equation}
 Y_{B-L} = \sum Y_{B-L}^a \quad \quad  \text{with} \quad Y_{B-L}^a =  \frac{45 N_D}{12 \pi^2 g_s}  \mu_{\Delta a} \,,
\end{equation}
where $g_s = 106.75$ denotes the effective number of degrees of freedom in the SM above the EW phase transition. 
SM sphaleron processes only pick up the asymmetry in the active sector, converting it to the baryon asymmetry we observe in the Universe today,
\begin{equation}
 Y_B = \frac{n_B - n_{\bar B}}{s} = \frac{28}{79} Y_{B - L} \,.
\end{equation}
Here $n_{B, \bar B}$ counts the number density of (anti-) baryons today, $s$  denotes the entropy density of the Universe and the baryon asymmetry of the Universe is measured to be $Y_B = (8.6 \pm 0.01) \times 10^{-11}$~\cite{Ade:2015xua}.{ For later reference we introduce also the asymmetry in the heavy neutrino sector, 
\begin{equation}
Y_N = \frac{1}{s} \int \frac{d^3 k}{(2 \pi)^3} f_F(k) \text{Tr}[R_N - R_{\bar N}] =  
\frac{135 \, \zeta(3)}{8 \, g_s \, \pi^4}
\text{Tr}[R_N - R_{\bar N}] \,,\label{YNDef}
\end{equation}
which in the absence of $\tilde L$-violating processes is identical to $Y_{B-L}$.}
Note that the definition (\ref{YNDef}) refers to quasiparticle occupation numbers and should therefore be applied in the basis where the effective Hamiltonian $H$ is diagonal, which does not necessarily coincide with the flavour basis where $M_M$ or $M_N$ are diagonal, as discussed in the following.

\subsection{Mass and interaction bases.}\label{sec:bases}
It is worthwhile to emphasise that some caution should be taken with respect to the flavour basis in which the above equations are defined. In general, neither the basis where $M_M$ nor the one where $M_N$ is diagonal correspond to the physical (quasi)particle mass basis. 

On the  one hand there is the $\mathcal{O}[\theta^2]$ correction in Eq.~(\ref{MN_Def}) from the Higgs expectation value which affects the physical masses at temperatures below $\sim 160$ GeV \cite{DOnofrio:2014rug}. In the $B-\bar{L}$ symmetry protected regime  the physical mass splitting at $T=0$ (given by the eigenvalues of $M_N$) can considerably deviate from the splitting between the eigenvalues of $M_M$. This effect, which was already pointed out in Ref.~\cite{Shaposhnikov:2008pf}, can be crucial for the generation of late time asymmetries (and hence DM production \cite{Laine:2008pg}) in the $\nu$MSM \cite{Canetti:2012kh}. Recent discussions of possible phenomenological implications can be found in e.g.\  Refs.~\cite{Drewes:2016jae,Antusch:2017pkq}, where it is also described how the $v(T)$ dependent term should be added to the effective Hamiltonian (\ref{eq:Heff}).
Moreover, the mixing induced by the temperature dependent Higgs field value $v(T)$ can have a significant impact on the generation of lepton asymmetries shortly before sphaleron freeze-out \cite{Eijima:2017anv}.
We ignore both of these effects in the following because we expect that they only lead to $\mathcal{O}[1]$ corrections in a limited part of the parameter region that we consider.

On the other hand there are corrections to the dispersion relations in a medium from forward scatterings \cite{Klimov:1981ka,Klimov:1982bv,Weldon:1989ys}, known as \emph{thermal masses} or \emph{matter potentials}, that affect the properties of SM particles \cite{Quimbay:1995jn} and heavy neutrinos \cite{Kiessig:2010pr}. These are responsible for the term $\propto F^\dagger F T/8$ in the effective Hamiltonian (\ref{eq:Heff}).
The physical heavy neutrino quasiparticles in the primordial plasma correspond to the eigenstates of the \emph{full} Hamiltonian (\ref{eq:Heff}), including the thermal correction. Since the relative size of the thermal and vacuum masses changes with temperature, the physical \emph{quasiparticle mass basis} rotates throughout the evolution of the Universe. At high temperatures, when the thermal masses dominate, it should be identified with the \emph{interaction basis} where $F^\dagger F$ is diagonal. 
At low temperatures it coincides with the \emph{vacuum mass basis} where $M_N$ is diagonal.
{It is important to note that the interpretation of the diagonal elements of the density matrix $R_{N}$ as measuring the corresponding number densities only holds if the density matrix is expressed in the quasiparticle mass basis.}

To which degree the rotation of the effective mass basis affects the generation of asymmetries depends on the model parameters. It is only relevant if the heavy neutrinos have reached sizeable occupation numbers while the temperature dependent contribution from the "thermal masses" to the splitting of the eigenvalues in the effective Hamiltonian (\ref{eq:Heff}) still dominates over the vacuum splittings $M_i^2 - M_j^2$. 
This happens quite generically for experimentally accessible heavy neutrinos because the approximate $B-\bar{L}$ symmetry that is required to make sizeable $F_{a i}$ consistent with light neutrino oscillation data implies that at least two $M_i$ are quasi-degenerate.  
We focus on this case in the following. For small mixing angles finite temperature effects usually only lead to sub-dominant corrections because the thermal masses are smaller and the mass splittings are in general larger. 

At high temperatures the $\tilde{L}$-conserving rates $\langle \gamma^{(i)} \rangle$ are parametrically larger than the $\tilde{L}$-violating rates $\langle \tilde{\gamma}^{(i)} \rangle$, and one can understand the behaviour in terms of the eigenvalues and eigenvectors of the matrices $F^\dagger F$ and $M_M^\dagger M_M$.
In the $B-\bar{L}$ conserving limit the matrix $F^\dagger F$ has three vastly different eigenvalues; one is roughly given by $\sum_a |F_a|^2$ while the other two are suppressed by $\epsilon_a^2$ and $\epsilon_a^{'2}$.
Hence, one interaction eigenstate ($\nu_{R{\rm s}}$, which is always part of the pseudo-Dirac spinor $\psi_N$) feels the full strengths $\sim F_a$ of the Yukawa interactions,  while the other two ($\nu_{R{\rm w}}$ and $\nu_{R 3}$ or combinations thereof) practically decouple.

Unless $\Mu'$ is very close to unity (i.e. $M_3 \simeq \bar{M}$), the mixing and rotation mainly occur between the components of the pseudo-Dirac spinor $\psi_N$ because the thermal corrections to $M_3^2$ are of order $\mathcal{O}[(\epsilon'_a F_a)^2]$. We shall consider this case first and neglect $\nu_{R 3}$ for the moment. In this case most of the discussion for the case $n=2$ in Ref.~\cite{Drewes:2016gmt} can directly be applied.
At high temperatures the states $\nu_{R {\rm s}}$ and $\nu_{R {\rm w}}$ defined in Eq.~(\ref{ChargeAssignments}) are approximately both, effective mass and interaction eigenstates.
$\nu_{R {\rm s}}$ picks up a large thermal mass $\sim \sum_a |F_a|^2 T^2$ and is produced at a large rate $\sim \sum_a |F_a|^2 \langle \gamma^{(i)} \rangle$, while the corrections to the mass of $\nu_{R {\rm w}}$ are only $\sim \sum_a |\epsilon_aF_a|^2 T^2$ and its production rate is only $\sim \sum_a |\epsilon_aF_a|^2 \langle \gamma^{(i)} \rangle$. 
Despite the large thermal mass splitting, there are no rapid oscillations between the two states because the effective heavy neutrino mass and interaction bases are almost aligned. 
If $\nu_{R{\rm s}}$ comes into equilibrium before the oscillations commence, then the BAU is generated in a single overdamped oscillation in the strong washout regime ("overdamped regime"). This is in contrast to the case of small mixing angles, where a large number of oscillations occur before the sphaleron freeze-out in the weak washout regime ("oscillatory regime"), see Ref.~\cite{Drewes:2016gmt} for details. 
At late times the feebly coupled state is driven to equilibrium by the overdamped oscillation and by the $\tilde{L}$-violating damping rates, which are  $\propto F^* F^T$ and not $\epsilon_a$-suppressed.

If one considers all three heavy neutrinos, then the situation can be much more complicated. 
For $n=2$ there are practically only two relevant time scales in the heavy neutrino sector: the time when the first heavy neutrino state reaches equilibrium (given by its thermal damping rate) and the frequency of the oscillations (given by the mass splitting).  
On the contrary, there are generally three equilibration and three oscillation time scales in the system with $n=3$.
In the simplest scenarios all frequencies and damping rates are well-separated, and no heavy neutrinos reach thermal equilibrium before the sphaleron freeze-out. In that case the separation of scales allows to treat each of the oscillations separately in a simplified $n=2$ model, similar to the treatment of light neutrino oscillations in the Sun or in the atmosphere.
However, for $\Mu\ll1$ and $\Mu'\sim 1$ all three states can mix with each other and complicated behaviour can arise. We illustrate this for a few example points in Section~\ref{sec:benchmarks}.

Finally we note that the charge $\tilde{L}$ in Eq.~(\ref{HelicityChargeDef}) should be defined in the rotating \emph{quasiparticle mass basis}. The relation between this basis and the \emph{vacuum mass basis}, however, depends not only on temperature, but also on the model parameters. For simplicity we use  in the following the \emph{vacuum mass basis} and the \emph{interaction basis} as approximations for the actual \emph{quasiparticle mass basis} at very low and very high temperatures, respectively.


\section{ {Most important new physical effects } \label{sec:mechanisms}}

The generation of a baryon asymmetry via the freeze-in mechanism is a complex nonequilibrium process that involves an interplay between coherent oscillations and decoherent scatterings, both of which can occur in a $\tilde{L}$-conserving or $\tilde{L}$-violating manner. For $n=3$ there is a large number of (possibly vastly different) times scales involved, and the phenomenology of the leptogenesis parameter space is very rich.
While our scan systematically explores this parameter space numerically, a qualitative analytic understanding of the results is highly desirable.

\subsection{Minimal \texorpdfstring{$n=2$}{n = 2} scenario without \texorpdfstring{$\tilde{L}$}{L}-violation.}
Let us first briefly review the case of only two heavy neutrinos, in order to highlight the qualitative differences in the full three neutrino case studied in this paper. Moreover, we  neglect for the moment $\tilde{L}$-violating processes.
In this case the system can be studied by analytic methods \cite{Akhmedov:1998qx,Asaka:2005pn,Abada:2015rta,Drewes:2016gmt}.
No CP-violation can arise in the heavy neutrino oscillations \cite{Drewes:2016gmt}, so that the CP-violation necessary to generate a lepton asymmetry must arise in the active sector and/or in the mixing between the active and sterile sectors.
In particular, washout processes play a crucial role in the generation of a net $L\neq 0$. In the weak washout regime\footnote{ 
For $n=2$, one can parametrically distinguish a \emph{weak washout regime} (or \emph{"oscillatory regime"}) in which the equilibration of both heavy neutrinos occurs after the freeze-out of weak sphaleron processes
(i.e. $|(R_N)_{ii}|, |(R_{\bar N})_{ii}|\ll 1$ for all $i$ and $T>T_{\rm sph}$)
and a \emph{strong washout regime} (or \emph{"overdamped regime"}) in which the occupation numbers of one heavy neutrino interaction eigenstate reach equilibrium before sphaleron freeze-out.
For weak washout the BAU scales as $\propto (m_P^2/|M_i^2-M_j^2|)^{2/3}$ \cite{Asaka:2005pn}, while the dependence in the strong washout regime is rather complicated \cite{Drewes:2016gmt}.
} an analytical expression for the lepton asymmetry was derived in Refs.~\cite{Akhmedov:1998qx,Asaka:2005pn} (see also~\cite{Abada:2015rta}). It was found to be proportional to $\sum_{i,a}(F_{ai})^\dagger \delta_a F_{ai}$ with
\begin{equation}
 \delta_a = \sum_{i > j} \text{Im} \left[ F_{a i} \left( F^\dagger F\right)_{ij} (F^\dagger)_{j a} \right] \,,
\end{equation}
which in particular vanishes in the flavour symmetric limit
\begin{equation}
 |F_{e i}| = |F_{\mu i }| = |F_{\tau i}| \ , \quad \forall \, i \,.
\end{equation}
In the strong washout regime an asymmetric coupling $\flav\ll1$ to the active flavours is typically necessary to protect the asymmetry from the strong washout processes in the sterile sector, see e.g.\ the discussion in Refs.~\cite{Drewes:2016gmt, Abada:2017ieq}, where
\begin{equation}\label{FlavourHierarchy}
		\flav \equiv \sum_{i} \frac{{\rm min}|F_{ai}|}{{\rm max}|F_{ai}|}\left( \frac{\sum_b|F_{bi}|^2}{\sum_{c,j}|F_{cj}|^2}\right)
		\simeq \frac{{\rm min}|F_{a}|}{{\rm max}|F_{a}|} .
\end{equation}
The ${\rm max}$ and ${\rm min}$ are to be understood with respect to the index $a$, and the term between parenthesis averages the sum with a weight proportional to the relative size of the Yukawa couplings for the right-handed neutrino $i$.
The approximate second equality holds only in the $B-\bar{L}$ conserving limit, where $F_a$ are the large entries in the parameterisation (\ref{FullNeutrinoMass}).
If one active flavour is coupled significantly weaker than the other generations, the asymmetry in this flavour can be preserved for a considerable time even when the right-handed neutrinos approach thermal equilibrium. 
For $n=2$ the heavy neutrino mixing pattern is strongly constrained by light neutrino oscillation data \cite{Ruchayskiy:2011aa,Asaka:2011pb,Hernandez:2016kel,Drewes:2016jae,Drewes:2018gkc}, and in particular $\flav > 5\times 10^{-3}$ \cite{Drewes:2018gkc}. This imposes an upper limit on the maximal $U_i^2$ for which leptogenesis is feasible: Leaving aside highly fine-tuned parameter choices, the two heavy neutrinos necessarily form a pseudo-Dirac spinor $\psi_N$ if their mixings are much larger than the estimate (\ref{NaiveSeesaw}). 
The larger Yukawa couplings $F_a \gg \epsilon_aF_a$ 
in (\ref{PseudoDiracL}) then drive the entire heavy neutrino sector towards equilibrium in an overdamped manner \cite{Drewes:2016gmt}, and the only way to protect the BAU from washout is a strong hierarchy $\flav \ll 1$. How strong this hierarchy must be to ensure the survival of some asymmetry until sphaleron freeze-out depends on the magnitude of the Yukawa couplings and masses (and hence $U_i^2$). As a result, the experimental constraint on $\flav$ from neutrino oscillation data imposes an upper limit on $U_i^2$ for a given $M_i$.

\subsection{\texorpdfstring{$n=3$}{n = 3} scenario without \texorpdfstring{$\tilde{L}$}{L}-violation.}
The situation is qualitatively different in the case of three right-handed neutrinos. 
One can distinguish three different new physical effects: 
\begin{enumerate}[{1)}]
\item \textbf{Larger flavour hierarchies are allowed.} \label{mechanism1} 
A hierarchy in the couplings of the heavy neutrinos to individual SM flavours ($\flav \ll1$) can protect a part of the lepton asymmetry from the washout even if the heavy neutrinos reach equilibrium ($R_N\simeq R_{\bar N}\sim \mathbb{1}$) if $\flav$ is small enough to keep one of the washout rates below the Hubble rate. 
In the scenario with $n=2$ the requirement to reproduce the light neutrino oscillation data practically requires  $\flav > 5\times 10^{-3}$ \cite{Drewes:2018gkc} (cf also~\cite{Hernandez:2016kel,Drewes:2016jae}). 
This hierarchy is not strong enough to protect the asymmetry from washout \cite{Antusch:2017pkq} for mixings $U_i^2$ near the current LHC bounds \cite{Sirunyan:2018mtv}.
As already pointed out in Ref.~\cite{Canetti:2014dka}, the relaxed lower experimental bound on $\flav$ in the scenario with $n=3$ \cite{Gorbunov:2013dta,Drewes:2015iva} allows to protect the BAU from washout for much larger $U_i^2$ if one SM flavour couples only very feebly to the pseudo-Dirac pair.
\item \textbf{Asymmetry in the heavy neutrino oscillations.} \label{mechanism2}  \label{page:listmechanism}
Contrary to the $n=2$ case discussed above, the $n=3$ case allows for a generation of a net asymmetry $L\neq 0$ during the heavy neutrino oscillations (even if $\tilde{L}$-violating effects are neglected), without requiring any flavour asymmetric Yukawa couplings. In Appendix~\ref{sec:perturbative} we explicitly derive the corresponding source terms entering the quantum kinetic equations by means of a perturbative expansion in the lepton asymmetries. We emphasise the presence of a new source term for the asymmetry in the heavy neutrino sector, which arises (for $n \geq 3$ only) from the first term in Eq.~\eqref{eq:Rnstart}. This allows for the generation of an asymmetry even in the absence of (flavour asymmetric) washout processes, contrary to the situation for $n = 2$~\cite{Drewes:2016gmt}. 
\item \textbf{Resonantly enhanced asymmetry.} \label{mechanism3} 
The produced asymmetry strongly depends on the heavy neutrino mass splitting and is resonantly enhanced if the splitting between two of the heavy neutrino masses is very small \cite{Asaka:2005pn}. 
In the primordial plasma the effective quasiparticle masses are given by the eigenvalues of the effective Hamiltonian (\ref{eq:Heff}). Due to the interplay between temperature dependent and independent terms in the effective Hamiltonian, the effective mass splittings are time dependent. As a result, a maximal resonant enhancement can be generated dynamically, even if the mass spectrum in vacuum is only moderately degenerate.\footnote{This effect is well-known within the $\nu$MSM \cite{Shaposhnikov:2008pf}, where it is crucial \cite{Canetti:2012kh} to ensure that the generation of asymmetries can occur twice during the history of the Universe, before sphaleron freeze-out (for baryogenesis \cite{Asaka:2005pn}) and afterwards (to generate the asymmetries required for resonant sterile neutrino Dark Matter production \cite{Shi:1998km}) for the same parameters. }
This is similar to the Mikheyev-Smirnov-Wolfenstein (MSW) effect that affects light neutrino oscillations in matter.
In contrast to the MSW effect for light neutrinos it does not require the presence of lepton chemical potentials because the Yukawa couplings can give different thermal masses to the neutrinos (while the light neutrinos' gauge interactions are flavour blind, so that different effective masses can only be realised through chemical potentials).
In particular, an (avoided) level crossing necessarily occurs for $\Mu'>1$, i.e., if the state $\nu_{R 3}$ with couplings $F_{a 3}\sim \epsilon'_a F_a$ has a vacuum mass $M_3 > \bar{M}$ larger than the pseudo-Dirac spinor $\psi_N$ with couplings $\sim F_a$. 
This is because the component $\nu_{R {\rm s}}$ of $\psi_N$ defined in (\ref{ChargeAssignments}) receives a comparably large thermal mass $\sim |F_a|^2 T^2$, which necessarily exceeds the effective mass of $\nu_{R 3}$ at sufficiently high temperature.
If this crossing occurs during the time when the asymmetry is generated, the resonant effect can maximally enhance it, even if the vacuum masses are only moderately degenerate.  In contrast, in the $B-\bar L$ protected regime of the $n=2$ case,  the interaction and Majorana mass bases have to be maximally misaligned to reproduce the small active neutrino masses, and hence any avoided level crossing comes with a mass gap which is typically too large to resonantly enhance the asymmetry. For $n=3$ with $\Mu'>1$ the level-crossing temperature $T_{\text{crossing}}$ can be estimated in the limit of approximate $B-\bar{L}$ symmetry ($|\epsilon_a|,|\epsilon'_a|,\Mu \ll 1$ in Eq.~(\ref{FullNeutrinoMass})), yielding
\begin{equation}\label{MSWestimate}
T_{\text{crossing}} \approx \frac{2 \sqrt{2} \M \sqrt{{\Mu'}^2-1}}{\sqrt{\sum_{a} \left| F_a \right|^2}} = 2.8 \times 10^{5} \text{ GeV} \left(\frac{\M}{\text{GeV}}\right) \frac{\sqrt{{\Mu'}^2-1}}{\sqrt{\sum_{a} \left| (F_a/10^{-5}) \right|^2}} \, .
\end{equation}
 
\end{enumerate}

\subsection{Effects of \texorpdfstring{$\tilde{L}$}{L}-violation. \label{subsec:Lviolation}}
In the weak washout or oscillatory regime, the BAU is generated at temperatures $T\gg M_i$. In this case $\tilde{L}$ is in good approximation conserved during the heavy neutrino oscillations. However, for $U_i^2$ that are large enough to be probed with the LHC, the couplings $F_a$ are generally large enough to drive part of the heavy neutrinos (in particular $\nu_{R {\rm s}}$) to equilibrium before sphaleron freeze-out. In this strong washout or overdamped regime, the asymmetry is often generated near the sphaleron freeze-out, and $\tilde{L}$ violating effects can in general not be neglected. This leads to several new effects.
\begin{enumerate}[{1)}]
\setcounter{enumi}{3}
\item \textbf{Direct $L$-generation in Higgs decays.} \label{mechanism4}  The rates for $\tilde{L}$-violating processes (in particular Higgs decays) are suppressed by $(M_i/T)^2$, but can directly generate a $L\neq 0$ at order $\mathcal{O}[F_{ai}^4]$.
This is in particular important in the case $n=2$, where the oscillations themselves cannot generate a $L\neq0$  
(cf.\ point \ref{mechanism2}) )
and leptogenesis always relies on the flavour asymmetric washout,
so that the BAU is necessarily of order $\mathcal{O}[F_{ai}^6]$.  It has been pointed out in Refs.~\cite{Hambye:2016sby,Hambye:2017elz} that the $\tilde{L}$ violating source $\sim\mathcal{O}[F_{ai}^4 (M_i/T)^2]$ can exceed the $\tilde{L}$-conserving source $\sim\mathcal{O}[F_{ai}^6]$ for certain parameter choices. 
For $n=3$ we expect this effect to be less relevant because a $L\neq0$ can already be produced in the oscillation in absence of $\tilde{L}$-violation, cf.\ Appendix \ref{sec:perturbative}. This latter effect is $\sim\mathcal{O}[F_{ai}^6]$~\cite{Drewes:2016gmt}, but not suffering from any $M_i/T$-suppression, it can be active over a much longer period of time.
We should stress that the above power counting only holds when the washout is weak enough that there is a clear separation between the times when the heavy neutrinos start to oscillate and when they come into equilibrium ("oscillatory regime"). If some heavy neutrino degrees of freedom reach equilibrium at early times ("overdamped regime"), the parametric dependence is different \cite{Drewes:2016gmt}.

\item \textbf{Damping of $\nu_{R{\rm w}}$.} \label{mechanism5}  In absence of $\tilde{L}$-violating processes the heavy neutrino damping rates are approximately proportional to $F^\dagger F$. In the $B-\bar{L}$ symmetric limit, one eigenvalue of this matrix is much larger ($\sim F_a^2$) than the other two ($\sim \epsilon_a^2F_a^2, \epsilon_a^{'2}F_a^2$).  
This means that the states $\nu_{R{\rm w}}$ and $\nu_{R 3}$ approach thermal equilibrium very slowly. In particular, $\nu_{R{\rm w}}$ is primarily driven to equilibrium via the overdamped oscillation with $\nu_{R{\rm s}}$ \cite{Drewes:2016gmt}. 
The heavy neutrino damping rate due to $\tilde{L}$-violating processes is proportional to $\sim F^T F^*$. This in particular means that $\nu_{R {\rm w}}$ is driven to equilibrium at a rate $\propto F_a^2 (\bar{M}/T)^2$, which can be much larger than the $\tilde{L}$-conserving rate $\propto \epsilon_a^2F_a^2$, {since the weakly coupled eigenstate of $F^T F^*$ in general does not coincide with $\nu_{R {\rm w}}$}. 
In contrast to that, the state $\nu_{R 3}$, which is not part of the pseudo-Dirac system $\psi_N$, remains feebly coupled  with equilibration rate $\propto \epsilon_a^{'2}F_a^2$ 
(unless $M_3\simeq \bar{M}$, in which case there can be significant mixing between $\nu_{R 3}$ and the other two states). 
\item \textbf{Washout efficiency.} \label{mechanism6}  If the $\tilde{L}$-violating processes are neglected, the washout of the helicity-based $\tilde{L}$ charges in the heavy neutrino sector enforces a simultaneous decrease of the $\tilde{L}$ charges in the active sector, suppressing the SM lepton number asymmetry $L$, i.e.\ $Y_{B-L}$. In the presence of $\tilde{L}$-violating processes this is not necessarily the case, allowing for a sizeable asymmetry in the active sector even if the washout in the sterile sector is effective. This results in a larger final BAU \cite{Antusch:2017pkq}. 

 \item \textbf{Equalising flavoured asymmetries.} \label{mechanism7}  When some of the heavy neutrinos come into equilibrium, the  $\tilde{L}$-conserving processes wash out the flavoured asymmetries $L_a$, but cannot erase the total SM asymmetry $L$ (and hence the BAU) unless all  charges in the heavy sector have been erased.
In this situation the washout of the total $L$ is driven by the $\tilde{L}$-violating processes. 
Since $L$ is in equal parts composed of $e$, $\mu$ and $\tau$ asymmetries, the direction in flavour space in which all $L_a$ are equal is only slowly erased, while deviations from $L_e=L_\mu=L_\tau$ are erased by the much faster $\tilde{L}$-conserving processes. 
Therefore the asymmetries in all SM flavours tend to be equal in this situation, cf.\ e.g.\ Fig.~\ref{fig:benchmark1a}. 
Note that the total $L$ would also be erased in absence of the $\tilde{L}$-violating processes once the sterile charges are driven to equilibrium because the total $\tilde{L}$ vanishes for our initial conditions.
\item \textbf{Direct $L$-generation through active-sterile mixing.} \label{mechanism8} So far we have mainly considered $\tilde{L}$-violating decays in the symmetric phase of the SM (in particular Higgs decays). Similar arguments apply if one includes $\tilde{L}$-violating scatterings in the symmetric phase of the SM. There is, however, moreover a brief period between the moment when the Higgs field develops a non-zero value $v(T)$ at $T\sim 160$~GeV and the sphaleron freeze-out at $T\sim 130$~GeV.
During this period the mixing between active and sterile neutrinos directly violates $\tilde{L}$ \cite{Eijima:2017anv}. We neglect this effect in the present work.
\end{enumerate}

\noindent In summary, the dynamics governing leptogenesis in the $n=3$ case is much more diverse than in the $n=2$ case. After quantifying the parameter space yielding successful leptogenesis in Secs.~\ref{sec:scanstrategy} and \ref{sec:Results}, we will illustrate some of the effects described above by means of a few exemplary points in Section~\ref{sec:benchmarks}.


\section{Strategy for the parameter scan \label{sec:scanstrategy}}

 \subsection{General strategy. \label{sec:randomisation}}
Our goal is to perform a systematic parameter scan to identify the range of the heavy neutrino properties that are consistent with all experimental constraints and can reproduce the observed BAU. A major obstacle is the high dimensionality of the parameter space. 
For $n$ heavy neutrinos, the seesaw model contains $7 n - 3$ free parameters in addition to those of the SM. Only $5$ of those (two mass splittings and three mixing angles) are constrained by light neutrino oscillation data. For $n=2$ it is possible to perform a complete parameter scan to clearly identify the boundaries of the region where leptogenesis is possible \cite{Canetti:2012vf,Canetti:2012kh,Drewes:2016jae,Antusch:2017pkq} or perform a Bayesian analysis \cite{Hernandez:2016kel}.
For $n=3$ the dimensionality of the parameter space is so high that even a systematic combination of all experimental constraints in the mass region under consideration here (without leptogenesis) is numerically challenging \cite{Drewes:2015iva}. If one includes the computation of the BAU, which requires solving the coupled differential equations~\eqref{eq:Rnstart} and \eqref{eq:mustart} for each parameter choice and is numerically much more demanding than imposing laboratory constraints, then it becomes practically impossible to explore the entire parameter space. However, from a phenomenological viewpoint, one is mostly interested in the projection of the viable parameter region on the  $M_i - U_{a i}^2$ planes. In Section~\ref{sec:Results} we present scatter plots in these planes which illustrate that, for masses $M_i$ below the electroweak scale,
the leptogenesis region covers the entire experimentally allowed range of mixings $U_{a i}^2$. Since both, the BAU and experimental constraints, depend smoothly on the model parameters that determine the $U_{a i}^2$, it seems physically reasonable that the entire region between the scattered points can be filled if the scan would run for infinitely long.
In the remainder of this section we explain how the parameter scan is performed.

It is well-known that leptogenesis is feasible in the $n=2$ model with values of $U_i^2$ at most four orders of magnitude above the estimate (\ref{NaiveSeesaw}), i.e. $U_i^2 < 10^{-6}~{\rm {GeV}}/M_i$, for any value of $M_i$ in the range considered here \cite{Drewes:2016jae}. Since the $n=2$ parameter space is a subset of the larger parameter space of the $n=3$ model under consideration here (in the limit $\epsilon'_a\to 0$), the same must apply in the present model. We are therefore primarily interested in studying leptogenesis with $U_i^2 > 10^{-6}~{\rm {GeV}}/M_i$.
The strongest constraints on the $N_i$ properties come from the seesaw relation (\ref{seesaw}) and light neutrino oscillation data.
The requirements to reproduce the observed data without fine-tuning for $U_i^2 > 10^{-6}~{\rm {GeV}}/M_i$ practically enforces the $B{\rm -}\bar{L}$ symmetry discussed in Section~\ref{sec:B-L}.
The parameterisation (\ref{FullNeutrinoMass}) in principle is ideal to explore this region, but the preproduction of the light neutrino parameters in a randomised scan is a search for the needle in the haystack due to the small error bars of these parameters and the complicated relations between model parameters and observables.
To keep the numerical effort at a feasible level, we adopt a three-step strategy that treats neutrino oscillation data different from other constraints:
\begin{enumerate}
\item For the generation of parameter points, we employ the Casas-Ibarra parameterisation \cite{Casas:2001sr}, see below. This parameterisation is not ideal to explore the $B{\rm -}\bar{L}$ symmetry protected region, but guarantees consistency with light neutrino oscillation data at the perturbative level. 
\item We then remove all points which are not consistent with the experimental constraints described in Section~\ref{sec:ExpConstr}.
\item For each remaining parameter choice we compute the BAU. We consider leptogenesis feasible if the BAU deviates from the observed value by less than a factor five.\footnote{{The experimental uncertainty on the BAU is much smaller than this. The larger range for an acceptable $Y_B$ adopted here reflects theoretical uncertainties  as well as the strong sensitivity of the final value of $Y_B$ on the model parameters - starting from a parameter point whose computed $Y_B$ deviates from the observed value by an ${\cal O}(1)$ factor we expect to be able to reproduce the observed value by minimally varying the input parameters.}} 
To improve the numerical performance we rewrite the quantum kinetic equations \eqref{eq:Rnstart} and \eqref{eq:mustart} as described in Appendix~\ref{sec:equations}.
We furthermore only track the first ten oscillations of the heavy neutrinos and then set off-diagonal elements of the density matrix to zero. We explicitly check that this does not induce any discontinuity in the evolution of the asymmetries. 
A similar procedure has been proposed in Ref.~\cite{Canetti:2010aw} and has been analytically verified in Ref.~\cite{Garbrecht:2011aw}.
\end{enumerate}

\noindent \textbf{Parametrisation.} In order to reproduce the observed neutrino masses and lepton mixing we adopt a generalisation of the Casas-Ibarra parameterisation~\cite{Casas:2001sr}, extended to include 1-loop corrections~(\ref{seesaw1loop}) to the light neutrino mass matrix~\cite{Lopez-Pavon:2015cga}. 
For small $\theta$ we may approximate the relation~\eqref{eq:tildeM} by\footnote{If the splitting between two eigenvalues of $M_M$ is smaller than the light neutrino mass differences, then the $\mathcal{O}[\theta^2]$ term in Eq.~(\ref{MN_Def}) can cause large deviations of $U_N$ from unity. However, in the $B-\bar{L}$ symmetric regime one still observes 
$|(F U_N)_{ai}|^2\simeq |F_{ai}|^2$
and hence
$|\theta_{ai}|^2\simeq |\Theta_{ai}|^2 = U_{ai}^2$ 
due to the structure of the $F$ and $M_M$ in  Eq.~(\ref{FullNeutrinoMass}).}
\begin{equation}
\label{eq:Mtildediag}
\tilde{M}_{ij}\simeq
M_i\delta_{ij}\left(1 - \frac{M_i^2}{v^2}l(M_i)\right) \equiv \tilde{M}^{\rm diag} \,,
\end{equation}
i.e., $\tilde{M}$ can be approximately expressed in terms of the entries of $M_N^\text{diag}$. 
In this formalism the Yukawa couplings are determined after having specified the low-energy neutrino oscillation data, the right-handed neutrino masses and a 3-dimensional orthogonal matrix $\mathcal{R}$,
\be\label{eq:F_CI} 
F = \frac{i}{v}\
U_\nu
\sqrt{m_\nu^{\rm diag}}
\ \mathcal{R}\ 
\sqrt{\tilde{M}^{\rm diag}}^{-1}M_M \,.
\ee
The matrix $\mathcal{R}$ can be parameterised as a product of three rotations,
\be 
\mathcal{R} = V_{23}(\omega_{23}) \ V_{13}(\omega_{13}) \ V_{12}(\omega_{12}),
\ee
where the angles $\omega_{ij}$ are in general complex numbers and
\be 
V_{12} (\omega_{12}) =
\left(
\begin{array}{ccc}
 \cos (\omega_{12} ) & \sin (\omega_{12} ) & 0
   \\
 -\sin (\omega_{12} ) & \cos (\omega_{12} ) &
   0 \\
 0 & 0 & 1 \\
\end{array}
\right),
\ee
and analogous definitions hold for $V_{13}$ and $V_{23}$. 
We work in the basis where the charged lepton Yukawa couplings are diagonal, so that $U_\nu$ can be identified with the 
Pontecorvo-Maki-Nakagawa-Sakata (PMNS)
matrix $U_\text{PMNS}$.

The full system is thus characterised by 13 real free parameters: 6 real numbers for the three imaginary angles $\omega_{ij}$, 3 heavy neutrino masses, 3 complex phases in the PMNS mixing matrix (one Dirac $\delta_\text{CP}$ and two Majorana $\alpha_{1,2}$) and the overall light neutrino mass scale. On the other hand, neutrino oscillation data fix (within experimental uncertainties) the mixing angles in the PMNS matrix and the mass differences in the light neutrino spectrum, although current data does not allow to disentangle between two possibilities for the ordering of light neutrino masses (normal ordering (NO) and inverted ordering (IO)).
We note however that global fits of neutrino oscillation data currently show a preference for NO at $3\sigma$~\cite{Capozzi:2018ubv,deSalas:2018bym}, and current experiments are starting to constrain the value of the Dirac phase $\delta_\text{CP}$~\cite{Abe:2017aap,Cao:2018vdk,NOvA:2018gge}.
In our scan we randomly generate Yukawa couplings $F$ accordingly to the relation~(\ref{eq:F_CI}), using the ranges of input parameters reported in Table~\ref{tab:parameters}.
{For the heavy neutrino mass splittings and the complex angles in ${\cal R}$, we randomly alternate between drawing our parameters from a linear versus a logarithmic distribution. This enables us to effectively sample the different regions of the parameter space, including the $B-\bar L$ protected regime as well as different flavour and mixing structures. }
The PMNS mixing angles and light neutrino mass splittings (as well as the PMNS Dirac phase $\delta_\text{CP}$) are allowed to vary in the $3\sigma$ ranges as determined by the NuFIT collaboration~\cite{Esteban:2016qun}.

\vspace{3mm}

\noindent \textbf{Targeted scans.} In order to efficiently explore the most interesting regions of the parameter space in a reasonable timescale, we run three different scans: \label{pageref:targetedscans}
\begin{itemize}
\item \textbf{generic} scan: all the generated  points complying with neutrino constraints and reproducing the observed BAU value are collected; 
\item \textbf{large mixing} targeted scan: the BAU value is only computed for points featuring a mixing $U^2>10^{-4}\ ({\text{GeV}}/M_2)^2$. This region is especially interesting because it can be probed by Belle~II, LHCb, ATLAS and CMS;
\item \textbf{low mass region} targeted scan: we only generate solutions where the lightest of the heavy neutrino masses ($M_1$) is lighter than $0.35 \text{ GeV}$. This region can be probed in the decay of kaons, for instance by T2K.
\end{itemize}

\begin{center}
\begin{table}[htb]
\begin{tabular}{|c|c|c|c|}
\hline
Parameter & Description & Range of values & Distribution \\
\hline
$m_\text{lightest}$ & Lightest neutrino mass & $\left[ 10^{-10} , 0.12 \right]$ eV & Log\\
$M_1$ & 1st heavy neutrino mass & $\left[ 0.1, 50 \right] $ GeV & Log\\
$ M_2$ & 2nd heavy neutrino mass & 
Random choice $ \left\{ 
\begin{array}{c}
\left[ M_1, 50 \text{ GeV} \right] \\
M_1 \frac{2+\Delta}{2-\Delta} 
\end{array} 
\right. $ & 
\begin{tabular}{c}
Log \\
Log on $\Delta$
\end{tabular}
\\
$ M_3$ & 3rd heavy neutrino mass & 
Random choice $ \left\{ 
\begin{array}{c}
\left[ M_2, 50 \text{ GeV} \right] \\
M_2 \frac{2+\Delta}{2-\Delta} 
\end{array} 
\right. $ & 
\begin{tabular}{c}
Log \\
Log on $\Delta$ \\
\end{tabular}
\\
$\Delta$ & Relative heavy neutrino mass splitting & $\left[ 10^{-10}, 2 \right]$ & Log\\
$\text{Re} \,  \omega_{ij}$ & Real component of $\mathcal{R}$ angles & 
Random choice $ \left\{ 
\begin{array}{c}
\left[ 0, 2\pi \right] \\
\left[ 10^{-10}, 2 \pi \right] 
\end{array} 
\right. $ &
\begin{tabular}{c}
Linear \\
Log \\
\end{tabular} \\
$\text{Im} \, \omega_{ij}$ & Imaginary component of $\mathcal{R}$ angles &
Random choice $ \left\{ 
\begin{array}{c}
\pm \left[ 0, 13 \right] \\
\pm \left[ 10^{-10}, 13 \right] 
\end{array} 
\right. $ &
\begin{tabular}{c}
Linear \\
Log \\
\end{tabular}\\
$\delta_\text{CP} $ & Dirac PMNS phase & $\begin{array}{c}
\left[144^\circ, 374^\circ \right] \text{ (for NO) }\\
\left[192^\circ, 354^\circ \right] \text{ (for IO) }
 \end{array} $
 & Linear \\
$\alpha_i$ & Majorana PMNS phases & $\left[ 0 , 2\pi \right]$ & Linear \\
\hline
\end{tabular}
\caption{Range of values and distribution of the free parameters sampled in the numerical scan. \emph{Random choice} means that, in the generation of each point, one of the described alternatives is randomly chosen. The heavy  neutrino masses $M_i$ are labelled following $M_1<M_2<M_3$.}
\label{tab:parameters}
\end{table}
\end{center}

\subsection{Further experimental constraints.}\label{sec:ExpConstr}
The realisations of the seesaw mechanism constructed as outlined in Section~\ref{sec:randomisation} (by construction compatible with the neutrino oscillation data) are compared to the following experimental constraints (cf.\ e.g.\ Ref.~\cite{Drewes:2015iva} for a more detailed discussion):
\begin{itemize}
\item The sum of light neutrino masses must remain below $\sum_i m_i < 0.12$ eV, as determined by the Planck collaboration~\cite{Aghanim:2018eyx}.
\item The neutrinoless double $\beta$ decay effective mass must remain below $m_{\beta\beta} < 0.165$ eV, as determined by the KamLAND-Zen collaboration~\cite{Shirai:2017jyz}. The computation includes the contribution of the light active as well as of the heavy sterile neutrinos (see~\cite{Bezrukov:2005mx,Blennow:2010th,Faessler:2014kka,Abada:2014vea,Drewes:2016lqo,Asaka:2016zib,Hernandez:2016kel,Abada:2017jjx,Babic:2018ikc}).
\item We impose constraints on the deviation from unitarity of the PMNS mixing matrix, as determined in~\cite{Drewes:2015iva,Fernandez-Martinez:2016lgt}.
\item We impose bounds from direct searches of heavy neutral leptons relevant in the considered mass range (i.e.\ $[0.1, 50]$ GeV), constraining the mixing of the new sterile neutrinos with the light active ones in the electron~\cite{Yamazaki:1984sj,CooperSarkar:1985nh,Bergsma:1985is,Badier:1986xz,Bernardi:1987ek,Akrawy:1990zq,Britton:1992xv,Adriani:1992pq,Baranov:1992vq,Adriani:1993gk,Abreu:1996pa,PIENU:2011aa,Liventsev:2013zz,Antusch:2015mia,Sirunyan:2018mtv}, muon~\cite{Hayano:1982wu,Yamazaki:1984sj,CooperSarkar:1985nh,Bergsma:1985is,Badier:1986xz,Bernardi:1987ek,Akrawy:1990zq,Adriani:1992pq,Adriani:1993gk,Vilain:1994vg,Gallas:1994xp,Abreu:1996pa,Vaitaitis:1999wq,Kusenko:2004qc,Liventsev:2013zz,Artamonov:2014urb,Antusch:2015mia,Aguilar-Arevalo:2017vlf,Sirunyan:2018mtv} and tau~\cite{Akrawy:1990zq,Adriani:1992pq,Adriani:1993gk,Abreu:1996pa,Astier:2001ck,Orloff:2002de,Antusch:2015mia} flavours.
\item We require an upper bound on the lifetime, requiring the sterile neutrinos to not be too long-lived in order to not spoil predictions from Big Bang Nucleosynthesis (BBN). We set a conservative upper bound of $0.1$ seconds, cf.\ Ref.~\cite{Ruchayskiy:2012si}. 
\end{itemize}

\subsection{Theoretical considerations: parameter volume and tuning.}
In addition to the experimental constraints listed above we apply a number of theoretical arguments.
\begin{itemize}
\item \textbf{Perturbative unitarity}. We require for each state that the corresponding decay width does not exceed half of the particle's mass~\cite{Chanowitz:1978mv,Durand:1989zs,Korner:1992an,Bernabeu:1993up,Fajfer:1998px,Ilakovac:1999md}.
\item \textbf{Perturbativity.} Although the parameterisation in Eq.~(\ref{eq:F_CI}) allows for an efficient exploration of the parameter space, the complex angles $\omega_{ij}$ cannot acquire arbitrary values: the magnitude of the Yukawa couplings $F$ grows exponentially with the modulus of the imaginary parts $\Im \ \omega_{ij}$, and too large couplings are excluded, either because they break the perturbative expansion leading to Eq.~(\ref{eq:mnu_full_low}), thus rendering the full parameterisation unreliable, or because they give rise to a strong dynamics.  
In our scan we allow for sizeable values of the imaginary angles, and explicitly diagonalise the full $6\times 6$ mass matrix (including 1-loop corrections) in order to verify the agreement with experimental data, excluding realisations that do not comply with them; moreover we require each entry in $|F|$ to be smaller than $4\pi$.
\item \textbf{Fine-tuning.}
In the exploration of the parameter space we do not impose any symmetry, but we allow the underlying parameters in the theory to vary as reported in Table~\ref{tab:parameters}, in order to generate symmetry protected as well as generic solutions. 
We then quantify a posteriori the level of fine-tuning for each solution, by defining the following quantity
\be\label{eq:fine_tuning} 
f.t. (m_\nu) = \sqrt{\sum_{i=1}^3 \left(\frac{m_i^\text{loop}-m_i^\text{tree}}{m_i^\text{loop}}\right)^2},
\ee
where $m_i^\text{loop}$ are the light neutrino masses computed at 1-loop level, while $m_i^\text{tree}$ are the same observables computed neglecting loop corrections.
The parameter $f.t.$ in Eq.~(\ref{eq:fine_tuning}) quantifies the importance of the loop corrections for reproducing the observed neutrino mass spectrum: the smaller it is the more neutrino masses are stable under radiative corrections, suggesting the presence of an underlying symmetry if Yukawa couplings are sizeably larger than the naive seesaw scaling $\left| F \right| \lesssim 10^{-7} \sqrt{\bar{M}/\text{GeV}}$.
\end{itemize}


\section{Results}\label{sec:Results}

In this section we discuss the results obtained performing the parameter scan described in Section~\ref{sec:scanstrategy}. Projecting the high-dimensional data set consisting of all parameter points meeting the experimental constraints (including the requirement of successful leptogenesis) on to different physically meaningful two-dimensional planes, we illustrate the qualitative new features arising in the $n=3$ case of `` freeze-in  leptogenesis''.

\subsection{The range of allowed mass and mixing. \label{subsec:mixing}} 
\begin{figure}
\centering
\vspace{-0.3cm}
\includegraphics[width=0.49\textwidth]{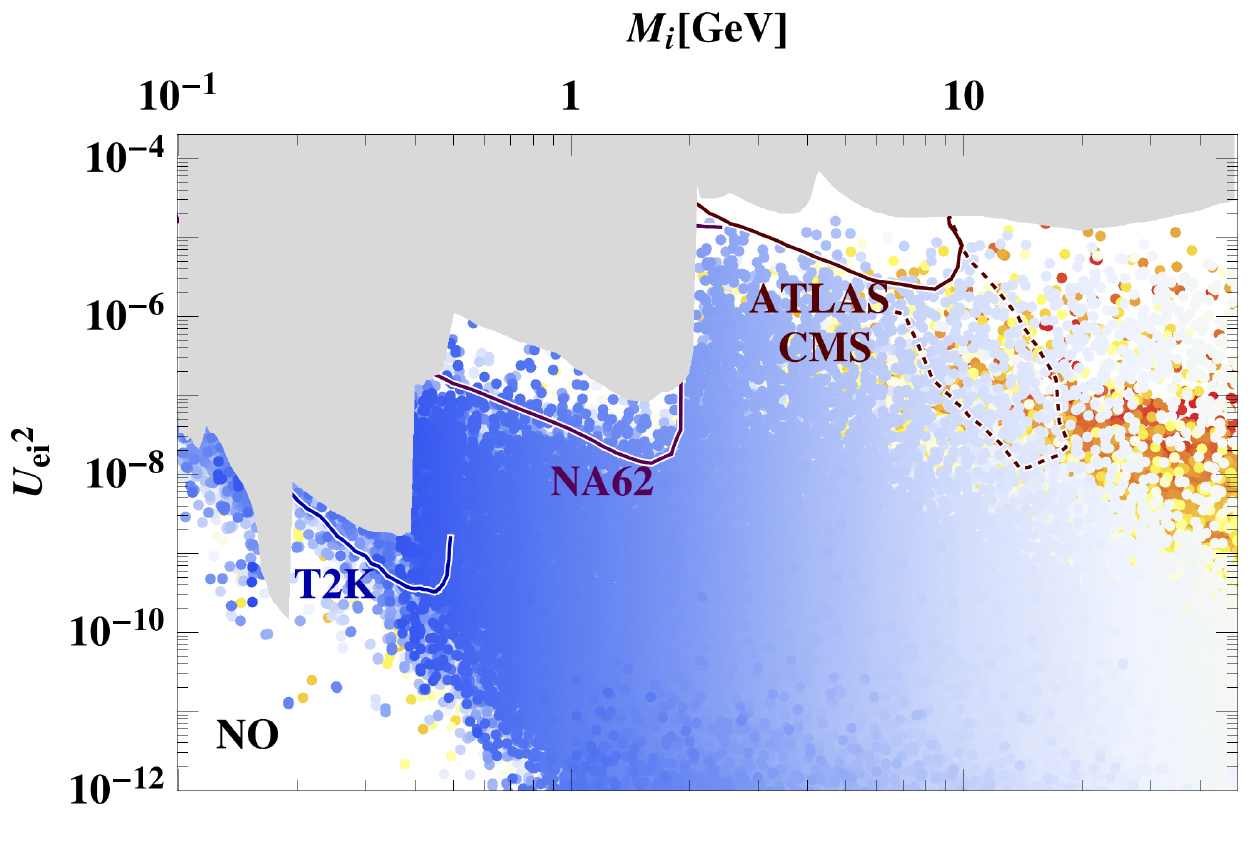}
\hspace{-0.3cm}\vspace{-0.35cm}
\includegraphics[width=0.5\textwidth]{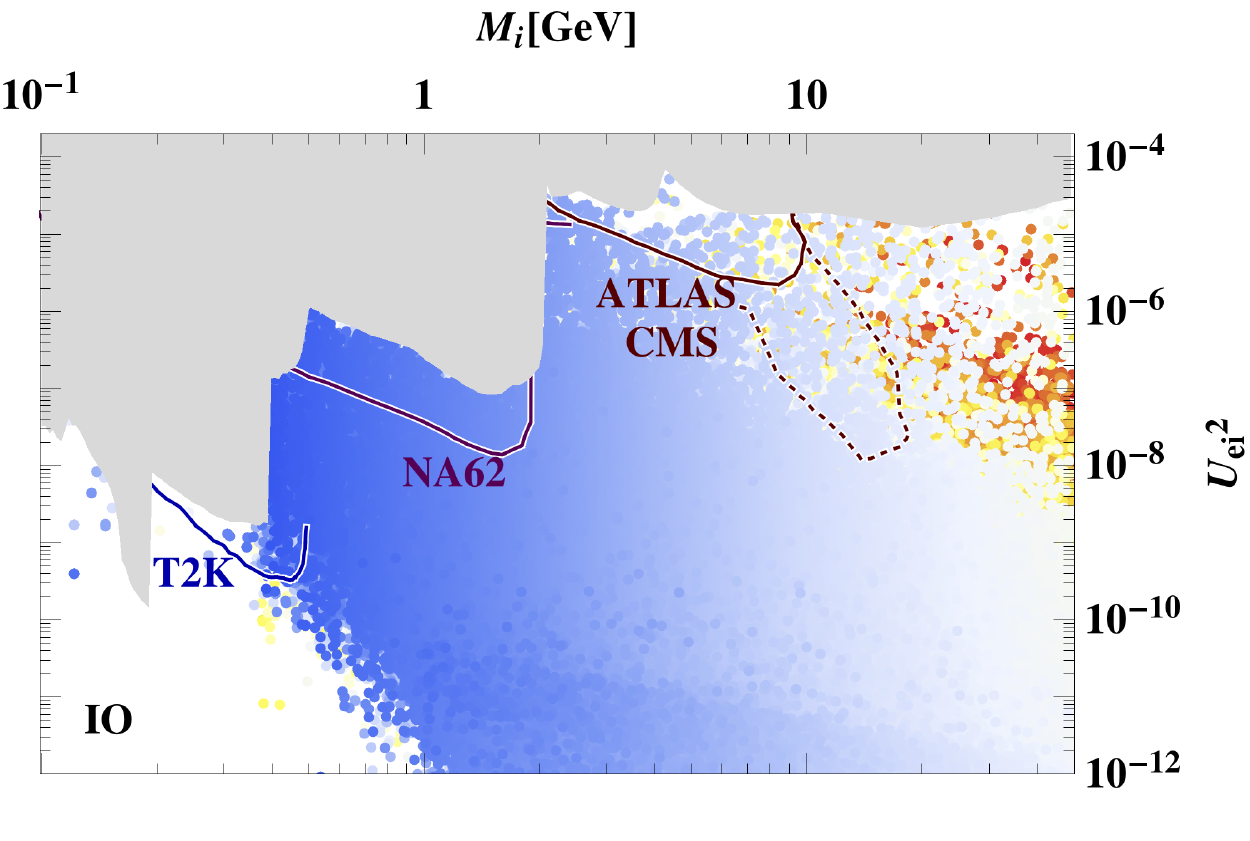}
\includegraphics[width=0.5\textwidth]{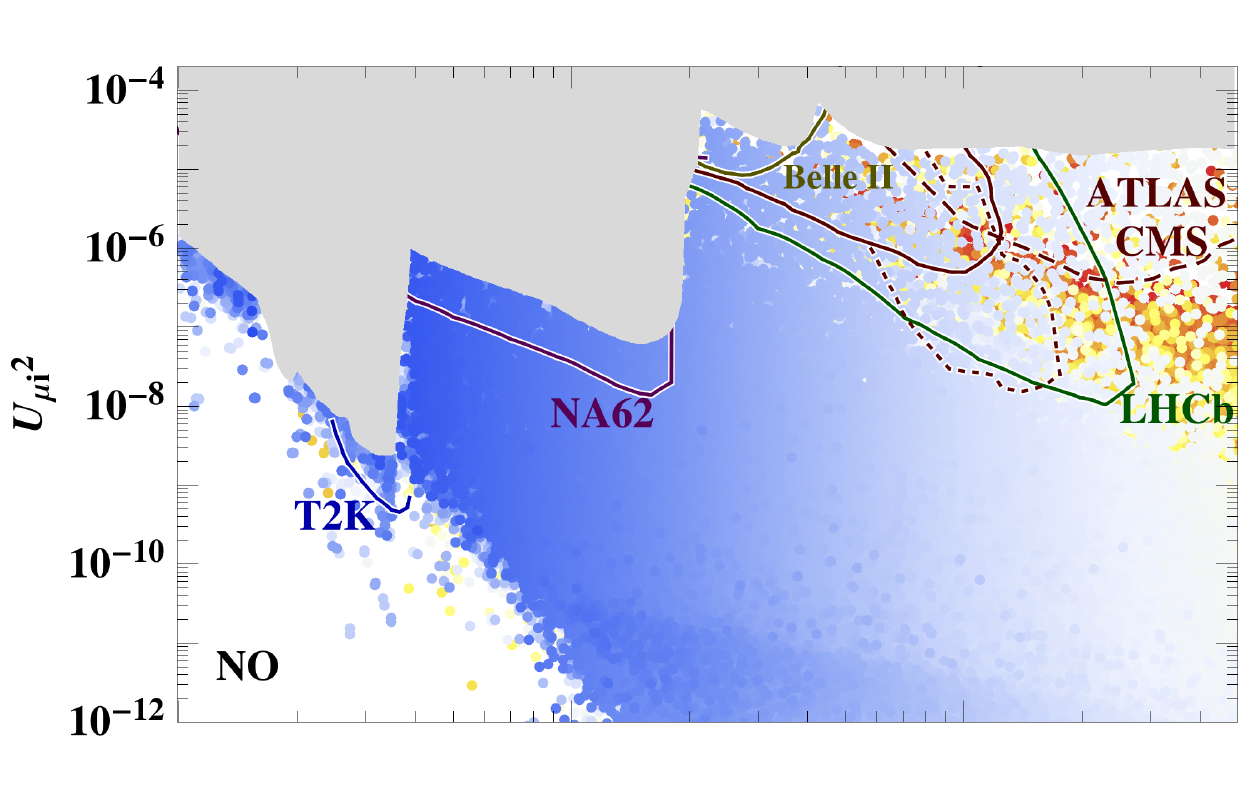}
\hspace{-0.3cm} \vspace{-0.35cm}
\includegraphics[width=0.5\textwidth]{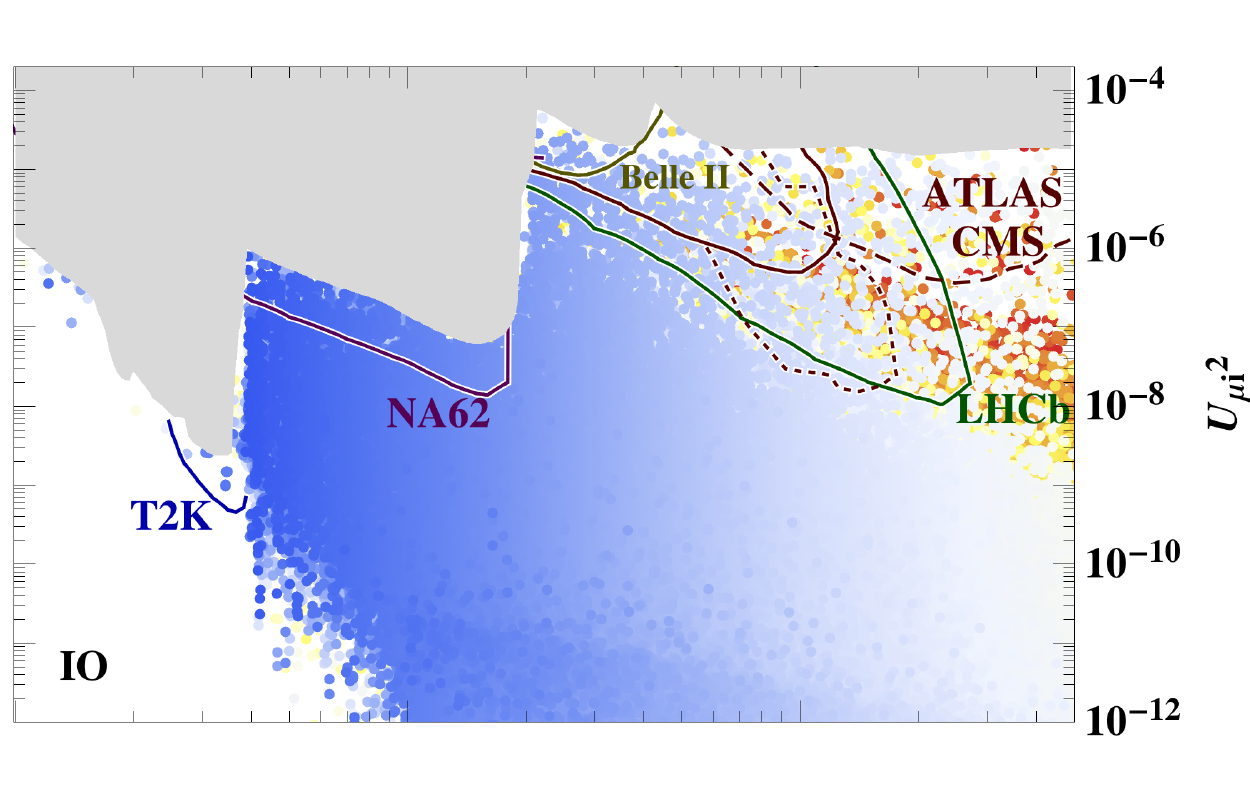}
\includegraphics[width=0.5\textwidth]{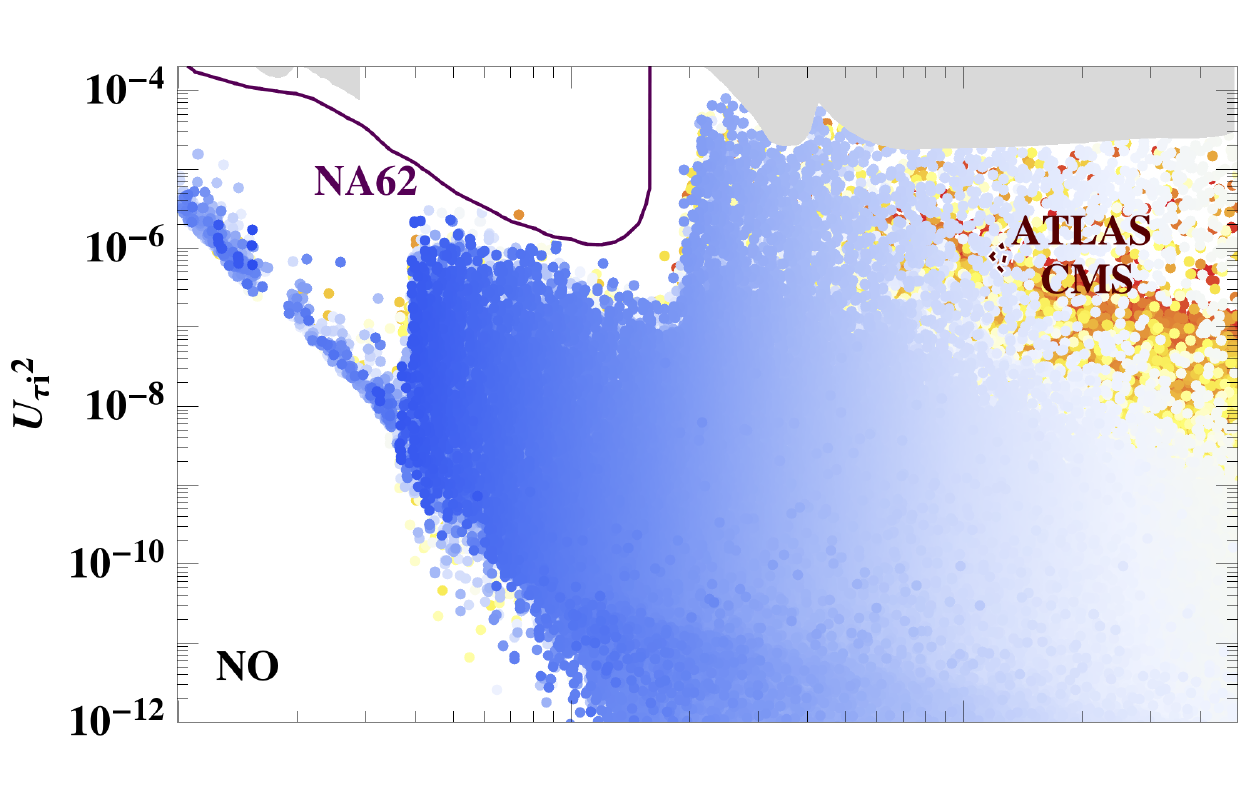}
\hspace{-0.3cm} \vspace{-0.35cm}
\includegraphics[width=0.5\textwidth]{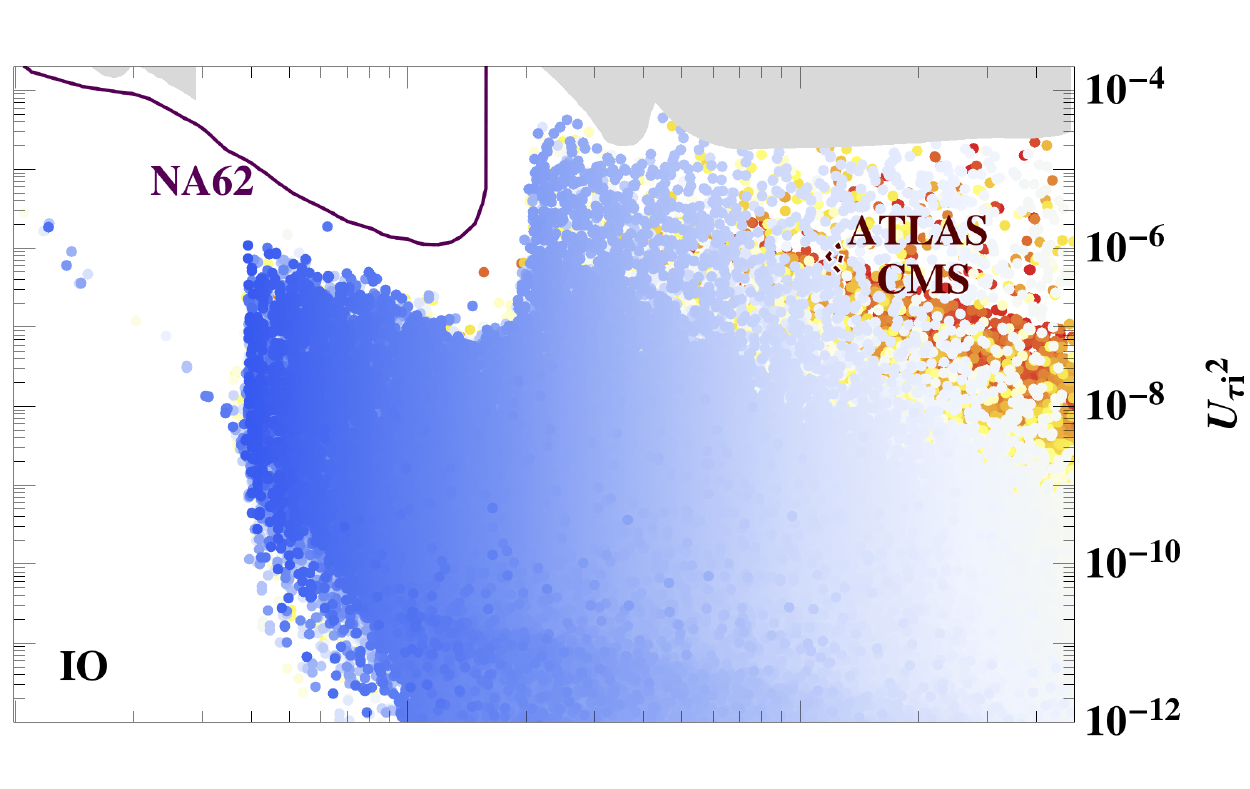}
\includegraphics[width=0.495\textwidth]{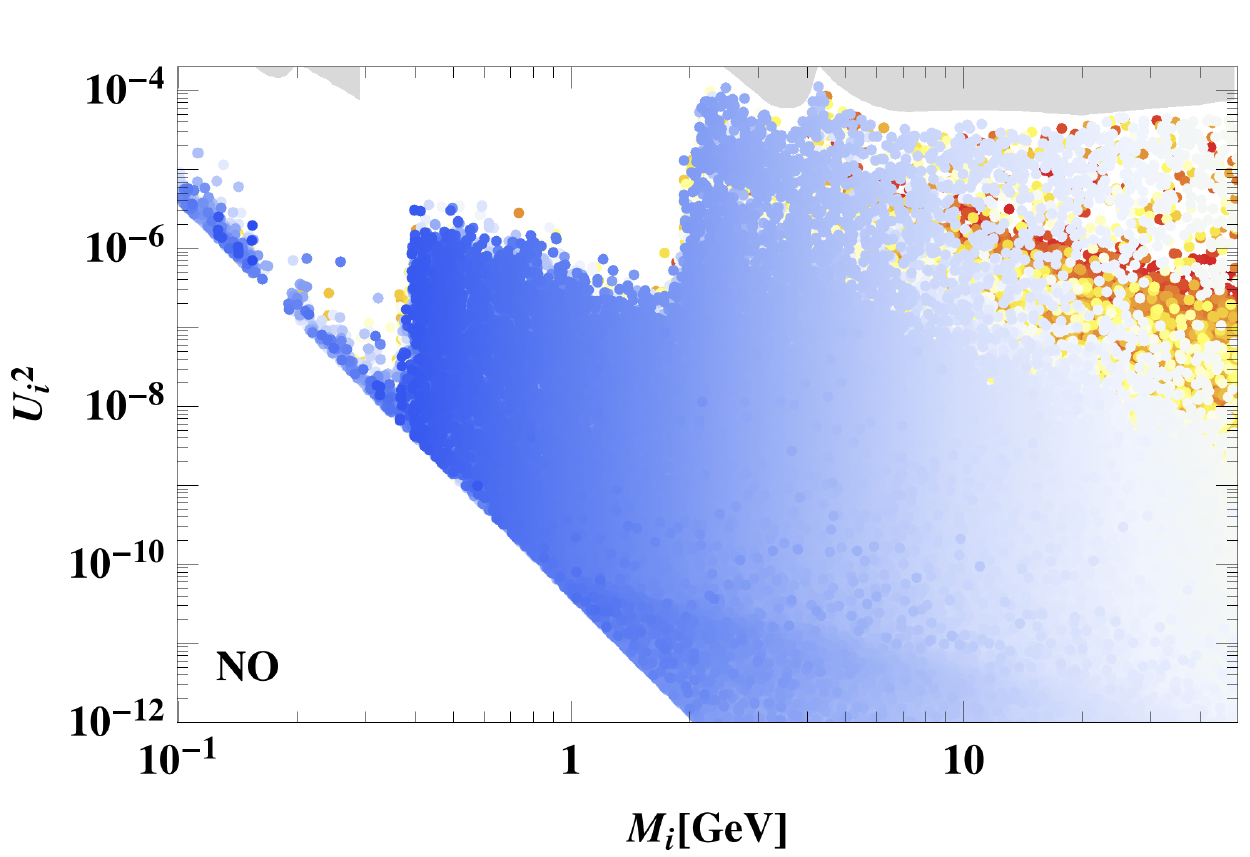}
\hspace{-0.3cm} \vspace{-0.35cm}
\includegraphics[width=0.5\textwidth]{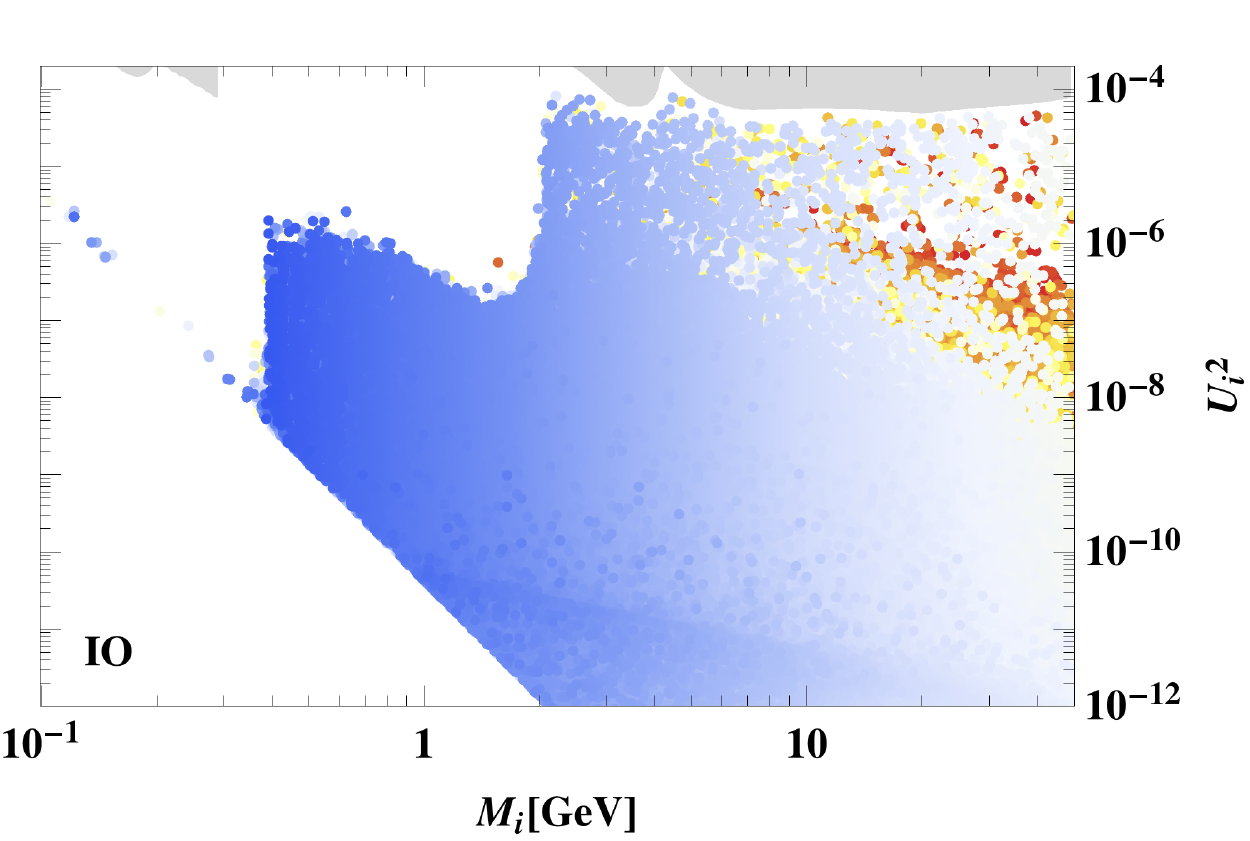}
\includegraphics[width=0.3\textwidth]{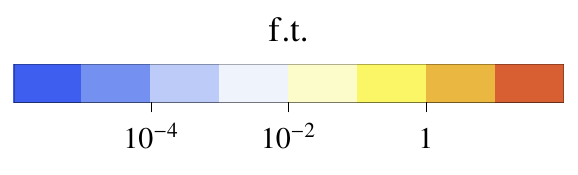}
\vspace{-0.3cm}
\caption{Active-sterile mixing for the viable BAU solutions as a function of the heavy  neutrino mass, for a normal (left) and an inverted (right) ordering in the light neutrino mass spectrum. From top to bottom: electron $U_{ei}^2$, muon $U_{\mu i}^2$, tau $U_{\tau i}^2$ and summed $U_i^2$ mixings. The grey region is excluded by direct searches of heavy neutral leptons (cf.\ Section \ref{sec:ExpConstr}), the lines show the expected sensitivities for the ongoing experiments T2K~\cite{Asaka:2012bb}, NA62~\cite{Drewes:2018gkc}, Belle II~\cite{Asaka:2016rwd}, LHCb~\cite{Antusch:2017hhu} with an integrated luminosity of 380 fb${}^{-1}$, and for ATLAS and CMS with an integrated luminosity of 300 fb${}^{-1}$. The latter include different proposed searches: \cite{Abada:2018sfh} (continuous line), \cite{Izaguirre:2015pga} (dashed line), \cite{Cottin:2018nms} (dotted line).}
\label{fig:mass_mixing}
\end{figure}

Figure~\ref{fig:mass_mixing} depicts the allowed range of active-sterile mixing after imposing all experimental constraints as a function of the heavy neutrino mass. 
We find that large mixing angles $U_{ai}^2$ right up to the current experimental bounds are allowed in the $n=3$ case across the entire mass range we consider.
This is in contrast to the model with $n=2$, where a gap of one order of magnitude was reported in \cite{Antusch:2017pkq} for $M_i\sim 5$~GeV that grows to about three orders of magnitude for $M_i\sim 50$~GeV. 
Moreover, we find points with very low fine-tuning (according to the criterion of Eq.~\eqref{eq:fine_tuning}) in the entire viable parameter space projected on to the mass-mixing plane.
This provides rich prospects for ongoing and planned experiments searching for sterile neutrinos in the GeV range. 
In particular, in contrast to the $n=2$ scenario, searches for prompt decays of $N_i$ at the LHC~\cite{Sirunyan:2018mtv,Aad:2015xaa,Ossowska:2018ybk} and Belle II~\cite{Liventsev:2013zz,Harrison:2015bja} can probe the viable leptogenesis parameter space for $n=3$. Moreover, in the region of large mixings and for $M_i$ below $\sim 20$ GeV, displaced vertex searches at the LHC \cite{Aad:2015xaa,Izaguirre:2015pga,Gago:2015vma,Antusch:2017hhu,Cottin:2018kmq,Abada:2018sfh,Drewes:2018xma} could see thousands of events, assuming that displacements in the mm range can be resolved. 
This would allow for a determination of the heavy neutrino flavour mixing pattern \cite{Caputo:2016ojx,Antusch:2017pkq}, which is crucial to test the hypothesis that these particles are responsible for leptogenesis \cite{Hernandez:2016kel,Drewes:2016jae}. For $n=2$ the sensitivity of such searches could barely touch the viable leptogenesis parameter space \cite{Chun:2017spz}, and it seems unlikely that the flavour mixing pattern can be measured at a level that allows to draw any conclusions. Hence, the perspectives to test low-scale leptogenesis are much better in the scenario with $n=3$.

A few comments on the distribution of the points in the scatter plots in Fig.~\ref{fig:mass_mixing} are in place. The main purpose of these plots is to illustrate that leptogenesis is feasible in the entire mass-mixing plane without fine-tuning in the sense of Eq.~(\ref{eq:fine_tuning}). The density of points within the allowed area should not be misinterpreted as a measure for any theoretical or experimental preference for particular values. Instead, it is primarily a result of the parameterisation (\ref{eq:F_CI}) and the randomisation procedure described in Section~\ref{sec:randomisation}. 
In particular, some of the most prominent features in the distribution of points appear because we performed a number of targeted scans as described on page~\pageref{pageref:targetedscans}. In addition to the variation in the density of points within the allowed region, there are also parts of the mass-mixing planes that appear to be empty. This does not necessarily imply that there are no viable parameter choices in these regions, but may also simply indicate that our scan failed to fully exploit these regions. 
For instance, the distribution of points above $M_i=2$ GeV suggests that leptogenesis is feasible for mixings all the way up to the experimental upper limit on the individual $U_{ai}^2$, but not all the way up to the experimental upper limit on the total $U_i^2$. We suspect that the reason is that it is difficult to find points where all three mixings $U_{a i}^2$ are maximal for one of the $N_i$ within the parameterisation (\ref{eq:F_CI}), while there is no reason why such points would not exist. Similarly, it is very difficult to explore the region of large $U_{\tau i}^2$ below $M_i=2$ GeV, while there is 
evidence that, {at least from the point of view of neutrino mass generation}, this region is experimentally allowed for $n=3$  \cite{Drewes:2015iva}. 
This is in contrast to the $n=2$ model, where the results presented in Ref.~\cite{Drewes:2016jae} indicate that this region is indeed ruled out by the combination of different constraints.
Finally, a similar problem arises in the determination of the lower bound on the mixings. 
While the light neutrino oscillation data and the requirement for the $N_i$ to decay before BBN both impose lower bounds on the $U_i^2$ that depend on $m_{\rm lightest}$ \cite{Gorbunov:2013dta,Drewes:2015iva},
neither of them can impose a lower bound on the individual $U_{ai}^2$ for $n=3$. The BBN constraint can always be avoided if the $N_i$ decays into a SM final state of different flavour, while the neutrino oscillation data can always be explained if another heavy neutrino provides the required mixing with the flavour $a$. \\

\begin{figure}[t]
\centering
\includegraphics[width=0.478\textwidth]{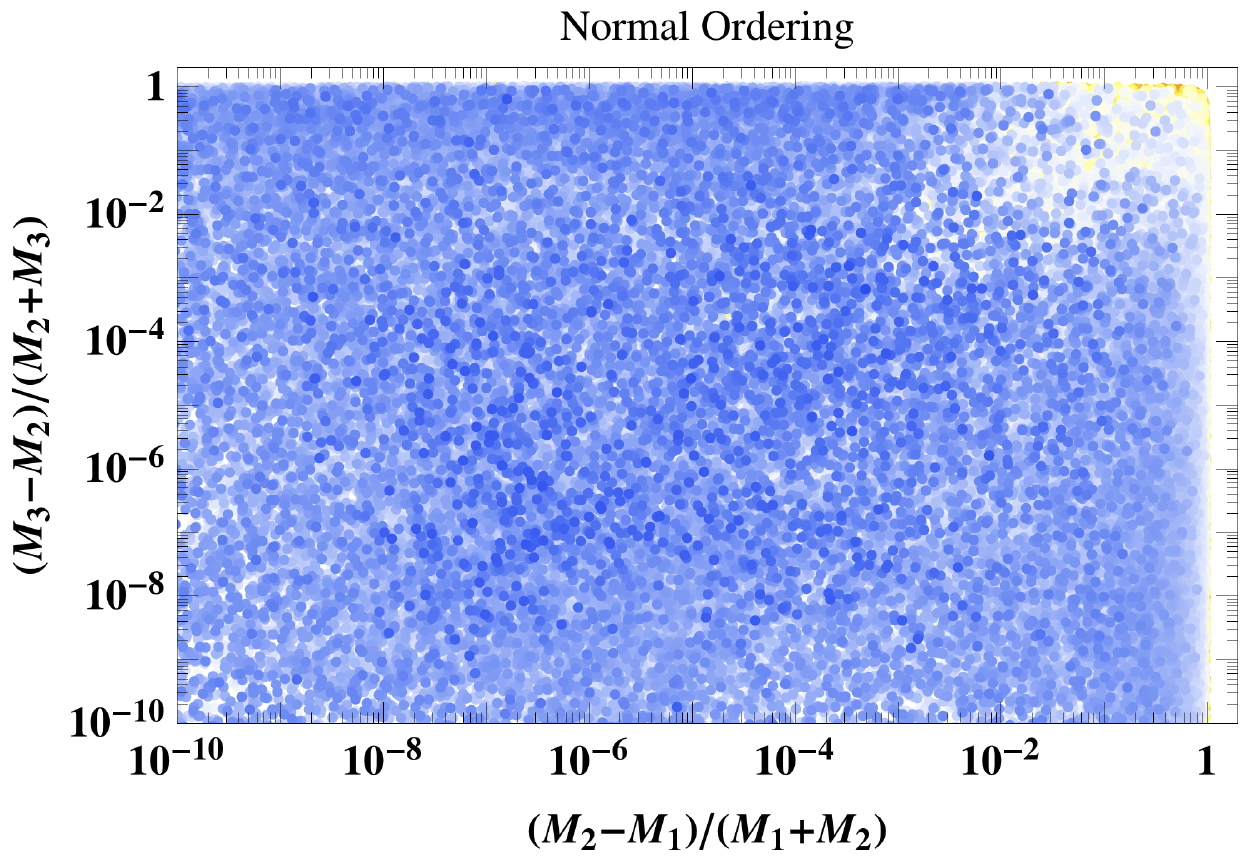}
\includegraphics[width=0.49\textwidth]{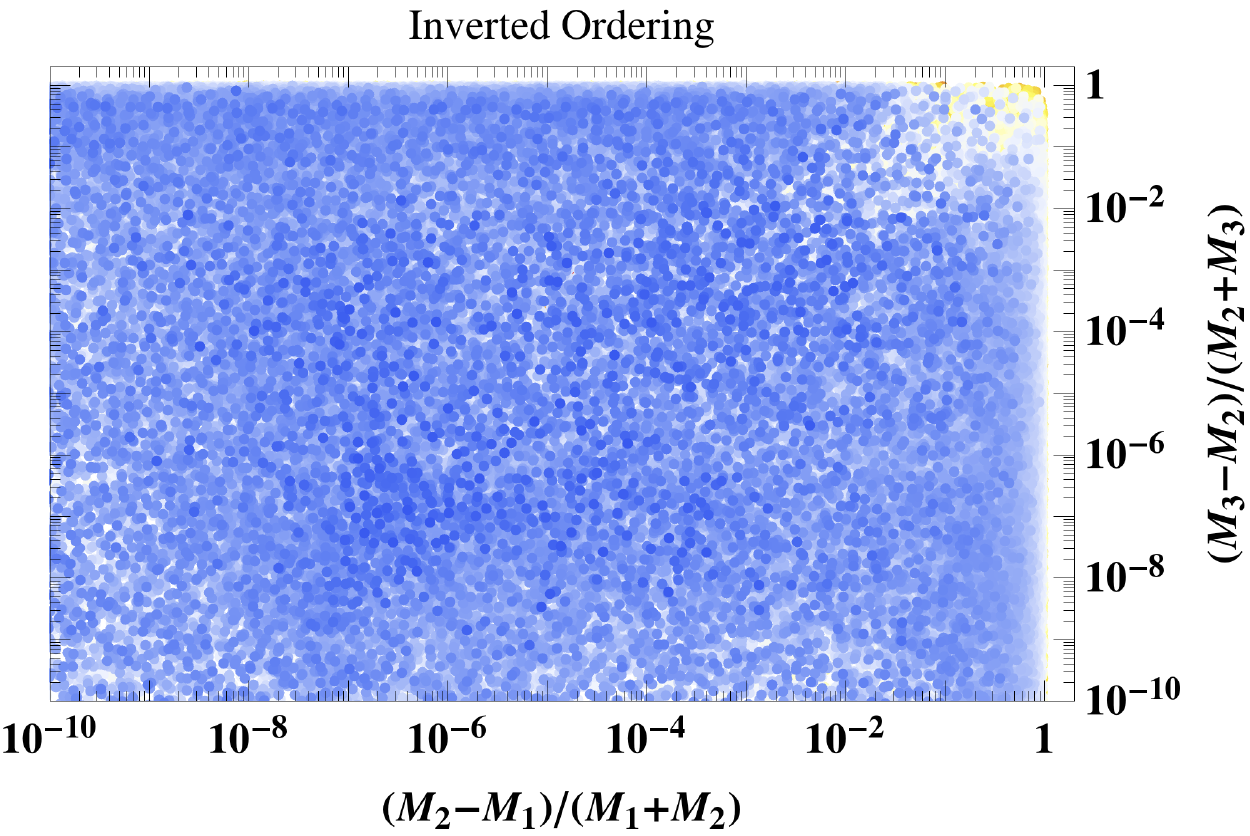}
\caption{Relative mass splittings for the viable BAU solutions in the model, for a normal (left panel) and inverted (right panel) ordering in the light neutrino mass spectrum. Colour coding as in Fig.~\ref{fig:mass_mixing}. }
\label{fig:degeneracy}
\end{figure}

\noindent \textbf{Relative mass degeneracy.} Fig.~\ref{fig:degeneracy} shows the parameter points of Fig.~\ref{fig:mass_mixing} projected onto a plane spanned by the two mass splittings among the heavy neutrinos.  
As discussed above, the density of the points carries little physical meaning. It is however remarkable that we find viable leptogenesis points in the entire parameter plane, for all possible hierarchies 
of the heavy neutrino masses, and covering a wide range of values for the physical mass differences. 
The regions along the top and right axes correspond to a situation with one pair of very degenerate neutrinos and a third neutrino with at least an ${\cal O}(1)$ hierarchy. 
This can be realised in the $B$$-$$\bar L$ symmetry protected regime for $\mu\ll1$ and $\mu'\sim1$ or physically equivalent configurations in which the labels of the $N_i$ are permutated. This region contains effective $n=2$ models if the third neutrino decouples. 
In the upper right corner both, $\mu$ and $\mu'$, are sizeable, and there is no protecting symmetry for the neutrino masses..
On the other hand, the central and bottom left area of Fig.~\ref{fig:degeneracy} is characterised by three very degenerate neutrinos, with in general all three of them contributing to leptogenesis. The low value of the fine-tuning (according to the criterion of Eq.~\eqref{eq:fine_tuning}) indicates that again this is a $B$$-$$\bar L$ protected region. Finally, in the top right corner we find the fully non-degenerate $N_i$ spectra, which can accommodate leptogenesis only at the cost of fine-tuning.

A further important difference between the $n=2$ and the $n=3$ case appears in the flavour structure, i.e.\ in the relative coupling strength of a given $N_i$ to the 3 active flavours, $U_{ei}^2/U_i^2 : U_{\mu i}^2/U_i^2 : U_{\tau i}^2/U_i^2$. For $n = 2$, the requirements of successful neutrino mass generation and leptogenesis limit the allowed values of these ratios, see Refs.~\cite{Hernandez:2016kel,Drewes:2016jae,Antusch:2017pkq}. On the contrary, for $n = 3$ we find parameter points yielding successful neutrino mass generation and leptogenesis in the entire parameter space. This provides a further interesting possibility to test  these leptogenesis mechanisms.

\subsection{Effect on neutrinoless double \texorpdfstring{$\beta$}{beta} decay.}  
It is well known that the exchange of $N_i$ with masses below the electroweak scale can make a significant contribution to the rate of neutrinoless double $\beta$ decay \cite{Bezrukov:2005mx,Blennow:2010th,Faessler:2014kka,Abada:2014vea} in the region where freeze-in leptogenesis is feasible \cite{Drewes:2016lqo,Asaka:2016zib,Hernandez:2016kel,Abada:2017jjx,Babic:2018ikc}.  
The decay rate is proportional to the quantity
\begin{equation}\label{mee}
m_{\beta\beta}=\left|
\sum_i (U_\nu)_{ei}^2m_i + \sum_i \Theta_{ei}^2M_i f_A(M_i)
\right|,
\end{equation}
where the first term comes from light neutrino exchange and the second one from $N_i$ exchange.
Here 
\begin{equation}\label{eq:0nubb_mass_contr}
f_A(M)\simeq \frac{\PP^2}{\PP^2+M^2}\,,
\end{equation}
where $\PP$ is the momentum exchange in the decay and depends on the isotope, cf.\ e.g.\ \cite{Faessler:2014kka}. In our analysis we use the numerical value $\PP = 125 \text{ MeV}$, resulting from an average over different decaying nuclei (see e.g.~\cite{Blennow:2010th}).
Using the estimate (\ref{NaiveSeesaw}) one would generically expect that the relative size of the two contributions is roughly given by $f_A(M_i)$. Since we found viable parameter points for which $U_i^2$ exceeds the estimate (\ref{NaiveSeesaw}) by several orders of magnitude, one may wonder whether leptogenesis with large mixing angles generally predicts that $m_{\beta\beta}$ greatly exceeds the standard contribution,
\begin{equation}\label{meeSS}
m_{\beta\beta}^\nu=\sum_i (U_\nu)_{ei}^2m_i \,,
\end{equation}
 from light neutrino exchange. 
However, large $U_i^2$ can only be achieved without fine-tuning if the light neutrino masses $m_i$ are protected by the $B-\bar{L}$ symmetry. This symmetry automatically suppresses $m_{\beta\beta}$ and sets the rate of neutrinoless double $\beta$ decay (as well as the light neutrino masses) to zero if the symmetry is exact. 
It is instructive to study how much "tuning" is required to obtain a large decay rate if the symmetry violating parameters are not exactly zero.
To see this explicitly we bring Eq.~(\ref{mee}) into  the form 
\begin{eqnarray}
m_{\beta\beta}&=&\left|
m_{\beta\beta}^\nu
+f_A(\bar{M})\sum_iM_i\Theta_{e i}^2
+\sum_iM_i\Theta_{e i}^2[f_A(M_i)-f_A(\bar{M})]
\right|\nonumber\\
&=&
\left|
[1-f_A(\bar{M})]m_{\beta\beta}^\nu
+(\delta m_\nu^{\rm 1loop})_{ee}f_A(\bar{M})
+\sum_iM_i\Theta_{e i}^2[f_A(M_i)-f_A(\bar{M})]
\right|,\label{meerewritten}
\end{eqnarray}
by using the unitarity relation $\sum_i m_i (U_\nu)_{a i}^2 + \sum_i M_i \Theta_{a i}^2 = (\delta m_\nu^{\rm 1loop})_{a a}$ (see Eq.~\eqref{eq:Mdiagonalization}). 
Further using Eq.~(\ref{1loopquantities}) and the fact that $\Theta\simeq\theta$ in the $B-\bar{L}$ conserving regime, we can recast this as  
\begin{eqnarray}
m_{\beta\beta}&=&\left|
[1-f_A(\bar{M})]m_{\beta\beta}^\nu 
+\sum_i M_i\theta_{e i}^2\left[ f_A(M_i) - f_A(\bar{M}) \left( 1 - \frac{M_i^2}{v^2}l(M_i) \right) \right]
\right|\label{meerewrittenagain}\,.
\end{eqnarray}
\begin{figure}[tb]
\centering
\includegraphics[width=0.478\textwidth]{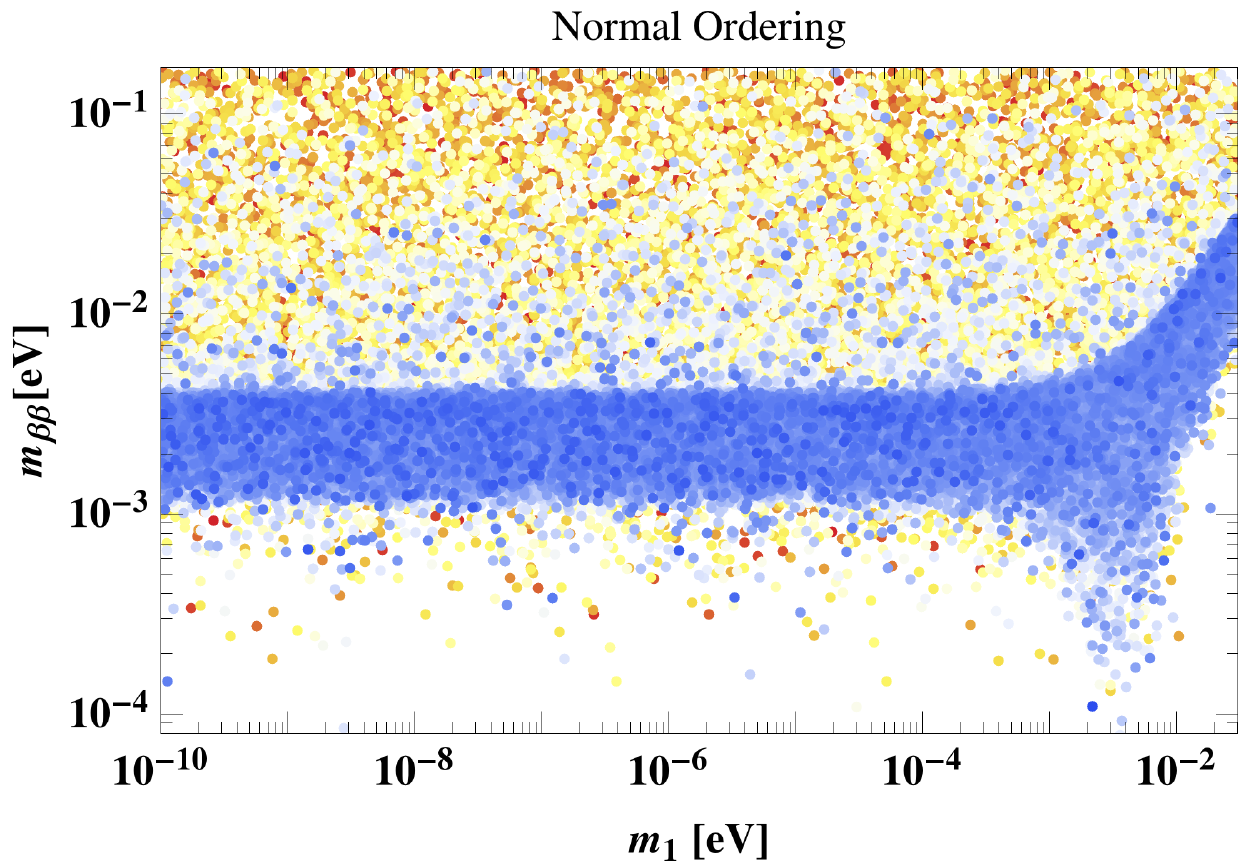}
\includegraphics[width=0.49\textwidth]{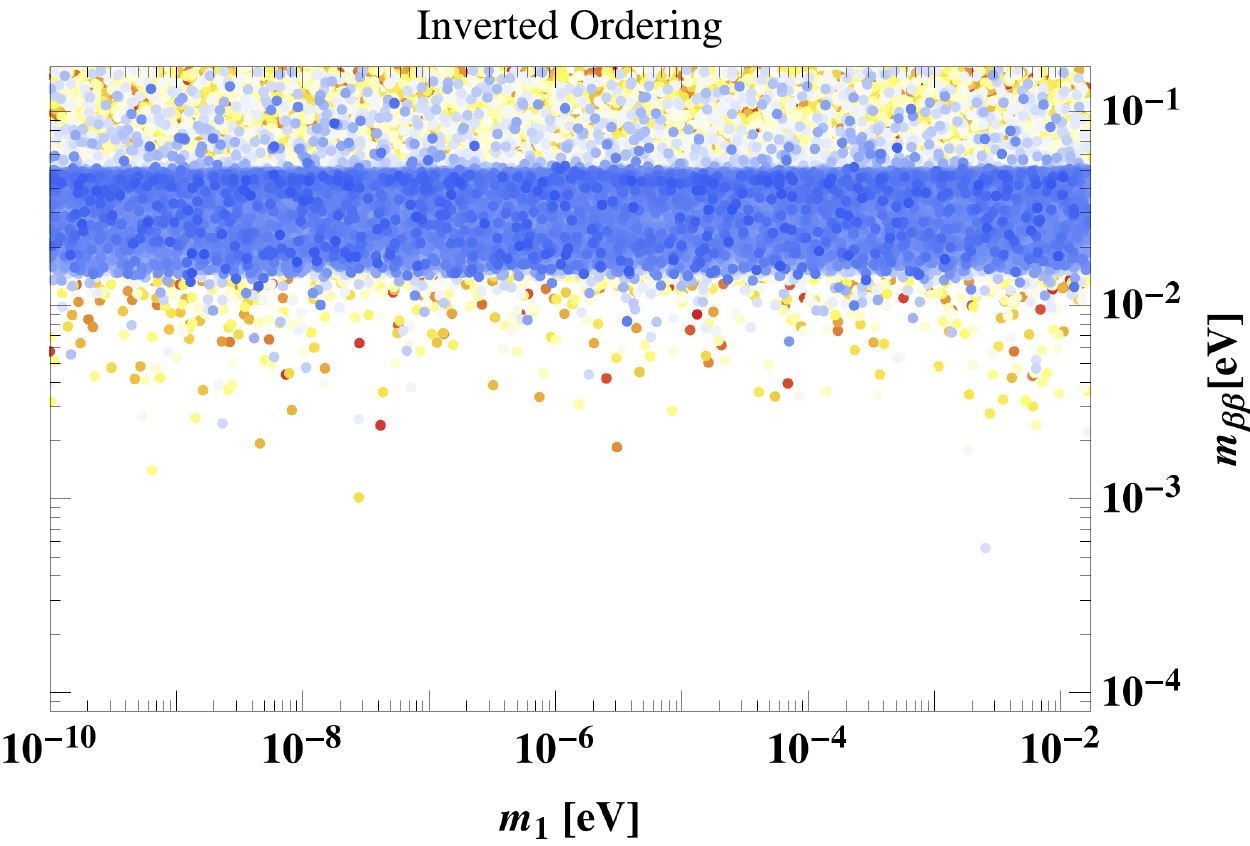}
\includegraphics[width=0.48\textwidth]{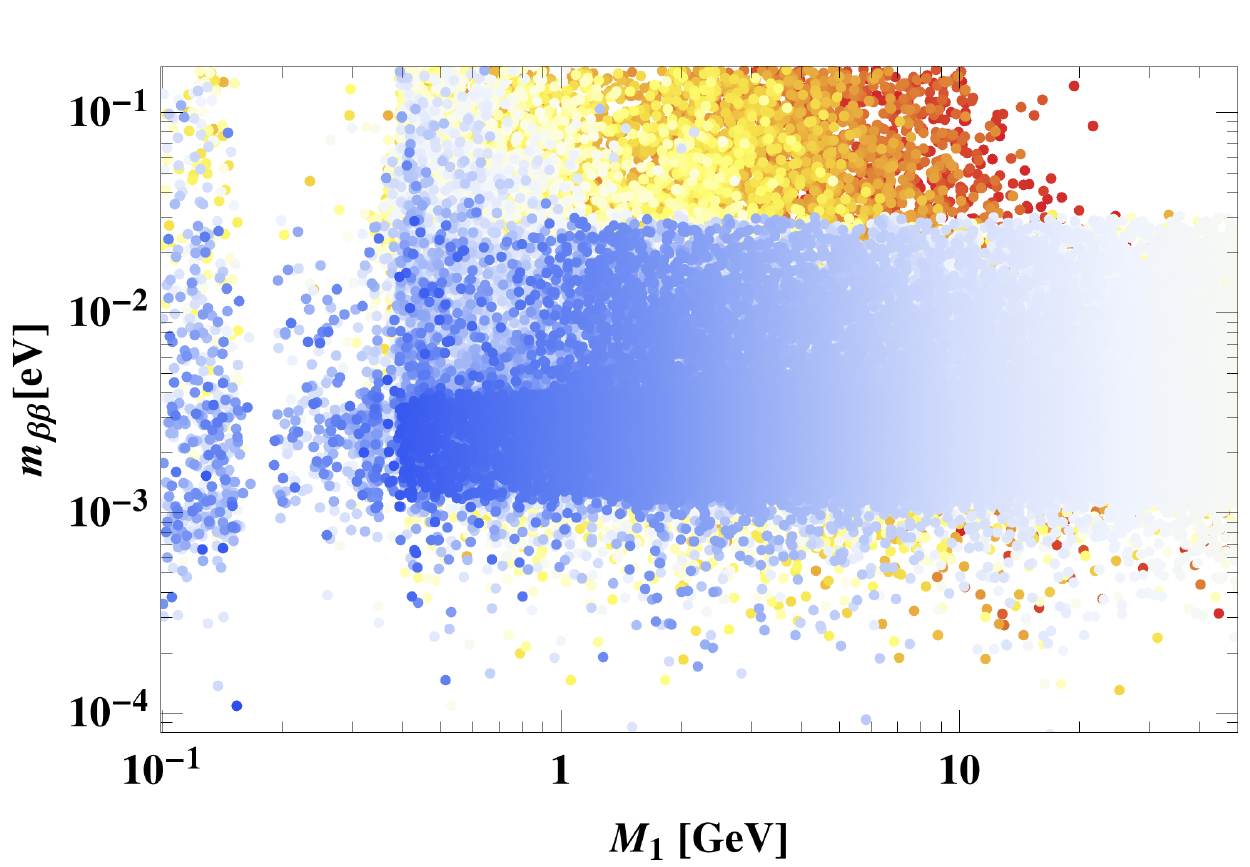}
\includegraphics[width=0.49\textwidth]{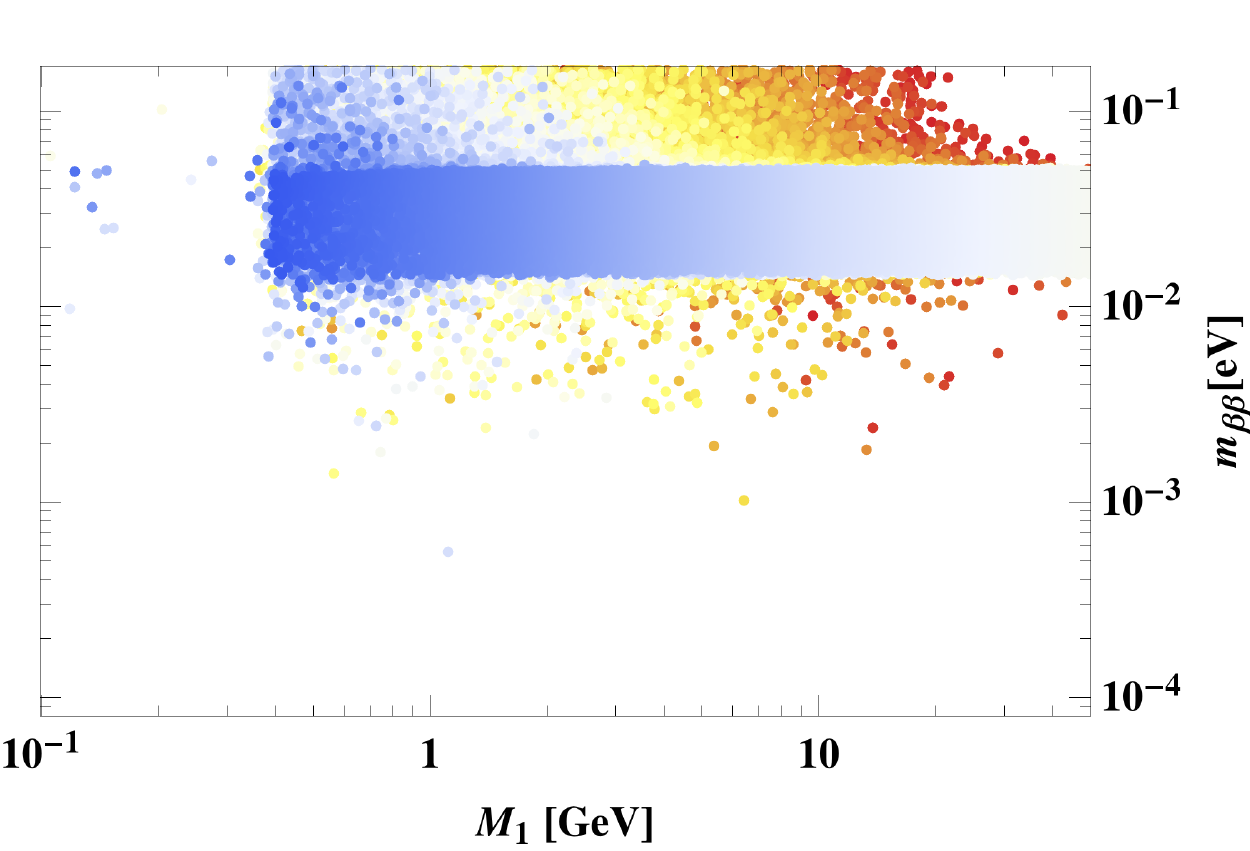}
\caption{Neutrinoless double $\beta$ decay effective mass values for the viable BAU solutions in the model as a function of the lightest neutrino mass (top panels) and of the lightest heavy neutrino mass (bottom panels), for a normal (left panels) and inverted (right panels) ordering in the light neutrino mass spectrum. 
The most prominent voids in the distribution of points inside the horizontal bands in the lower panel are the result of constraints on the $U_{ei}^2$ from direct experimental searches. 
Note that for normal ordering the SM model contribution $m_{\beta \beta}^\nu$ can be arbitrarily small, which is reflected by the blue band in the top left panel extending downwards at $m_1 \sim \text{few} \times 10^{-3}$~eV.  The low density of points in this region is a result of the sampling in our scan. 
Colour coding as in Fig.~\ref{fig:mass_mixing}.}
\label{fig:0nubb}
\end{figure}
The first term in this equation is always smaller than the standard prediction, so large contributions can at most come from the second term.\footnote{
It is straightforward to show that all eigenvalues of $m_\nu$ vanish if the parameters $\epsilon_a, \epsilon'_a$ and $\Mu$ in Eq.~(\ref{FullNeutrinoMass}) are set to zero, which implies that also $m_{\beta\beta}^\nu$ exactly vanishes in this limit.}
The contribution from $N_3$ is proportional to $\theta_{e 3}^2 = (v/\bar{M})^2\times F_e^2(\epsilon'_e/\Mu')^2$.
A priori this term looks potentially large in low-scale seesaw models because of the factor $(v/\bar{M})^2$ and because of the second power of $\Mu'$ in the denominator, which threatens to cancel the suppression from the $\epsilon'_e$ in the numerator. However, there are also at least two powers of $\Mu'$ in the numerator, one from $M_3 = \Mu' \bar{M}$ and one from expanding the $f_A(M_i)-f_A(\bar{M})$ in the tree-level contribution (the loop contribution comes with a prefactor $(M_3/v)^2 = \Mu^{' 2} (\bar{M}/v)^2$ anyway). What remains is a contribution $\propto F_e^2\epsilon_e^{' 2} v^2/\bar{M}\times f_A(\bar{M})^2 \bar{M}^2/\PP^2$, 
which should be compared to the contribution $\propto F_e^2 \epsilon_e^{' 2}  v^2/(\Mu' \bar{M})$  that $N_3$ makes to the neutrino masses. 
The current upper limit on the sum of neutrino masses \cite{Aghanim:2018eyx} is comparable to the limit on $m_{\beta\beta}$ \cite{Agostini:2018tnm}, hence a large contribution to the decay rate could only be achieved at the cost of cancellations in $m_\nu$ and/or $m_{\beta\beta}$ that are not explained by the $B-\bar{L}$ symmetry. 
For $\Mu'\ll 1$ 
the contributions from $N_1$ and $N_2$ are individually large because of the much larger mixing angle. However, due to the $B-\bar{L}$ symmetry, they interfere destructively, which is manifest in the imaginary unit $i$ in Eq.~(\ref{FullNeutrinoMass}).
To estimate their contribution, we expand to linear order in the $B-\bar{L}$ violating parameters
\begin{align}
 m_{\beta \beta}=\Bigg | [1-f_A(\bar{M})]m_{\beta\beta}^\nu + 2 \bar M  f_A(\bar{M}) \Bigg[  \Mu F_e^2  \left(\frac{v^2}{\PP^2} f_A(\bar{M}) - \frac{l(\bar{M})}{2} -   \frac{\bar{M}}{2} \frac{\partial l(\bar M)}{\partial \bar M}\right) + \epsilon_e F_e^2  l(\bar{M})\Bigg] \Bigg| \,.
\end{align}
This may again be compared to the contribution $\propto 2 F_e^2  (-2 \epsilon_e +  \Mu) v^2/\bar{M}$ that $N_1$ and $N_2$ make to neutrino masses through their mixing with $\nu_{L e}$. The term $\propto \epsilon_e F_e^2$ 
in $m_{\beta\beta}$ is always smaller than its counterpart in $m_\nu$ for $\bar{M}<v$,
while the term $\propto \Mu F_e^2 $ is parametrically of comparable size. 
Hence, for generic choices of the parameters that are not dictated by the symmetry, the current neutrino oscillation data clearly disfavours large contributions to $m_{\beta\beta}$ from the heavy neutrinos. Using the expression (\ref{meeSS}) for $m_{\beta\beta}^\nu$ and the numerical values of the mixing $(U_\nu)_{ei}$, the same argument suggests that the contribution from the $N_i$ exchange to $m_{\beta\beta}$ is comparable or smaller than that from light neutrinos.
Based on this, one can estimate that $m_{\beta\beta}$ should not greatly exceed the standard prediction $|m_{\beta\beta}^\nu|$ unless the model parameters are either highly tuned to cause accidental cancellations amongst the contributions involving different SM flavours in $m_\nu$, or there exist additional flavour structures/symmetries that lead to such cancellations.
There is, however, one way to avoid this conclusion that has already been discussed for $n=2$ in Refs.~\cite{Drewes:2016lqo,Hernandez:2016kel,Asaka:2016zib}: large deviations from $m_{\beta\beta}^\nu$ can be obtained in a technically natural way if the $M_i$ are of the same order as $\PP$.
This results from the combination of two factors. On the one hand the contribution of heavy neutrinos is maximal if their masses are comparable with the exchanged virtual momentum $\PP$, cf.\ Eqs.~(\ref{mee}, \ref{eq:0nubb_mass_contr}); on the other hand loop corrections to the light neutrino parameters are proportional to the heavy neutrino masses, cf.\ Eq.~(\ref{eq:blocks_mass_matrix}).

This is confirmed by the results shown in Fig.~\ref{fig:0nubb}. The plot confirms the claim from Ref.~\cite{Drewes:2016lqo} that leptogenesis in the $n=3$ low-scale seesaw model is compatible with both, a rate of neutrinoless double $\beta$ decay that is much larger or much smaller than the standard prediction $|m_{\beta\beta}^\nu|$. However, a much larger rate tends to require a considerable tuning in the sense of Eq.~(\ref{eq:fine_tuning}). 
The lower panels in Fig.~\ref{fig:0nubb} confirm that sizeable contribution from the heavy neutrinos to $m_{\beta\beta}$ can be achieved together with a low fine-tuning (in the sense of Eq.~\eqref{eq:fine_tuning}) for masses of order $\mathcal{O}(100 \text{ MeV})$. 

\section{Benchmark points \label{sec:benchmarks}}

Figures \ref{fig:mass_mixing}-\ref{fig:degeneracy} illustrate the larger viable parameter space with $n=3$ compared to $n=2$. Due to the effects \ref{mechanism1}) - \ref{mechanism8}) listed in Section~\ref{sec:mechanisms}, the range of parameters for which both, the BAU and light neutrino oscillation data, can be explained increases in all possible directions.
\begin{itemize}
\item Both, larger and smaller mixings $U_i^2$ can be made consistent with baryogenesis and neutrino mass generation, cf.\ Fig.~\ref{fig:mass_mixing}. In the entire mass range studied here, the upper limit on $U_i^2$ is practically given by experimental constraints. That is, for any value of $U_i^2$ that is allowed by experiments, one can find a set of model parameters for which baryogenesis is feasible. This considerably improves the perspectives for current and planned experiments to test the mechanism of baryogenesis. At the same time, there is no lower bound on the individual $U_i^2$. This is in contrast to the case with $n=2$, where the estimate (\ref{NaiveSeesaw}) practically acts as a "floor" for experimental searches.
\item The constraints on the heavy neutrino mass spectrum are relaxed. In particular, no mass degeneracy is needed to generate the BAU.
\item The constraints on the flavour mixing pattern, i.e., the relative size of the heavy neutrino couplings to different SM flavours, are relaxed.
\end{itemize}
Some of these effects have been predicted in the past.
For instance, in Ref.~\cite{Canetti:2014dka} it was argued that the relaxed constraints on the flavour mixing parameter $\flav$ should allow for baryogenesis with much larger $U_i^2$ than for $n=2$. 
The fact that baryogenesis with $n=3$ is feasible for non-degenerate heavy neutrino spectra was discussed in detail in Ref.~\cite{Drewes:2012ma}. 
Different aspects of the $\tilde{L}$-violation have been discussed in Refs.~ \cite{Hambye:2016sby,Ghiglieri:2017gjz,Ghiglieri:2017csp,Antusch:2017pkq,Eijima:2017anv,Eijima:2018qke}.
However, it turns out that the behaviour for $n=3$ in general is much richer than anticipated in these works. In general, the evolution of charges is governed by a complex interplay of several amongst the effects \ref{mechanism1}) - \ref{mechanism8}) in Section~\ref{sec:mechanisms}, and one cannot uniquely relate the viability of a particular parameter choice to any individual of these mechanisms. 
It is nevertheless instructive to illustrate some of the most important physical effects for a few selected benchmark points. The parameters of these points are summarised in Table~\ref{tab:bench_parameters}.

\begin{center}
\begin{figure}[t]
\includegraphics[width = \textwidth]{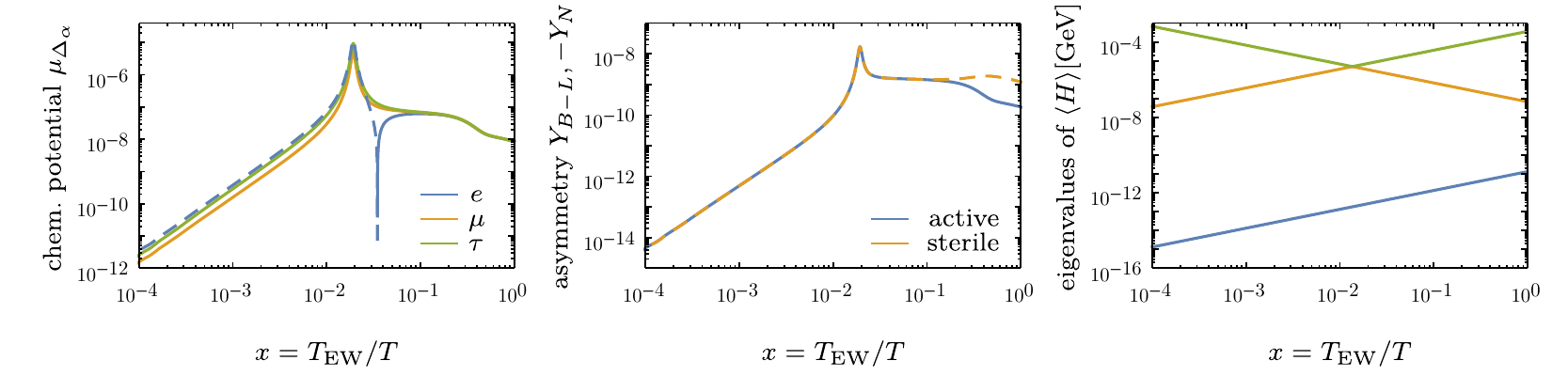}
\caption{Benchmark point I). Left: asymmetries in the individual active flavours. Center: sum of asymmetries in the active (blue) and sterile (orange) flavours. Right: eigenvalues of the effective Hamiltonian $\langle H \rangle$. 
Continuous (dashed) lines indicate positive (negative) values.}
\label{fig:benchmark1a}
\end{figure}
\end{center}

\begin{figure}[t]
\includegraphics[width = \textwidth]{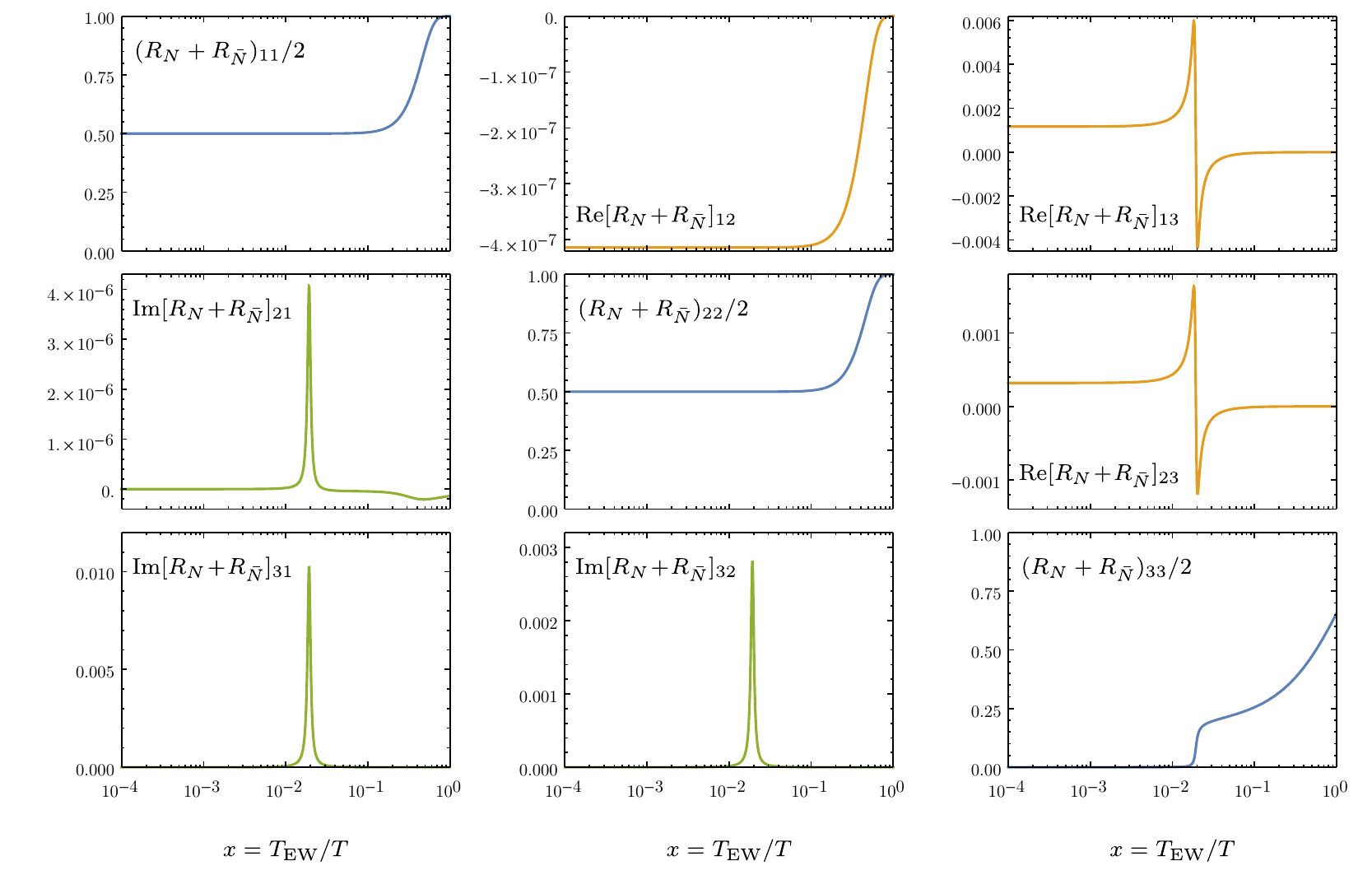}
\caption{Entries of the density matrix for benchmark point I) in the basis where $M_M$ is diagonal. Note that this basis does not correspond to the physical quasiparticle mass basis, c.f.\ Section~\ref{sec:bases}, and the interpretation of the diagonal elements as physical quasiparticle occupation numbers is only valid if the effective masses are dominated by $M_M$, {i.e.\ at low temperatures.}
}
\label{fig:benchmark1b}
\end{figure}

\paragraph{Benchmark point I): Resonant enhancement due to level crossing.} 
The first parameter point we consider is given by the choice
 \begin{eqnarray}
F= \left(
 \begin{array}{ccc}
 ( -2.0- i \ 7.9) \times 10^{-5} & (7.9 -i\ 2.0) \times 10^{-5} & (1.8 -i\ 9.5)\times 10^{-8} \\
  (2.7 -i\ 1.3)\times 10^{-5} & (1.3 +i\ 2.7)\times 10^{-5} & (4.6 -i\ 2.8)\times 10^{-8} \\
  (-2.9 -i\ 0.4) \times 10^{-5} & (0.4 -i\ 2.9)\times 10^{-5}& (-4.0 + i\ 1.0)\times 10^{-8} \\
 \end{array}
 \right) \,,
\end{eqnarray}
\begin{eqnarray}
\M = 2.70 \text{ GeV},\quad  \Mu=5.59\times 10^{-10},\quad \Mu'=1.02 \,.
 \end{eqnarray}
It features a degenerate heavy neutrino mass spectrum (c.f.\ Eq.~\eqref{FullNeutrinoMass}) 
and couples the heavy neutrinos with roughly the same strength to all SM flavours, $\flav = 0.36$. 
The level of fine-tuning in the sense of Eq.~(\ref{eq:fine_tuning}) is very low, $f.t. (m_\nu)=7.7\times 10^{-5}$, thanks to an approximate $B-\bar{L}$ symmetry ($|\epsilon_a| \leq 4.1 \times 10^{-7}$, $|\epsilon'_a| \leq 1.3 \times 10^{-3}$). 
Due to the small mass splitting amongst all three $N_i$, the generation of the BAU occurs in the \emph{overdamped regime}, i.e., the flavour eigenstate $\nu_{R{\rm s}}$ reaches thermal equilibrium before the heavy neutrino oscillations start (cf.\ e.g.\ \cite{Drewes:2016gmt} for a detailed discussion).
The BAU is resonantly enhanced by an (avoided) level crossing in the eigenvalues of the effective Hamiltonian, i.e., effect \ref{mechanism3}) in Section~\ref{sec:mechanisms}. 
This can be seen in the right panel of Fig.~\ref{fig:benchmark1a}.
The precise moment of the resonance well agrees with the simple estimate (\ref{MSWestimate}), $x_{\text{crossing}}\approx 1.4 \times 10^{-2}$. 
The resonant production of asymmetries is clearly visible in the middle and left panels of Fig.~\ref{fig:benchmark1a}, as well as in the off-diagonal elements of the density matrix in Fig.~\ref{fig:benchmark1b}.
In the middle panel of Fig.~\ref{fig:benchmark1a} one can identify the moment when $\tilde{L}$-violating processes kick in as the point where the orange and blue lines start to deviate from each other. As explained in point \ref{mechanism7}), the asymmetries in the SM flavours are rapidly equalised by $\tilde L$-conserving processes and are then protected from washout as long as $\tilde L$-violating processes are inefficient.

\FloatBarrier

\begin{center}
\begin{figure}[t]
\includegraphics[width = \textwidth]{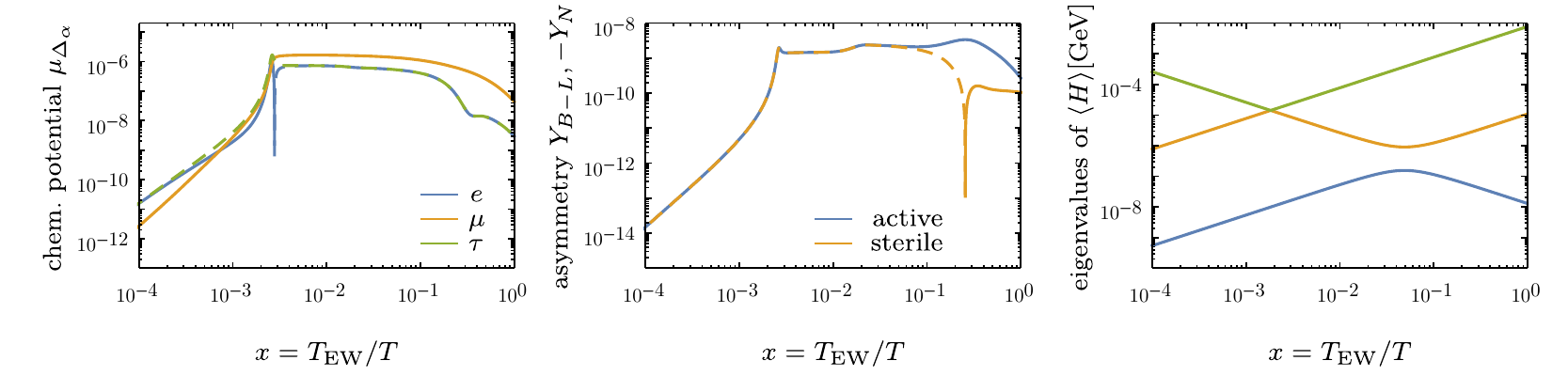}
\caption{Benchmark point II). Left: asymmetries in the individual active flavours. Center: sum of asymmetries in the active (blue) and sterile (orange) flavours. Right: eigenvalues of the effective Hamiltonian $\langle H \rangle$. 
Continuous (dashed) lines indicate positive (negative) values.
}
\label{fig:benchmark2a}
\end{figure}
\end{center}

\paragraph{Benchmark point II) : Flavour hierarchy and resonant enhancement.} 
Next we consider the parameter choice 
 \begin{eqnarray}
 F=
  \left(
\begin{array}{ccc}
 (2.8  -i\ 0.4)\times 10^{-5} & (0.4+i\ 2.8) \times 10^{-5} & (0.4 - i\ 1.3) \times 10^{-8} \\
( -3.0-i\ 0.8)\times 10^{-7}  & (0.8-i\ 3.0)\times
   10^{-7}  & (-3.9 +i\ 5.7)\times 10^{-8} \\
 (-4.9  +i\ 0.4)\times 10^{-5} & (-0.4 -i\ 4.9) \times 10^{-5} & (-3.4+i\ 5.0)\times 10^{-8}\\
\end{array}
 \right)\,,
\end{eqnarray}
\begin{eqnarray}
\M = 5.20 \text{ GeV},\quad  \Mu=6.16\times 10^{-5},\quad \Mu'=1.08 \,.
 \end{eqnarray}
This point is similar to the first one in the sense that there is also an approximate $B-\bar{L}$ symmetry and the masses of all three heavy neutrinos are quite degenerate.  
The fine-tuning is somewhat higher, but with $f.t. (m_\nu)=3.2\times 10^{-2}$ still small.
It also leads to overdamped behaviour and exhibits two avoided level crossings between the state that corresponds to $\nu_{R{\rm s}}$ at high $T$ and the two other states.\footnote{The temperature of the stronger level crossing agrees with the the estimate (\ref{MSWestimate}), $x_{\text{crossing}}\approx 1.8 \times 10^{-3}$.}    
The first one resonantly enhances the asymmetry production, while the second one is much weaker and occurs when $\tilde{L}$-violating processes are already relevant and two of the heavy neutrinos have reached equilibrium.
The main difference, however, lies in the strongly hierarchical flavour structure, $\flav=6.2\times 10^{-3}$. This prevents the 
equalising of all SM flavours by effect \ref{mechanism7})
 because the muon flavour couples only very feebly to the other charges.
This helps to avoid washout in spite of the fact that two heavy neutrino degrees of freedom reach equilibrium around $x\simeq 0.02$ as a result of effect \ref{mechanism5})  by the $\tilde{L}$-violating processes. 
Both level crossings lead to a re-distribution of charges, which is visible in the left panel of Fig.~\ref{fig:benchmark2a}, but only the second one leads to a sign change in the sterile charges. It is worthwhile noting that the zero crossing of the total sterile charge caused by the \textit{second level crossing} 
does not enforce a zero crossing of the total active charge $L$ due to effect~\ref{mechanism6}). 

\begin{figure}[t]
\includegraphics[width = \textwidth]{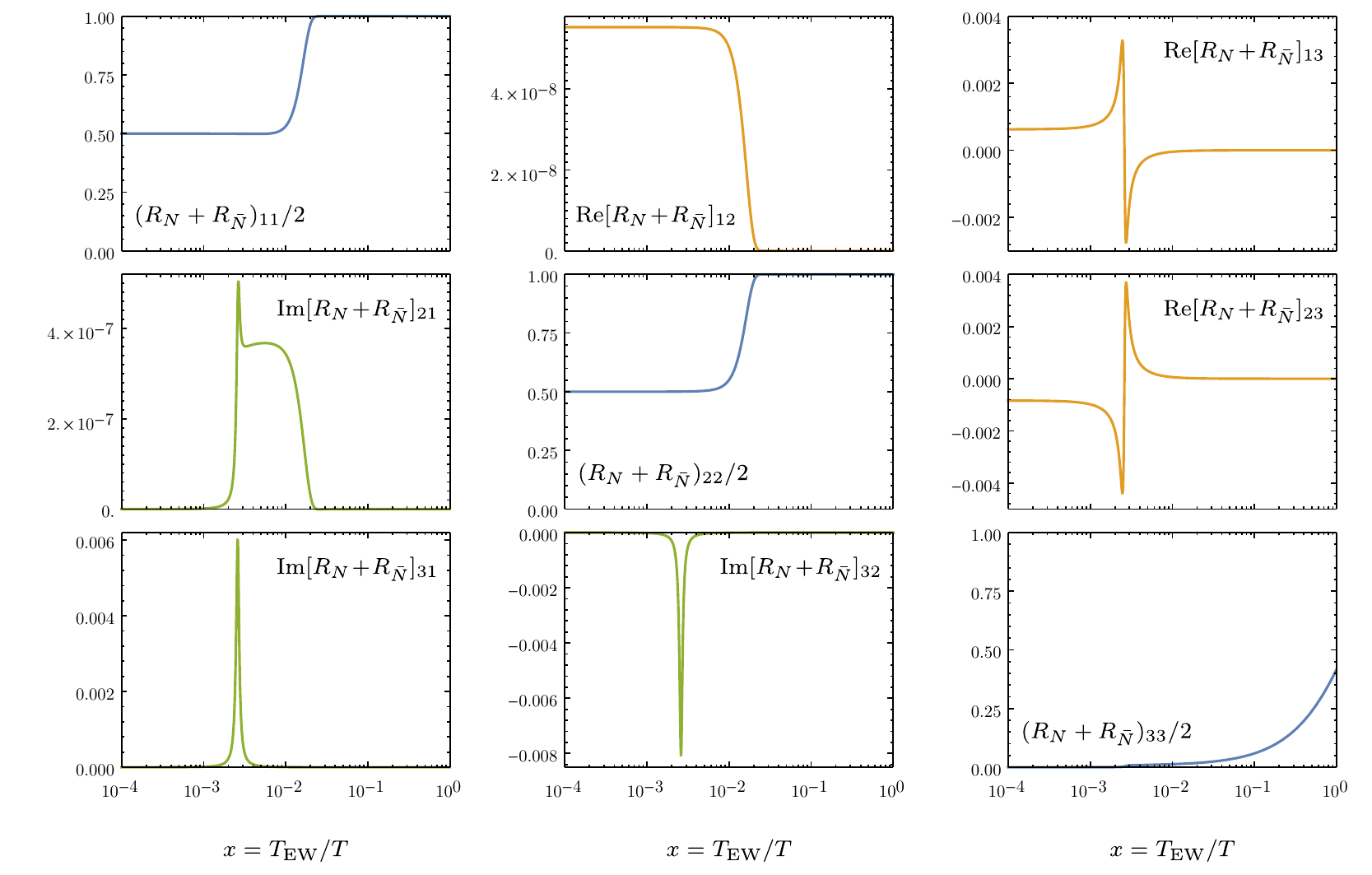}
\caption{Entries of the density matrix for benchmark point II). }
\label{fig:benchmark2b}
\end{figure}

\FloatBarrier

\paragraph{Benchmark point III): Large mass splittings.} 
The final point that we consider is given by
 \begin{eqnarray}
F= \left(
 \begin{array}{ccc}
  (-3.2-i\ 4.5)\times 10^{-8} & (-1.1-i\ 1.7)\times 10^{-7}& (-2.4+i\ 1.6)\times 10^{-7} \\
  (1.7-i\ 5.9)\times 10^{-7} & (0.6-i\ 2.1)\times 10^{-6} & (-2.9-i\ 0.8)\times 10^{-6} \\
  (4.4-i\ 3.0)\times 10^{-7} & (1.5-i\ 1.1)\times 10^{-6} & (-1.5-i\ 2.1)\times 10^{-6} \\
 \end{array}
 \right) \,,
\end{eqnarray}
\begin{eqnarray}
\M = 1.85 \text{ GeV},\quad  \Mu=5.49\times 10^{-1},\quad \Mu'=2.34 \,.
 \end{eqnarray}
Loop corrections remain comparably small, $f.t. (m_\nu)=0.14$, in spite of the fact that the parameters $\Mu$ and $\Mu'$ are not small. There is a moderate flavour hierarchy $\flav=9.5\times 10^{-2}$.
The evolution of charges corresponds to the standard mild washout scenario. The point serves as an example that leptogenesis can be realised with $\mathcal{O}(1)$ mass splitting without resorting to extreme fine-tuning. 

\FloatBarrier

\begin{center}
\begin{figure}[t]
\includegraphics[width = \textwidth]{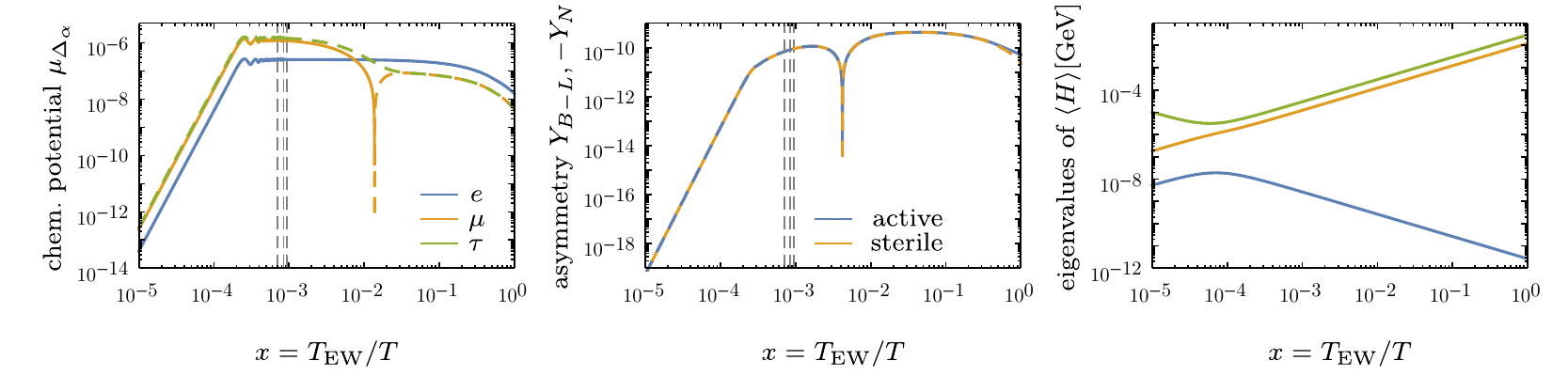}
\caption{Benchmark point III). Left: asymmetries in the individual active flavours. Center: sum of asymmetries in the active (blue) and sterile (orange) flavours. Right: eigenvalues of the effective Hamiltonian $\langle H \rangle$. 
Continuous (dashed) lines indicate positive (negative) values. The dashed vertical lines indicates the points in time when the oscillations among the heavy neutrinos are switched off, see main text.
 }
\label{fig:benchmark3a}
\end{figure}
\end{center}

\begin{figure}[t]
\includegraphics[width = \textwidth]{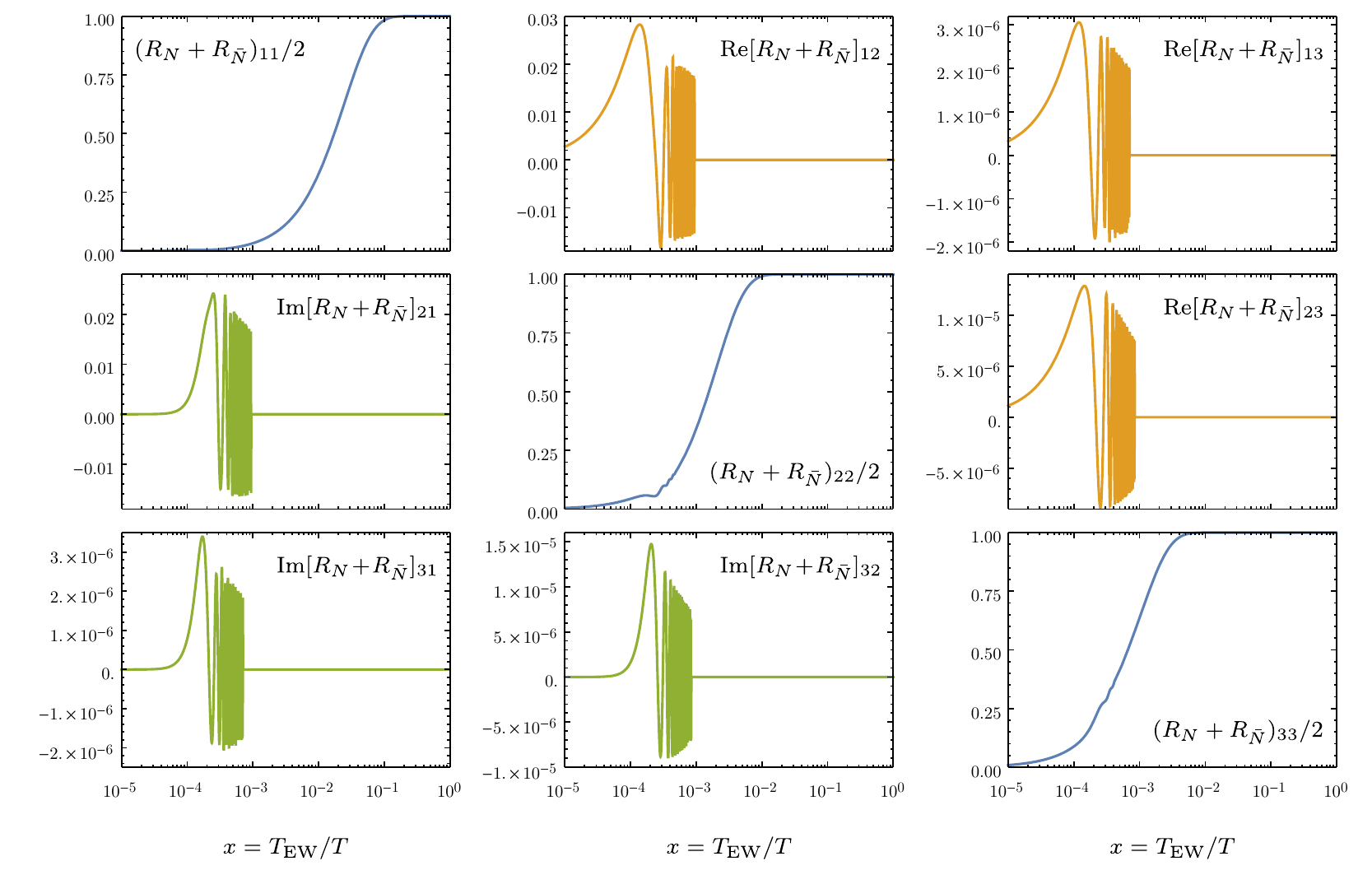}
\caption{ Entries of the density matrix for benchmark point III). 
The off-diagonal elements of the density matrix are set to zero (i.e.\ oscillations are switched off) after 10 completed oscillations; we checked that this has no significant effect on the asymmetries because they "average out" at later times.
}
\label{fig:benchmark3b}
\end{figure}

\begin{table}[htb]
\centering
\begin{tabular}{|c||c|c|c|}
\hline
Benchmark & I & II & III \\
\hline
Ordering & Inverted & Normal & Normal \\
\hline
$m_\nu^{\text{lightest}}$ & $6.93\times 10^{-7}$ eV & $1.32\times 10^{-9}$ eV & $1.01\times 10^{-3}$ eV \\
\hline
$\text{Re} \  \omega_{12}$ & $2.95\times 10^{-1}$ & $5.84$ & $5.80$ \\
\hline
$\text{Im} \  \omega_{12}$ & $7.36$ & $7.72$ & $-5.09\times 10^{-8}$\\
\hline
$\text{Re} \  \omega_{13}$ & $1.05\times 10^{-6}$ & $2.83 $ & $1.48\times 10^{-9}$ \\
\hline
$\text{Im} \ \omega_{13}$ & $-2.65\times 10^{-2}$ & $-1.53\times 10^{-1} $ & $4.81\times 10^{-6}$ \\
\hline
$\text{Re} \  \omega_{23}$ & $2.87\times 10^{-8}$ & $4.43\times 10^{-8}$ & $1.81$ \\
\hline
$\text{Im} \  \omega_{23}$ & $-8.93\times 10^{-1}$ & $4.05\times 10^{-4}$ & $-4.46$ \\
\hline
$ {\delta_\text{CP}, \alpha_1, \alpha_2}$ & $193^\circ \,, \; 148^\circ \,, \;78^\circ$ & $198^\circ \,, \; 300^\circ \,, \;74^\circ$ & $285^\circ \,, \; 33^\circ \,, \;36^\circ$ \\
\hline
\end{tabular}
\caption{Input parameters for the discussed benchmark points.}
\label{tab:bench_parameters}
\end{table}

\section{Conclusions \label{sec:conclusions}}

The ARS mechanism~\cite{Akhmedov:1998qx} for "freeze-in leptogenesis" is a remarkable and testable idea to implement leptogenesis within a minimal extension of the Standard Model by adding heavy neutrinos with masses below the electroweak scale. In this paper we perform the first systematic investigation of the ARS mechanism with three right-handed neutrinos ($n = 3$), extending previous analyses which encompassed only two right-handed neutrinos actively participating in leptogenesis ($ n = 2 $). For $n = 2$ there are only two characterstic time scales associated with the heavy neutrinos - the oscillation period of the two neutrinos, set by their mass difference, and the thermalisation rate, set by their coupling to the SM. On the contrary, for $n = 3$, a much richer phenomenology arises. As we show in this work, this does not only enlarge the parameter space, enhancing the possibility of a detection in present collider experiments due to a large mixing with the SM neutrinos (as anticipated in~\cite{Canetti:2014dka}), but moreover we find qualitatively new mechanisms to generate the lepton asymmetry, which do not have a counterpart in the $n = 2$ analysis.

The most striking of these qualitatively new effects is a resonant generation of a lepton asymmetry associated with an (avoided) level crossing of the effective mass eigenvalues of the three heavy neutrinos. As is well known from the analysis of the $n = 2$ case, the generation of a lepton asymmetry is enhanced for a small mass splitting within the neutrino pair. In the case of three neutrinos, a tiny mass splitting can occur dynamically through thermal corrections to the mass eigenstates, which induce a level crossing in the eigenvalues of the effective Hamiltonian. If this occurs 
when the respective heavy neutrinos have already been produced in significant numbers, but have not yet reached full equilibrium,
then the lepton asymmetry is resonantly enhanced. This enables successful leptogenesis with only a mild degeneracy in the vacuum masses of the heavy neutrinos and without any fine-tuning in the flavour mixing pattern.

Moreover, compared to the $n = 2$ case, we find richer flavour structures leading to successful leptogenesis. With only two right-handed neutrinos, successful leptogenesis with mixing angles that may be accessible with the LHC prefers a  hierarchical flavour structure, where the generated asymmetry can be protected from washout when stored in a very weakly coupled SM flavour, whereas the requirement to reproduce the observed neutrino oscillation data sets an upper bound on the flavour hierarchy. This tension forbids large mixings between the heavy and the SM neutrinos, making experimental tests challenging. In the case of three heavy neutrinos, these constraints are relaxed in two ways. Firstly, the additional parameter freedom due to the additional state allows to comply with the neutrino oscillation data while simultaneously allowing for a large flavour hierarchy. Secondly, we demonstrate that contrary to the $n = 2$ case, a lepton asymmetry can be generated even with flavour democratic couplings, due to a new term in the kinetic equations which only arises for $n \geq 3$. Consequently, and again contrary to the $n = 2$ case, we find large mixing between the heavy and the SM neutrinos, right up to the current bounds, to be compatible with both neutrino oscillation data and successful leptogenesis.

We point out that this large mixing, as well as the formation of pseudo-Dirac pairs of right-handed neutrinos, is natural in the context of an approximate global $B - \bar L$ symmetry, where $B$ denotes the SM baryon number and $\bar L$ denotes a generalised lepton number under which also the right-handed neutrinos are charged. With this in mind, we define `fine-tuned' solutions as parameter points for which the radiative one-loop contributions to the light neutrino masses are large compared to the tree-level contributions. In this sense, we find that experimentally accessible large mixing is possible without any fine-tuning, whereas an enhancement of the neutrinoless double $\beta$ decay rate is possible only at the cost of fine-tuning unless the heavy neutrino masses are rather close the the momentum exchange in the process. Furthermore, the resonant generation of a lepton asymmetry due to a level crossing of the mass eigenvalues occurs quite generically in the regime protected by the $B - \bar L$ symmetry, since one state of the pseudo-Dirac pair receives large thermal corrections whereas the quasi decoupled third right-handed neutrino does not. Moreover, the participation of the quasi decoupled heavy neutrino automatically protects the generated asymmetry from subsequent washout. All this renders the $B - \bar L$ symmetry protected regime particularly interesting for ARS leptogenesis.

At high temperatures far above the heavy neutrino mass scale, the different helicities of the right-handed neutrinos are conserved quantum numbers. Approaching the EW phase transition, this approximation breaks down, allowing for `lepton number violating' ($\tilde L$-violating) processes. Although active for only a fairly short period of time, these processes can significantly alter the predicted lepton asymmetry. We highlight the different physical processes at work, showing that they can both enhance or reduce the final asymmetry.

In summary, we find that leptogenesis invoking the oscillations of three right-handed neutrinos just before the EW phase transition comes with some qualitative and quantitative differences to the well-studied $n = 2$ case. New channels of leptogenesis lead to an enhanced lepton asymmetry. The viable parameter space, reproducing both the observed neutrino oscillation data and the baryon asymmetry of the Universe, projected onto the mass versus active-sterile mixing plane shows promising opportunities for ongoing experiments, such as NA62, T2K, Belle II and the LHC. 
This calls for a more detailed study of some of the effects that we have neglected here.
These include effects of the electroweak transition (temperature dependent Higgs field value, gradual sphaleron freeze-out, particle masses generated by the Higgs mechanism and the $\tilde{L}$-violation due to the active-sterile mixing),
the full momentum dependence of the equations, a fully systematic perturbative computation of the $\tilde L$-violating rates, as well as, a verification of the
validity of the gradient expansion (which justifies the usage of the density matrix equations) during the level crossing.

\section*{Acknowledgements}
We thank Bj\"orn Garbrecht, 
Jacopo Ghiglieri, Jacobo Lopez-Pavon and Inar Timiryasov for helpful discussions.
A.A.  acknowledges support within the framework of the
European Union's Horizon 2020 research and innovation programme under
the Marie Sklodowska-Curie grant agreements No 690575 and No 674896. M.L. acknowledges support from the European Union's Horizon 2020 research and innovation programme under the Marie Sklodowska-Curie grant agreement No 750627.
J.K. acknowledges support of the DFG  cluster  of  excellence  ’Origin  and  Structure  of
the Universe’ (www.universe-cluster.de),  by the Collaborative Research Center SFB1258, the Collaborative Research Center SFB1258 of the Deutsche Forschungsgemeinschaft and the ERC-AdG-2015 grant 69489.

\appendix


\appendix

\section{Notation for the quantum kinetic equations}\label{sec:equations}

We provide in this appendix additional details on the system of differential equations used in our numerical scan to compute the baryon asymmetry, starting from the original set of Boltzmann equations given in Eqs.~\eqref{eq:Rnstart} and \eqref{eq:mustart}, see Ref.~\cite{Abada:2017ieq} for more details.

Let us consider first the equation for the abundances of the heavy neutrinos, described by $R_N$ and $R_{\bar N}$. Their oscillation processes are described by the term proportional to $ \left[\langle H \rangle,R_{N,\bar N}\right]$ where the Hamiltonian $ H $ can be split as $ H = H_0 + V_N$ with  $V_N$ denoting the effective potential and $ H_0$ denoting the vacuum Hamiltonian, 
\begin{align}
  H_0 = \frac{1}{2k}\, \text{diag}(M_1^2, M_2^2, M_3^2) \mapsto \frac{1}{2k} \, \text{diag}(0, \Delta M_{12}^2, \Delta M_{13}^2) \,, \quad \quad  {V_N = \frac{N_D}{16} \, \frac{T^2}{k} }F^\dagger F \,, 
\end{align}
where $\Delta M_{ij}^2 = M_j^2 - M_i^2$, and in the last expression  we have dropped the part of the matrix proportional to the unity matrix, since it drops out in the commutator of Eq.~\eqref{eq:Rnstart}. Thermal averaging yields
\begin{align}
  \langle H_0 \rangle & = \frac{\pi^2}{36 \,  \zeta(3)\,  T} \, \text{diag}(0, \Delta M_{12}^2, \Delta M_{13}^2) \,, \\
 \langle V_N \rangle & = \frac{ \pi^2 N_D}{288 \, \zeta(3)} T F^\dagger F 
 \,.
\end{align}
For the numerical solution of the system of Boltzmann equations it is convenient to adopt $x=T_{\rm EW}/T$ as new time variable. The change of variables is described by the  relation
\begin{equation}
 dt = \frac{M_0}{T_{EW}^2} x \, dx \quad \text{with } M_0 = 7.12 \times 10^{17}~\text{GeV}\,. \label{eq:dtdx}
\end{equation}
It will be convenient to introduce a new parameterisation of the effective potential,
\begin{equation}
W_N = \frac{M_0}{T_{EW}^2} x \, \langle V_N \rangle = \frac{\pi^2}{144 \, \zeta(3)} \frac{M_0}{T_{EW}} F^\dagger F \,,  \label{eq:WN} 
\end{equation}
and analogously,
\begin{eqnarray}
 W_{\rm N,LNV} &=& \frac{\pi^2}{144 \, \zeta(3)} \frac{M_0}{T_{EW}}M F^\dagger F M  \,,\label{eq:WNLNV}\\
 o_\mu &=& \frac{\pi^2}{144 \, \zeta(3)}\frac{M_0}{T_{\rm EW}}F^{\dagger}\mu F  \,, \label{eq:omu}\\
 o_{\mu,\rm LNV} &=& \frac{\pi^2}{144 \, \zeta(3)}\frac{M_0}{T_{\rm EW}}M F^{\dagger}\mu F M \,.
\end{eqnarray}
The analogous terms for the equations for $R_{\bar N}$ are  obtained by setting $F \rightarrow F^{*}$ and $\mu \rightarrow -\mu$.
Performing the change of variables $t \rightarrow x$ also in the equations for the chemical potentials $\mu_{\Delta_a}$, we finally obtain the system:
\bee
\frac{d R_N}{dx} &=& i \comm{R_N}{W_N}+3 i x^2 \comm{R_N}{r} - \phi^{(0)} \acomm{R_N}{W_N}-\tilde{\phi}^{(0)} \acomm{R_N}{W_{N,LNV}}\non
&+& 2 \phi^{(0)} W_N + 2 \tilde{\phi}^{(0)} W_{N,LNV} + \phi^{(1a)} o_\mu - \tilde{\phi}^{(1a)} o_{\mu,LNV}\non
& + & \frac{1}{2}  \phi^{(1b)} \left\{  o_\mu , R_N \right\}-\frac{1}{2}  \tilde{\phi}^{(1b)} \left\{  o_{\mu,LNV} , R_N \right\}, \label{eq:RNscan} \\
\frac{d R_{\bar{N}}}{dx} &=& i \comm{R_{\bar{N}}}{W_N^T}+3 i x^2 \comm{R_{\bar{N}}}{r} - \phi^{(0)} \acomm{R_{\bar N}}{W_N^T}-  \tilde{\phi}^{(0)} \acomm{R_{\bar N}}{W_{N,LNV}^T}\non
& + & 2 \phi^{(0)} W_N^T +2 \tilde{\phi}^{(0)} W_{N,LNV}^T+\phi^{(1a)} o_{\bar{\mu}}-\tilde{\phi}^{(1a)} o_{\bar{\mu},LNV} \non
& + &   \frac{1}{2}  \phi^{(1b)} \left\{  o_{\bar \mu} , R_{\bar{N}} \right\}-\frac{1}{2}  \tilde{\phi}^{(1b)} \left\{  o_{\bar \mu} , R_{\bar{N}} \right\}, \\
\frac{d {\mu_\Delta}_{a}}{dx} &=& \frac{1}{32}\frac{M_0}{T_{EW}} \left[ -\phi^{(0)} \left(F R_N F^{\dagger}-F^{*} R_{\bar N} F^T\right)_{aa} +\phi^{(1a)}  \left(F F^{\dagger}\right)_{a a} {\mu}_a \right. \non 
& + &  \frac{\phi^{(1b)}}{2}   \left(F R_N F^{\dagger}+F^{*} R_{\bar N}F^T\right)_{a a} {\mu}_{a} \non
&+& \tilde{\phi}^{(0)} \left(F M R_N M F^{\dagger}-F^{*} M R_{\bar N} M F^T\right)_{aa} -\tilde{\phi}^{(1a)}  \left(F M^2 F^{\dagger}\right)_{a a} {\mu}_a  \non 
& - &  \frac{\tilde{\phi}^{(1b)}}{2} \left.  \left(F M R_N M F^{\dagger}+F^{*} M R_{\bar N} M F^T\right)_{a a} {\mu}_{a} \right] \, ,\label{eq:muscan}
\eee
where the functions $\phi^{(i)}$ are related to the thermally averaged rates $\langle \gamma^{(i)} \rangle$ by:
\begin{align}
 \phi^{(0)} & = \frac{144 \, \zeta(3)}{N_D \pi^2 T } \langle \gamma^{(0)} \rangle \label{eq:phi0} \\ 
 & = \frac{1}{16 \pi N_D} \left[c_Q^{(0)} h_t^2+c_{LPM}^{(0)}+(3 g^2+g^{\,2})\left(c_V^{(0)}+\log\left(\frac{1}{3 g^2+g^{'\,2}}\right)\right)\right] \, ,\nonumber\\
\phi^{(1a)} & \equiv   \frac{144 \, \zeta(3)}{\pi^2 T} \langle \gamma^{(1a)} \rangle \nonumber \\
& = \frac{1}{{32} \pi} \left[c_Q^{(1a)} h_t^2+c_{LPM}^{(1a)}+(3 g^2+g^{\,2})\left(c_V^{(1a)}+\log\left(\frac{1}{3 g^2+g^{'\,2}}\right)\right)\right] \,, \\
\phi^{(1b)} & \equiv - \frac{144 \, \zeta(3)}{\pi^2 T} \langle \gamma^{(1b)} \rangle \nonumber \\
&  =\frac{1}{{64} \pi} \left[c_Q^{(1b)} h_t^2+c_{LPM}^{(1b)}+(3 g^2+g^{\,2})\left(c_V^{(1b)}+\log\left(\frac{1}{3 g^2+g^{'\,2}}\right)\right)\right] \, .
\end{align}
The coefficients $c_X^{(i)}$ are given in Eq.~\eqref{tab:ci}. The corresponding expressions for $\tilde{\phi}^{(i)}$ are obtained by replacing $\langle \gamma^{(i)} \rangle$ by  $\langle \tilde \gamma^{(i)} \rangle$.

The two terms $[R_N,W_N]$ and $[R_N,r]$ in Eq.~\eqref{eq:RNscan} (and the corresponding terms in the equation for $R_{\bar N}$) represent the  terms originating from the potential $V_N$ and the vacuum Hamiltonian $H_0$, respectively, with
\begin{align}
 r \equiv \text{diag}(0, r_2^{ 3}, r_3^{ 3}) \,, \qquad  r_i \equiv \frac{T_{L,i}}{T_{\rm EW}} \,, \qquad T_{L,i}\equiv{\left(\frac{\pi^2}{108 \, \zeta(3)}M_0 \Delta M^2_{1i}\right)}^{1/3}\,,
 \label{eq:ri}
\end{align}
where $T_{L,i}$ denotes the typical leptogenesis temperatures associated to the oscillations between the ``1st'' and ``ith'' ($i=2,3$) heavy neutrino eigenstate.

\section{Perturbative expansion \label{sec:perturbative}}

In this Appendix we perform a perturbative expansion of the system of Boltzmann equations in terms of  the chemical potentials in the active sector $\mu_a$, following the procedure outlined in \cite{Abada:2017ieq} for $n=2$. This allows us to gain an analytical understanding of some of the main processes involved in the $n = 3$ ARS leptogenesis and to identify the qualitative differences with respect to the case of a single pair of quasi mass-degenerate neutrinos. In particular, we will discover an additional (leading order) source term for the lepton asymmetry, enabling successful leptogenesis in the absence of flavour asymmetric Yukawa couplings (see point \ref{mechanism2}) on page \pageref{page:listmechanism}).

While we do not employ this formalism for the main parameter scan of this paper, we have confirmed for a range of parameter points that it accurately reproduces the results of the full equations. For simplicity and in order to facilitate the comparison with the results of~\cite{Abada:2017ieq}, we will omit in this appendix the $\tilde L$-violating terms. 

\subsection{0th order in the chemical potential}
To leading order in $\mu_a$ Eq.~\eqref{eq:Rnstart} reads,
\begin{equation}
\frac{d R_N^{(0)}}{dt}=  -i \left[\langle H \rangle ,R_N^{(0)}\right]-\frac{1}{2}\langle \gamma^{(0)} \rangle \left \{ F^{\dagger}F, R_N^{(0)}-I\right \} \,,
\label{eq:Rnstart0}
\end{equation}
with $\langle H \rangle = \langle H_0 \rangle + \langle V_N \rangle$ introduced in Appendix~\ref{sec:equations}. Performing unitary rotations of this equation we will in the following identify the different physical effects involved in the generation of a lepton asymmetry.\footnote{These `basis changes' obtained by unitary rotations of the density matrix should not be confused with the different basis discussed in Section~\ref{sec:seesaw_model}, which are obtained by rotating the spinors $(\nu_i, N_i)$.}
The first step consists in defining an `oscillation' basis, in which the neutrino oscillations driven by the vacuum Hamiltonian $H_0$ are removed. This is done by performing a rotation of the form $\tilde{R}_{N}=E^{\dagger} R_N E$ with $E(t)$ defined as \cite{Canetti:2010aw}
\begin{equation}
 E(t) \equiv \exp\left(- i \int_{t_0}^t \langle H_0 \rangle dt'\right) = \text{diag}\left(1, e^{- i r_2^3 x^3}, e^{-i r_3^3 x^3}\right)\,, \label{eq:E}
\end{equation}
with the typical oscillation temperatures encoded in the parameters $r_i$, see Eq.~\eqref{eq:ri}. With this,
\begin{align}
 \frac{d \tilde R_N}{dt} &  = \dot E^\dagger R_N E + E^\dagger \dot R_N E + E^\dagger R_N \dot E  \nonumber \\
 & = i [\langle H_0 \rangle, \tilde R_N] + E^\dagger \frac{d R_N}{dt} E  \nonumber \\
  &  =- i [\langle \tilde V_N \rangle, \tilde R_N ] - \frac{1}{2} \langle \gamma^0 \rangle \{\tilde{F^\dagger F}, \tilde R_N - 1\} \,,
\end{align}
and Eq.~\eqref{eq:Rnstart0} can be written as
\begin{equation}
 \frac{d \tilde R_N}{dx} = - i [\tilde W_N, \tilde R_N] - \phi^{(0)} \{ \tilde W_N, \tilde R_N \} + 2 \phi^{(0)} \tilde W_N \,,
 \label{eq:RN02}
\end{equation}
with $\tilde W_N = E^\dagger W_N E$ and $\phi^{(0)}$  introduced in Eq.~\eqref{eq:phi0}. It will be convenient to introduce a third basis, which we refer to as `interaction' basis, in which $\tilde W_N$ is diagonal. This is accomplished by means of the unitary matrix $U$,
\begin{equation}
 U^\dagger \tilde W_N U = 
{\frac{\pi^2}{144 \, \zeta(3)}}  \, 
 \frac{M_0}{T_{EW}} (F E U)^\dagger F E U = \text{diag}(y_1, y_2, y_2) \equiv Y \,.
 \label{eq:U}
\end{equation}
Since $E$ and $U$ are unitary, the eigenvalues $y_i$ of $\tilde W_N$ are proportional to those of $F^\dagger F$, and in particular time-independent and real. Motivated by this we construct the time independent part $U_c$ of $U$ as
\begin{equation}
{U_c = E \, U \quad \Rightarrow \quad } U = E^\dagger U_c = \text{diag}(1, e^{i r_2^3 x^3}, e^{i r_3^3 x^3}) U_c \,.
\end{equation}
To switch between the oscillation and flavour basis we introduce
\begin{align}
 U^\dagger \frac{d U}{dx}  =\left(E^\dagger U_c\right)^\dagger \left(\frac{d E^\dagger}{dx} U_c\right) 
   = x^2 3 i U_c^\dagger \, \text{diag}(0, r_2^3, r_3^3) U_c  
  \equiv x^2 D \label{eq:D}  \,,
\end{align}
where we note that the matrix $D$ is anti-hermitian and time-independent. Denoting the leading order density matrix of the right-handed neutrinos in the interaction basis by $S_N^0$,  $S_N^0 = U^\dagger \tilde R_N U$,
we finally find
\begin{align}
 \frac{d S_N^0}{dx} & = - x^2 [D, S_N^0] - i [Y, S_N^0] - \phi^{(0)} \{ Y, S_N^0 \} + 2 \phi^{(0)} Y \label{eq:SN0} \\
 &= S_N^0(x) ((i - \phi^{(0)}) Y + x^2 D) - ((i + \phi^{(0)})Y + x^2 D) S^0(x) + 2 \phi^{(0)} Y  \,, \label{eq:RN03}
\end{align}
as in the case of two right-handed neutrinos.

To obtain the corresponding equation for $S_{\bar N}$ we need to replace the Yukawa coupling by its complex conjugate, $F \mapsto F^*$. Denoting the quantities in the $S_{\bar N}$ equation with overbars, this implies
\begin{align}
 \bar Y = Y  \,, \quad \bar E = E  \,, \quad \bar U_c = E \bar U = E^* U^* = U_c^* \,, \quad  \bar D = D^T\,.
 \label{eq:antineutrinos}
\end{align}
Here the first equality follows since $Y$ contains the real eigenvalues of $\widetilde{W}_N \propto E^\dagger F^\dagger F E $. The second is trivial since no powers of $F$ are involved in the definition of $E$, and the third follows from
\begin{align}
Y \propto \bar U^\dagger (E^\dagger F^T F^* E) \bar U 
     =   ( \bar U \, E F^\dagger F E^* \bar U^*)^T  
   =  (U^\dagger E^\dagger F^\dagger F E \, U)^T \,.
\end{align}
Finally the fourth equality in Eq.~\eqref{eq:antineutrinos} follows from $\bar D = \bar U^\dagger (d \bar U/dx)$ with $\bar U = E^\dagger U_c^*$. Note that in the case of two right-handed neutrinos the matrix $D$ is symmetric\footnote{
Consider a general unitary $2 \times 2$ matrix
\begin{equation}
 U = e^{- i \varphi/2} \begin{pmatrix} 
                        e^{i \varphi_1} \cos\theta & e^{i \varphi_2} \sin \theta \\
                        - e^{- i \varphi_2} \sin\theta & e^{- i \varphi_1} \cos \theta
                       \end{pmatrix}\,.
\end{equation}
An explicit computation shows that the quantity $D \sim i U^\dagger \text{diag}(0,1) \,U$ has purely imaginary, symmetric off-diagonal elements if and only if $\varphi_1 = \varphi_2$. Since there is a free phase in each column of $U$, this condition can always be met. In the case of 3 right-handed neutrinos, this freedom of choosing the phases of the columns is not sufficient to make $D$ symmetric for a generic $3\times3$ unitary matrix $U$.} and hence $\bar D = D^T = D$. For $n > 2$, $D$ is anti-hermitian but not symmetric, so this simplification does not apply.
With this, the  equation for the opposite helicity ($\bar N$) neutrinos reads
\begin{align}
 \frac{d S_{\bar N}^0}{dx} & = - x^2 [D^T, S_{\bar N}^0] - i [Y, S_{\bar N}^0] - \phi^{(0)} \{ Y, S_{\bar N}^0 \} + 2 \phi^{(0)} Y \label{eq:SbN0}\\
 &= S_{\bar N}^0(x) ((i - \phi^{(0)}) Y + x^2 D^T) - ((i + \phi^{(0)})Y + x^2 D^T) S^0(x) + 2 \phi^{(0)} Y  \,. \label{eq:RN03b}
\end{align}
Defining
\begin{equation}
 S^0 = \frac{1}{2}(S_N^0 + S_{\bar N}^0) \,, \quad \Delta S_-^0 = S_N^0 - S_{\bar N}^0 \,,
\end{equation}
and noting that $D$ is anti-hermitian, $D^T = - D^*$, implying
\begin{align}
 (\left[ D, S_N\right] - \left[D^T, S_{\bar N} \right]) = \left[ \text{Re}(D), S_+ \right] + i \left[ \text{Im}(D), \Delta S_- \right]\,,
\end{align}
we find
\begin{align}
 \frac{d S^0}{dx} & = - i x^2 [\text{Im}(D), S^0] - i [Y, S^0] - \phi^{(0)} \{ Y, S^0 \} + 2 \phi^{(0)} Y - \frac{1}{2} x^2 [\text{Re}(D), \Delta S^0_-]  \,, \label{eq:S0} \\
 \frac{d \Delta S_-^0}{dx} & = - 2 x^2 [ \text{Re}(D), S^0 ] - i x^2 [ \text{Im}(D), \Delta S_-^0] - i [Y, \Delta S_-^0] - \phi^{(0)} \{ Y, \Delta S_-^0 \} \,. \label{eq:DS0}
 \end{align}
We highlight two crucial differences to the case of only two right-handed neutrinos. Firstly, in the case of two right-handed neutrinos the freedom of phase rotations allows us to impose $D^T = D$ and hence the equations for $N$ and $\bar N$ in the interaction basis at leading order are identical (cf.~Eqs.~\eqref{eq:SN0} and \eqref{eq:SbN0}). Consequently, in this case the first term on the right-hand side of Eq.~\eqref{eq:DS0} is absent, and  $\Delta S_-^0 = 0$ is a solution to Eq.~\eqref{eq:DS0}.
{This reflects that for appropriate initial conditions, the reduced number of $CP$-violating phases for $n = 2$ impedes the generation of asymmetries in the sterile sector (see also Appendix~D of Ref.~\cite{Drewes:2016gmt}).} 
On the contrary, in the case of three right-handed neutrinos this is no longer the case, leading to $\Delta S_-^0 \neq 0$ already at leading order. Secondly, in the case of two right-handed neutrinos, the last term in Eq.~\eqref{eq:S0} is absent. One might be tempted to discard this term, since it is proportional to a (small) asymmetry, however  at early times when the oscillations are large, the off-diagonal terms of $\Delta S_-^0$ can in fact be rather large. We note that in particular in the case of (mildly) hierarchical Yukawa couplings this term can be crucial to obtain the correct thermalisation time scales of the different right-handed neutrino species.

\subsection{1st order in the chemical potentials}

\subsubsection*{Sterile sector}

In the oscillation basis, Eq.~\eqref{eq:Rnstart} reads
\begin{equation}
 \frac{d \widetilde{R_N}}{dt} = - i \left[ \widetilde{\langle V_N \rangle}, \widetilde{R_N}\right] - \frac{1}{2} \langle \gamma^{(0)} \rangle \left\{\widetilde{F^\dagger F}, \widetilde{R_N} - 1 \right\} - \frac{1}{2} \langle \gamma^{(1b)} \rangle \left\{ \widetilde{F^\dagger \mu F}, \widetilde{R_N} \right\} + \langle \gamma^{(1a)} \rangle \widetilde{F^\dagger \mu F}\,,
\end{equation}
where as above $\tilde X = E^\dagger X E$. Using Eq.~\eqref{eq:dtdx}, as well as the functions $\phi^{(i)}$ and the $o_{\mu}, \bar o_\mu$ defined in Appendix~\ref{sec:equations}, this becomes
\begin{align}
  \frac{d \widetilde{R_N}}{dx} = - i \left[ \widetilde{W_N}, \widetilde{R_N}\right] - \phi^{(0)}  \left\{\widetilde{W_N}, \widetilde{R_N} \right\} + 2 \phi^{(0)} \widetilde{W_N} + \frac{1}{2} \phi^{(1b)} \left\{ \widetilde{o_\mu}, \widetilde{R_N} \right\} +\phi^{(1a)} o_\mu \,.
\end{align}
Switching to the interaction basis, $S_N = U^\dagger \widetilde{R_N} U$ with $U$ introduced in Eq.~\eqref{eq:U},  this yields
\begin{align}
 \frac{d S_N}{dx} & = - \left[ x^2 D, S_N \right] + U^\dagger \frac{\widetilde{R_N}}{dx} U  \nonumber \\
 & = - x^2 \left[D, S_N \right] - i \left[ Y, S_N \right] - \phi^{(0)} \left\{ Y, S_N \right\} + 2 \phi^{(0)} Y + \frac{1}{2}  \phi^{(1b)} \left\{ U^\dagger \widetilde{o_\mu} U, S_N \right\} +\phi^{(1a)} U^\dagger \widetilde{o_\mu} U \,,
\end{align}
and 
\begin{align}
  \frac{d S_{\bar N}}{dx}  = - x^2 \left[D^T, S_{\bar N} \right] - i \left[ Y, S_{\bar N} \right] - \phi^{(0)} \left\{ Y, S_{\bar N} \right\} + 2 \phi^{(0)} Y 
   + \frac{1}{2}  \phi^{(1b)} \left\{ \bar U^\dagger \widetilde{ \bar o_\mu} \bar U, S_{\bar N} \right\} +\phi^{(1a)} \bar U^\dagger \widetilde{ \bar o_\mu} \bar U \,.
\end{align}
We now switch variables to
\begin{equation}
 S_+ = S_N + S_{\bar N} = 2 S_0 + \Delta S_+ \,, \quad S_- = S_N - S_{\bar N} = \Delta S_- \, ,
\end{equation}
with $S_0$  determined by Eq.~\eqref{eq:RN03}. The equation for $\Delta S_-$ reads
\begin{eqnarray}
 \frac{d \Delta S_-}{dx}
  & = & - i x^2 \left[ \text{Im}(D), \Delta S_- \right]  - i \left[ Y, \Delta S_- \right] - \phi^{(0)} \left\{ Y, \Delta S_- \right\}  \nonumber \\
  & & - x^2 \left[ \text{Re}(D),  S_+ \right] 
  + \frac{1}{2} \phi^{(1b)} \left\{ O_\mu, S_0 \right\} + \phi^{(1a)} O_\mu\,,  \label{eq:DSm} 
\end{eqnarray}
where we have dropped the subleading term proportional to $\mu \Delta S_-$ in the $\phi^{(1b)}$ term and
\begin{align}
 O^+_\mu \equiv U_c^\dagger o_\mu U_c + U_c^T o_\mu U_c^* \,.
\end{align}
As indicated above, in the context of our perturbative expansion, the leading order term driving the asymmetry in the sterile sector is the first term in the second line in Eq.~\eqref{eq:DSm}, which is present already at 0th order but is absent for $n=2$.

To good approximation, we may set $S_+ \simeq 2 S_0$ in the first term of the second line of Eq.~\eqref{eq:DSm}. In this approximation, the equation of motion for $\Delta S_+$ decouples, and the equations of motion describing the sterile sector are \eqref{eq:S0} and \eqref{eq:DSm}. In the case of only two sterile neutrinos, this is in fact an exact result to first order in $\mu_a$. For completeness, we give here also the equations for $\Delta S_+$:
\begin{eqnarray}
  \frac{d \Delta S_+}{dx}
  & = & - i x^2 \left[ \text{Im}(D), \Delta S_+ \right]  - i \left[ Y, \Delta S_+ \right] - \phi^{(0)} \left\{ Y, \Delta S_+ \right\} \nonumber \\
  & & - x^2 \left[ \text{Re}(D),  \Delta S_- \right] + \frac{1}{2} \phi^{(1b)} \left\{ O^+_\mu, S_0 \right\} + \phi^{(1a)} O^+_\mu  \,.
\end{eqnarray}
Note that contrary to Eq.~\eqref{eq:DSm} (see also discussion below Eq.~\eqref{eq:DS0}), the equation for $\Delta S_+$ has no source term in the limit $\Delta S_- , \mu_a \rightarrow 0$, justifying the approximation $S_+ \simeq 2 S_0$ above.

\subsubsection*{Active sector}
The starting point for the equation of the active sector is Eq.~\eqref{eq:mustart}. Replacing the $\langle \gamma^{(i)} \rangle $ with $\phi^{(i)}$ and using Eq.~\eqref{eq:dtdx}, this can be rephrased as
\begin{align}
 16 N_D \frac{T_{EW}}{M_0} \frac{d \mu_{\Delta_a}}{dx} = & \left[- \frac{N_D}{2} \phi^{(0)} (F R_N F^\dagger - F^* R_{\bar N} F^T) {+} \phi^{(1a)} \mu F^\dagger F \right. \nonumber \\  & \left. + \frac{1}{2} \phi^{(1b)} \mu (F R_N F^\dagger + F^* R_{\bar N} F^T) \right]_{a a}\,.
\end{align}
With $R_N = U_c S_N U_c^\dagger$, $R_{\bar N} = U_c^* S_{\bar N} U_c^T$ we may write
\begin{align}
 F R_{ N} F^\dagger & = F U_c (S_0 + \frac{1}{2} \Delta S_-) U_c^\dagger F^\dagger \,, \label{eq:aux1}\\
 F^* R_{\bar N} F^T & = F^* U_c^* (S_0 - \frac{1}{2} \Delta S_-) U_c^T F^T \,. \label{eq:aux2}
\end{align}
{Both these expressions are hermitian matrices, implying that the diagonal components are real, i.e.\
\begin{align}
 (F R_{ N} F^\dagger -  F^* R_{\bar N} F^T)_{a a} &= \text{Re}\left[ F U_c (2 \,i  \text{Im}[S_0] + \text{Re}[\Delta S_-]) U_c^\dagger F^\dagger \right]_{a a}\\
 ( F R_{ N} F^\dagger +  F^* R_{\bar N} F^T)_{a a} &= 2 \, \text{Re} \left[ R U_c \text{Re}[S_0] U_c^\dagger F^\dagger \right]_{a a} + {\cal O}(\Delta S_-) \,.
 \end{align}
Defining
\begin{align}
 S^\text{aux} & =  2 \, i \, \text{Im}[S_0] +  \text{Re}[\Delta S_-] \,, 
\end{align}
we obtain}
\begin{align}
  16 N_D \frac{T_{EW}}{M_0} \frac{d \mu_{\Delta_a}}{dx} =  \left[- \frac{N_D}{2} \phi^{(0)} (F U_c {S^\text{aux}} U_c^\dagger F^\dagger) {+} \phi^{(1a)} \mu F^\dagger F +  \phi^{(1b)} \mu (F  U_c {\text{Re}[S_0]} U_c^\dagger F^\dagger) \right]_{a a}\,.
  \label{eq:mu}
\end{align}
Note that the asymmetry in the sterile sector sources an asymmetry in the active sector through $S^\text{aux}$. More precisely, in the absence of $\tilde L$-violating terms, the final asymmetries in the active and sterile sectors are of equal magnitude but opposite sign.

In summary, all processes relevant for ARS leptogenesis involving three right-handed neutrinos are well described by the system of differential equations~\eqref{eq:S0}, \eqref{eq:DSm} and \eqref{eq:mu}. The results obtained from this simplified system agree up to percent-level with the results obtained by solving the original system~\eqref{eq:Rnstart} and \eqref{eq:mustart} in the absence of LNV processes. 

\section{Approximate analytical solution describing the level crossing}

In this Appendix we present the approximate solution to the {leading order} right-handed neutrino number density evolution, Eq.~\eqref{eq:Rnstart0}, in the $B- \bar L$ symmetric limit. We introduce the notation
\begin{align}
	\Delta H = W_N + 3 x^2 r\,, \qquad \Gamma = \phi^{(0)} W_N \,,
\end{align}
{for the effective Hamiltonian term and the production term, respectively.}
After approximately diagonalising the pseudo-Dirac block, the equilibration matrix takes the form:
\begin{align}
W_N \approx
\frac{\pi^2}{144 \zeta(3)}\frac{M_0}{T_{EW}}|F_a|^2
\begin{pmatrix}
1 & 0 &  \epsilon^{\prime *}_a \\
	0 &  |\epsilon_a|^2 &   \epsilon_a \epsilon^{\prime *}_a\\
	\epsilon^\prime_a  &  \epsilon^\prime_a \epsilon_a^* &  |\epsilon^{\prime}_a|^2 
\end{pmatrix}\,.
\end{align} 
For $\Mu \ll |\Mu^{\prime 2}-1|$, the effective Hamiltonian term can be approximated by:
\begin{align}
	r = \frac{\pi^2}{108 \zeta(3)T_{\rm EW}^3} M_0
	\begin{pmatrix}
		0&0&0\\
		0&0&0\\
		0&0& \bar{M}^2 (\Mu^{\prime 2}-1)
	\end{pmatrix} + \mathcal{O}(\Mu)  \,.
\end{align}

Assuming that the off-diagonal correlations are either oscillating quickly, or overdamped, their mean value approaches:
\begin{align}
	R_{N\,ij} \approx \frac{\Delta H_{ii} - \Delta H_{jj} + \frac{i}{2} (\Gamma_{ii}+ \Gamma_{jj})}{(\Delta H_{ii} - \Delta H_{jj})^2 + (\Gamma_{ii}+\Gamma_{jj})^2/4}
	\left[ \Delta H_{ij} (R_{N\,ii}-R_{N\,jj}) + \frac{i \, \Gamma_{ij}}{2}  (R_{N\,ii} + R_{N\,jj}-2) \right] + \mathcal{O}(R_{N\,k,l}),
\end{align}
with $k\neq l \neq i \neq j$.
In principle, the Yukawa couplings $F_a$ can be large enough to cause early equilibration of the sterile neutrinos. In that case, the diagonals of the density matrix $R_{N}$ are approximately given by:
\begin{align}
	R_{N\,ss} &= 1+\frac{|\Delta H_{s3} + \frac{i}{2}\Gamma_{s3}|^2}{(\Delta H_{ss}- \Delta H_{33} )^2 + (\Gamma_{ss}/2)^2} (R_{N\,33}-1)+\mathcal{O}(\epsilon)^3\,,\\\notag
	R_{N\,ww} &= \mathcal{O}(x^2 \mu^2) + \mathcal{O}(\epsilon^2) \,,
\end{align}
where we have neglected the equilibration through mixing of the pseudo-Dirac pair.
Note that the equations are given in the interaction basis, where the subscript $\rm{s}$ corresponds to the strongly coupled state $\nu_{R{\rm s}}$ and $\rm{w}$ to the weakly coupled one $\nu_{R\rm{w}}$.
The number density of the heaviest right-handed neutrino is governed by the equation:
\begin{align}
	 \frac{d R_{N\,33}}{d x}= - (R_{N\,33}-1)& \left[ \Gamma_{33}  + \Gamma_{11} \frac{|\Delta H_{13}|^2- |\Gamma_{13}|^2}{(\Delta H_{11}- \Delta H_{33} )^2 + (\Gamma_{11}/2)^2} \right. +  \\ \nonumber
	 & \quad +
	 \left. (\Delta H_{11}-\Delta H_{33}) \frac{\Re\left(\Delta H_{13}\Gamma_{13}^*\right)}{(\Delta H_{11}- \Delta H_{33} )^2 + (\Gamma_{11}/2)^2}\right]  \, .
\end{align}

\bibliographystyle{JHEP}
\bibliography{ARS_refs,references_e,references_mu,references_tau,referencesInspirehep}{}

\end{document}